\documentclass[a4paper,usenatbib,fleqn]{mnras}
\usepackage[T1]{fontenc}
\usepackage{ae,aecompl}
\usepackage{pdflscape}
\usepackage{graphicx}
\usepackage{longtable}
\usepackage{lscape}
\usepackage{amssymb}
\usepackage{footnote}
\usepackage{subfigure}
\usepackage{amsmath}
\usepackage{txfonts}
\usepackage{url}
\usepackage{breakurl}
\usepackage{helvet}
\usepackage{textcomp,rotating}
\title{A multi-wavelength analysis of a collection of short-duration GRBs observed between 2012-2015}

\author[S. B. Pandey et al.]
{S. B. Pandey,$^{1}$\thanks{E-mail: shashi@aries.res.in, bbzhang@nju.edu.cn} 
Y. Hu,$^{2}$
A. J. Castro-Tirado,$^{2,3}$
A. S. Pozanenko,$^{12,13,36}$ 
R. S\'anchez-Ram\'irez,$^{2,38}$
\newauthor J. Gorosabel,\dag$^{2,4,5}$
S. Guziy,$^{2,6,7}$
M. Jelinek,$^{2,9}$
J. C. Tello,$^{2}$
S. Jeong,$^{2,8}$
S. R. Oates,$^{2,41}$
\newauthor B. -B. Zhang,$^{2,39,40}$
E. D. Mazaeva,$^{12}$
A. A. Volnova,$^{12}$
P. Yu. Minaev,$^{12}$
H.~J. van Eerten,$^{42}$
\newauthor M. D. Caballero-Garc\'ia,$^{9,2}$
D. P\'erez-Ram\'irez,\dag$^{10}$
M. Bremer,$^{11}$
J.-M. Winters,$^{11}$
I. H. Park,$^{8}$
\newauthor A. Nicuesa Guelbenzu,$^{14}$
S. Klose,$^{14}$
A. Moskvitin,$^{15}$
V. V. Sokolov,$^{15}$
E. Sonbas,$^{16}$
A. Ayala,$^{2}$
\newauthor J. Cepa,$^{17}$
N. Butler,$^{18}$
E. Troja,$^{19}$
A. M. Chernenko,$^{12}$
S. V. Molkov,$^{12}$
A. E. Volvach,$^{20}$
\newauthor R. Ya. Inasaridze,$^{21,43}$
Sh. A. Egamberdiyev,$^{22}$
O. Burkhonov,$^{22}$ 
I. V. Reva,$^{23}$
K. A. Polyakov,$^{24}$
\newauthor A. A. Matkin,$^{25}$
A. L. Ivanov,$^{26}$
I. Molotov,$^{37}$
T. Guver,$^{27}$
A. M. Watson,$^{44}$
A. Kutyrev,$^{19}$
\newauthor W. H. Lee,$^{44}$
O. Fox,$^{30}$
O. Littlejohns,$^{18}$
A. Cucchiara,$^{19}$
J. Gonzalez,$^{44}$
M. G. Richer,$^{28}$
\newauthor C. G. Rom\'an-Z\'u\~niga,$^{28}$
N. R. Tanvir,$^{29}$
J. S. Bloom,$^{30}$
J. X. Prochaska,$^{31}$
N. Gehrels,\dag$^{19}$
\newauthor H. Moseley,$^{19}$
J. A. de Diego,$^{44}$
E. Ram\'irez-Ruiz,$^{31}$
E. V. Klunko,$^{32}$
Y. Fan,$^{33}$ 
X. Zhao,$^{33}$
\newauthor J. Bai,$^{33}$ 
Ch. Wang,$^{33}$ 
Y. Xin,$^{33}$
Ch. Cui,$^{34}$
N. Tungalag,$^{35}$
Z.-K. Peng,$^{40}$
Amit Kumar,$^{1}$
\newauthor Rahul Gupta,$^{1}$
Amar Aryan,$^{1}$
Brajesh Kumar,$^{1}$
L. N. Volvach$^{20}$
G. P. Lamb$^{29}$
A. F. Valeev$^{15}$}

\begin{document}
\vspace{-5mm}
\date{Accepted ------------, Received ------------; in original form ------------}
\pubyear{2018}

\label{firstpage}
\pagerange{\pageref{firstpage}--\pageref{lastpage}}
\maketitle
\label{firstpage}
\vspace{-5mm}
\begin{abstract}
We investigate the prompt emission and the afterglow properties of short-duration gamma-ray burst (sGRB) 130603B and 
another eight sGRB events during 2012-2015, observed by several multi-wavelength facilities including the GTC 10.4\,m 
telescope. Prompt emission high energy data of the events were obtained by $INTEGRAL$-SPI-ACS, $Swift$-BAT and {\it Fermi}-GBM 
satellites. The prompt emission data by $INTEGRAL$ in the energy range of 0.1--10 MeV for sGRB 130603B, sGRB 140606A, 
sGRB 140930B, sGRB 141212A and sGRB 151228A do not show any signature of the extended emission or precursor activity and 
their spectral and temporal properties are similar to those seen in case of other short bursts. For sGRB 130603B, our new 
afterglow photometric data constraints the pre jet-break temporal decay  due to denser temporal coverage. For sGRB 130603B, 
the afterglow light curve, containing both our new as well as previously published photometric data is broadly consistent 
with the ISM afterglow model. Modeling of the host galaxies of sGRB 130603B and sGRB 141212A using the LePHARE software 
supports a scenario in which the environment of the burst is undergoing moderate star formation activity. From the inclusion 
of our late-time data for 8 other sGRBs we are able to: place tight constraints on the non-detection of the afterglow, 
host galaxy or any underlying ``kilonova'' emission. Our late-time afterglow observations of the sGRB 170817A/GW170817 
are also discussed and compared with the sub-set of sGRBs.
\end{abstract}


\section{Introduction}
Short-duration gamma-ray bursts (sGRBs) were originally classified using the $Konus$ catalog
\citep{Mazets1981} which preceded the wider realization that sGRBs likely are 
binary compact mergers \citep{Narayan1992, Nakar2007} based on various observed properties like
duration, fluence etc. as described in \citet{Kouveliotou1993, Bromberg2013}. 
During the era of the Neil Gehrels {\it Swift} observatory, arcsec X-ray Telescope (XRT) localizations enabled the discovery of 
the first afterglow of sGRB 050509B \citep{Gehrels2005, Castro-Tirado2005} and subsequently other observed features like extended emission (EE) at {\it Swift} Burst Alert Telescope (BAT) energies, temporally extended variable X-ray emission 
suggesting late time central engine activity either due to merger of two neutron stars 
(NS-NS) or a neutron star and a stellar mass black hole (NS-BH) as possible progenitors \citep{Eichler1989, Narayan1992, Usov1992, Zhang2001, Troja2007, Rowlinson2013, Avanzo2014, Gibson2017, Desai2018}. 
The physical nature of the EE, observed in some of the sGRBs,  is not yet resolved. It could be 
connected with the beginning of the afterglow phase \citep{Minaev2010}, the activity of a magnetar, 
formed during merger process \citet{Metgzer2008} or viewing angle effects \citep{Barkov2011}.
The prompt emission properties of sGRBs: such as relatively harder spectra (higher E$_{\rm peak}$) and 
nearly zero spectral lag \citep{Gehrels2006, Zhang2009}; discriminate sGRBs from long GRBs (lGRBs). 
sGRBs have also been speculated as a potential key to understand gravitational wave sources 
and the nucleosynthesis of elements over the history of the Universe
\citep{Berger2014, KumarZhang2015, Abbott2017a, Abbott2017b}. 

\noindent
More than 90 afterglows of sGRBs have been detected at various wavelengths\footnote
{http://www.astro.caltech.edu/grbox/grbox.php} exhibiting diverse 
properties \citep{LeeRamirez-Ruiz2007, Gehrels2009, Berger2014}. 
Afterglows of sGRBs are in general less luminous, less energetic and favor typically lower
circumburst densities than those seen in the case of 
lGRBs \citep{Kann2011, Nicuesa2012, Berger2014}. Despite intensive efforts, this leads to a lower 
detection rate for sGRBs:
$\sim$ 75 \% in X-rays, $\sim$ 33 \% in optica-NIR and only a handful in the radio \citep{Berger2014}. In comparison to 
long ones, sGRBs are observed to occur at over a lower and narrower redshift range 
 (z $\sim$ 0.1 - 1.5) and both early and late-type galaxies have been 
identified as hosts \citep{Fong2013}. 
Afterglow observations of sGRBs also indicate that these bursts have a range of jet-opening 
angles \citep{Burrows2007, Kann2011, Nicuesa2012, Fong2013, Zhang2015, Troja2016, Lamb2018, margutti2018} and have systematically 
larger radial offsets from the host galaxies \citep{Fong2013, Tunnicliffe2014} in turn supporting compact binary merger as possible progenitors 
\citep{Bloom2002, Zhang2007, Troja2008, Zhang2009, Salvaterra2010}. Optical afterglows of sGRBs are generally 
fainter in comparison to those observed in the case of lGRBs, implying the need for fast and deep afterglow 
observations using moderate to large size telescopes. 

\noindent
Study of sGRBs now extends beyond understanding just about their explosion mechanisms, progenitors and environments. 
These explosions are now key to improve our understanding about multi-messenger astronomy and to search for new compact binary 
mergers as gravitational wave (GW) sources. It has been proposed that during the compact binary merger process, 
radioactive decay of heavy elements could give rise to a supernova-like 
feature, termed  ``macronovae'' or ``kilonovae'' \citep{LiPaczynski1998, 
Kulkarni2005, Hotokezaka2013, Kasen2015} having a component of thermal emission caused by radioactive decay of elements through
r-process nucleosynthesis. 
So far, tentative ``kilonova'' like signatures have been identified in only a few cases including sGRB 050709 
\citep{Jin2016}, sGRB 060614 \citep{Yang2015}, 
sGRB 080503A \citep{Perley2009}, sGRB 130603B \citep{Hotokezaka2013, Tanvir2013}, sGRB 150101B \citep{Fong2016, Troja2018}, 
sGRB 160821B \citep{Kasliwal2017} and recently 
sGRB 170817A/GW170817/AT 2017gfo \citep{Abbott2017a, Abbott2017b}. Discovery of the ground-breaking event called 
sGRB 170817A/GW170817/AT 2017gfo has opened new windows in the understanding of 
gravitational waves: their electromagnetic counterparts \citep{Abbott2017a, Albert2017}, 
and their likely contribution to heavy element nucleosynthesis in the 
nearby Universe \citep{LattimerSchramm1974, Piran2013, Pian2017}. \\
Multi-wavelength observations of a larger sample of nearby sGRBs 
and ``kilonovae'' features like GW170817/sGRB 170817A/AT 2017gfo are crucial to establish whether 
compact binary mergers are the progenitors \citep{Kasen2015} for all such events \citep{Abbott2017a, Abbott2017b}
and to put a constraint on the electromagnetic counterparts and number density of gravitational 
wave sources in near future \citep{LiPaczynski1998, ShibataTaniguchi2011, Loeb2016}. 

\noindent
In this paper, we present results based on prompt emission data from {\em INTEGRAL}, {\em Swift}, {\it Fermi} and 
multi-wavelength follow-up afterglow observations of 9 sGRBs. The data-set were mostly not published yet and were 
observed by various different size optical and NIR ground-based telescopes including the 10.4\,m Gran Canarias Telescope (GTC). 
Observations of these 9 bursts including sGRB 170817A were collected during 2012-2018 as a part of a large 
multi-wavelength collaboration. Our analysis of new data for the sub-set of sGRBs mainly focused towards constraining 
prompt emission, afterglow and host galaxy properties and adding value towards known physics behind these cosmic explosions.
We also attempt to compare the observed properties of the sub-set of sGRBs with new class of less-studied but associated events
called ``Kilonovae''. The paper is organized as follows: in sections 2 and 3 we 
present our own temporal and spectral analysis of the afterglow and host galaxy data of GRB 130603B alongside the published 
ones, in section 4 and in Appendix ``A'' we discuss the results of prompt emission and multi-band afterglow observations 
of the other 8 sGRBs, and in section 5 we present late time GTC observations of sGRB 170817A/GW170817/AT 2017gfo and 
compare the observed properties with the sub-set of the bursts presently discussed. 
Finally, in section 6 we summarize the conclusions drawn from the analysis 
of all the sGRBs. 
In this paper, the notation $F_\nu(t) \propto t^{-\alpha} \nu^{-\beta}$
is used, where $\alpha$ is the flux temporal decay index and $\beta$ is the spectral index.
Throughout the paper, we use the standard cosmological 
parameters, $H_0=71~\rm km~s^{-1}Mpc^{-1}$, $\Omega_{M}=0.27$, $\Omega_{\Lambda}=0.73$. 

\section{sGRB 130603B, multi-wavelength observations}

sGRB 130603B was discovered on 2013 June 3 at 15:49:14 UT by 
{\it Swift-}BAT \citep{Barthelmy2013, Melandri2013}, and by 
$Konus-Wind$ \citep{Golenetskii2013}. The $\gamma$-ray 
light-curve of GRB 130603B consists of a single group of pulses with a duration of 
$T_{90}$ = 0.18$\pm$0.02 s (15--350 keV; \citealt{Barthelmy2013}). The 
$Konus-Wind$ fluence of the burst is (6.6$\pm$0.7)$\times10^{-6}\rm ~erg ~ cm^{-2}~$ 
(20 to $10^{4}$ keV), with a peak energy of 660$\pm$100 
keV \citep{Golenetskii2013}. The reported measured value of E$_{\rm iso,\gamma} \sim 2.1\times10^{51}$ erg, places the burst well above the E$_{\rm peak}$-E$_{\rm iso}$ locus for long GRBs in the Amati diagram 
\citep[][also Fig. 6]{Amati2008}. Such behavior is often observed for short bursts \citep{Minpoz2019}. 

\noindent
sGRB 130603B shows negligible spectral lag \citep{Norris2013}, typical for short bursts. 
Many authors \citep[e.g.][]{Hakkila2011, Minaev2014} have found a strong correlation between pulse 
duration and spectral lag: longer pulses have larger lags. The correlation is similar both for sGRBs 
and lGRBs. As sGRBs typically consist of shorter pulses than long ones, they have less 
significant spectral lags in general. GRB light curves often consist of several pulses including 
highly overlapping ones: spectral and temporal properties of individual pulses may be 
not adequately resolved \citep{Chernenko2011}. By performing spectral lag analysis via the superposition of several 
overlapping pulses, one can obtain an unpredictable result 
because each pulse has unique spectral and temporal properties \citep{Minaev2014}. As a result, one can find negligible or 
negative lag under certain conditions even if each pulse has a positive (but unique) lag 
\citep[for details see][]{Minaev2014}. sGRB 130603B consists of several very short and overlapped 
pulses, so, its negligible spectral lag may be connected with short duration of pulses while 
performing spectral lag analysis for superposition of several pulses.

\subsection{SPI-ACS {\em INTEGRAL} Observations}
sGRB 130603B was also triggered by the INTEGRAL Burst Alert System (IBAS) system operating with 
spectrometer for {\em INTEGRAL}- anti-coincidence system (SPI-ACS) (Fig. 1). 
SPI-ACS {\em INTEGRAL} has very high effective area (up to 0.3 $m^{2}$) in energy range $>$ 100 keV 
and stable background at timescales of hundreds of seconds \citep{Minaev2010}, which makes 
SPI-ACS a suitable instrument to study light curves of short hard GRBs and especially to search 
for weak signals from their precursors and EE components. The off-axis angle of 
sGRB 130603B to the SPI-ACS axis is 103 degrees, which is almost optimal for detection, 
making sGRB 130603B one of the brightest short bursts 
ever registered by SPI-ACS. Nevertheless we do not find statistically significant EE
in the SPI-ACS data (Inset in Fig. 1, in terms of peak flux at 50 ms time scale), which is in agreement with results obtained from {\it Swift-}BAT 
in the softer energy range of 15-150 keV \citep{Norris2013}. There is also no evidence for a 
precursor in SPI-ACS data during timescales from 0.01s up to 5s, in agreement with the previous results \citep{Troja2010, Minaev2017, Minpoz2018}.

\noindent
In \citet{ViganoMereghetti2009}, it was shown that one SPI-ACS count corresponds on average to $\sim 10^{-10}$ ~erg ~cm$^{-2}$ in the (75, 1000) keV range, for directions orthogonal to the satellite pointing axis. 
Using the conversion factor, we can roughly estimate the flux values in the (75, 1000) keV range for 
GRBs observed by SPI-ACS. The fluence estimation of sGRB 130603B in SPI-ACS is $\sim$ 31000 
counts or S$_{EE} \sim 3.1\times10^{-6}$ ~erg ~cm$^{-2}$ in the (75, 1000) keV range, which is in agreement 
with Konus-Wind observations \citep{Golenetskii2013}.
At a time scale of 50s, the upper limit on EE activity for sGRB 130603B is $\sim$ 7100 counts 
(S$_{EE} \sim$ 7$\times10^{-7}$ ~erg ~cm$^{-2}$) at the 3$\sigma$ significance level, the corresponding 
upper limit on precursor activity at a time scale of 1s, is $\sim$ 1000 counts (S$_{EE} \sim$ 10$^{-7}$ ~erg ~cm$^{-2}$), 
both are in the (75, 1000) keV range.

\begin{figure}
\centering
\includegraphics[width=\columnwidth]{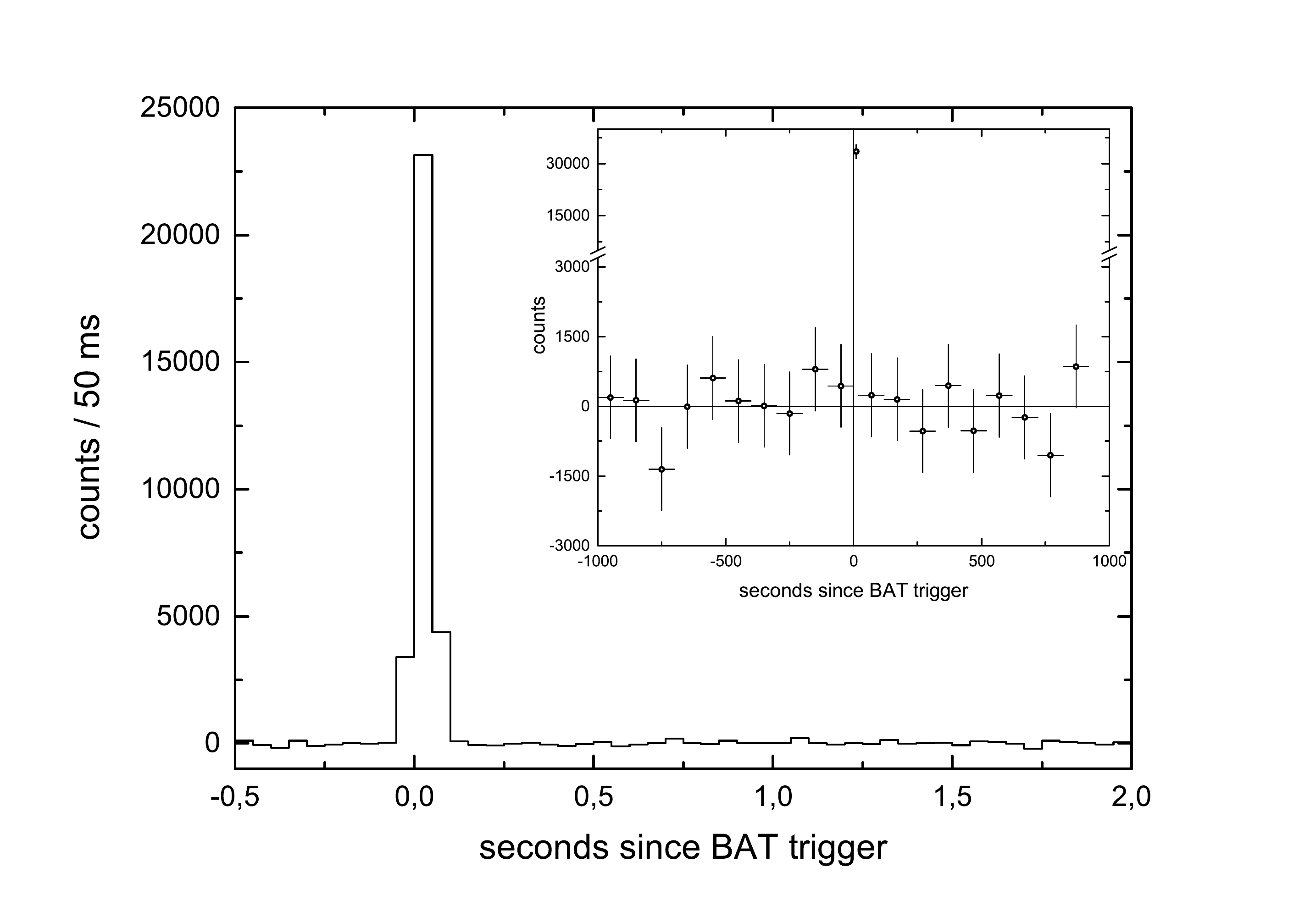}
\caption{\label{light} 
Background subtracted light curve of sGRB 130603B of {\em INTEGRAL} SPI-ACS
in the energy range 0.1-10 MeV with 50 ms time resolution. The x-axis 
shows time since BAT trigger. Inset: light curve with time resolution of 100 s.} 
\end{figure}
  
\subsection{Optical-IR photometric Observations}
As a part of this collaboration, photometric observations of the optical-IR afterglow and the host galaxy were 
performed using several facilities worldwide, including 1.0\,m telescope at the Tubitak National
Observatory (Antalya, Turkey); the 1.5\,m telescope at Observatorio de Sierra Nevada 
(Granada, Spain); the AS-32 0.7\,m telescope at Abastumani Astrophysical Observatory Georgia; the Reionization 
And Transients Infra-Red  RATIR camera at 
the 1.5\,m telescope of the San Pedro Martir observatory; the 2.0\,m Liverpool telescope at La Palma;
AZT-22 1.5\,m at the Maidanak observatory Uzbekistan; the Centro Astron\'omico Hispano-Alem\'an (CAHA) 3.5\,m located 
in Almeria (Spain); the newly 
commissioned 3.6\,m Devasthal Optical Telescope (DOT) at 
Aryabhatta Research Institute of Observational Sciences (ARIES) Nainital, India 
and with the 10.4\,m Gran Telescopio Canarias (GTC), located at the observatory of Roque de los 
Muchachos in La Palma (Canary Islands, Spain), equipped with the Optical System for Imaging and 
low-Intermediate-Resolution Integrated Spectroscopy (OSIRIS) instrument.
Our observations by the 1.0\,m telescope at the Tubitak, starting $\sim$ 0.122d after the 
burst are the earliest reported ground-based observations so far for sGRB 130603B. 
All optical-NIR data were processed using DAOPHOT software of NOAO's \textsc{iraf} 
package\footnote{http://iraf.noao.edu/}, a general purpose software system for the reduction 
and analysis of astronomical data.
The photometry was performed in comparison to nearby standard stars and image subtraction was 
applied whenever it was required to subtract the host galaxy contribution as exaplained in \citet{AlardLupton1998}. 
The unfiltered observations made with the AbAO AS-32 telescope 
have been considered equivalent to $r$-band as the quantum efficiency of the
detector is at a maximum around $r$-band frequencies. The final AB magnitudes of the afterglow 
and the host galaxy in different pass-bands as a part of the present analysis are listed in Table 1.

\normalsize
\begin{table}
\caption{Broad-band optical-IR photometric observations of
the GRB 130603B afterglow and its host galaxy (h) presented in the AB-magnitude system. 
The values are not corrected for extinction and are tabulated in order of time in days (d) 
since the burst. The quoted values of limiting magnitude are 3$\sigma$.}
\begin{center}
\scriptsize
\begin{tabular}{cccll} \hline \hline
t-t0,mid(d)&exp(s)&Afterglow/&pass-band&Telescopes\\
&  &      Host magnitudes   &   &        \\
\hline
0.1222&150$\times$10&20.15$\pm$0.17&R${_c}$&Tubitak 1.0\,m\\
0.1959&300$\times$10&21.37$\pm$0.25&$clear$&AS-32 0.7\,m\\
0.2024&300$\times$4&21.10$\pm$0.27&I${_c}$&OSN 1.5\,m \\
0.3360&50&21.29$\pm$0.02&$i$&GTC 10.4\,m \\
0.5196&3020.0&22.12$\pm$0.81&Y&RATIR 1.5\,m \\
0.5196&3020.0&20.37$\pm$0.28&H&RATIR 1.5\,m \\
0.5347&2818.0&21.64$\pm$0.34&Z&RATIR 1.5\,m \\
0.5347&2818.0&20.94$\pm$0.38&J&RATIR 1.5\,m \\
0.5405&6960.0&22.30$\pm$0.20&$r$&RATIR 1.5\,m \\
0.5405&6960.0&21.98$\pm$0.20&$i$&RATIR 1.5\,m \\
1.1141&150$\times$2+200$\times$8&21.34$\pm$0.50&R${_c}$&Tubitak 1.0\,m\\
1.1160&180$\times$14& $>$ 22.64&$clear$&AS-32 0.7\,m \\
2.0937&180$\times$10& $>$ 22.92&R${_c}$&Maidanak 1.5\,m\\
2.1489&200$\times$5&$>$ 21.14&R${_c}$&Tubitak 1.0\,m \\
2.2803&300$\times$5&20.69$\pm$0.15 (h)&I${_c}$&OSN 1.5\,m\\
5.1143&180$\times$23&$>$ 22.56&$clear$&AS-32 0.7\,m \\
16.2691&300$\times$10&20.69$\pm$0.06 (h)&$i$&LT 2.0\,m \\
19.2650&60$\times$15&19.69$\pm$0.13 (h)&K${_s}$&CAHA 3.5\,m \\
19.2323&60$\times$15&20.06$\pm$0.09 (h)&J&CAHA 3.5\,m \\
19.2481&60$\times$15&19.68$\pm$0.13 (h)&H&CAHA 3.5\,m \\
19.2155&60$\times$15&20.11$\pm$0.07 (h)&Z&CAHA 3.5\,m \\
32.2411&50$\times$4&22.01$\pm$0.03 (h)&$g$&GTC 10.4\,m \\
32.2471&50$\times$4&20.97$\pm$0.01 (h)&$r$&GTC 10.4\,m    \\
32.2511&50$\times$4&20.65$\pm$0.02 (h)&$i$&GTC 10.4\,m \\
35.5168&469.8&20.88$\pm$0.41 (h)&Y&RATIR 1.5\,m \\
35.5168&469.8&20.84$\pm$0.30 (h)&H&RATIR 1.5\,m \\
35.5168&335.6&20.39$\pm$0.19 (h)&Z&RATIR 1.5\,m \\
35.5168&335.6&20.49$\pm$0.43 (h)&J&RATIR 1.5\,m \\
35.5162&960.0&21.26$\pm$0.12 (h)&$r$&RATIR 1.5\,m \\
35.5162&960.0&20.79$\pm$0.09 (h)&$i$&RATIR 1.5\,m \\
1387.84&300.0$\times$2&22.13$\pm$0.05 (h)&$B$& 3.6\,m DOT\\
1387.86&300.0$\times$2&20.72$\pm$0.02 (h)&R${_c}$& 3.6\,m DOT\\
\hline
\hline
\end{tabular}
\end{center}
 \end{table}

\subsection{Spectroscopic Observations}

A spectroscopic redshift at the location of the afterglow was obtained by 
several groups including \citet{Xu2013}, 
\citet{Foley2013}, \citet{deUgartePostigo2013} and \citet{Cucchiara2013a}. 
As a part of the present study, spectroscopic observations were  
performed to measure the redshift of sGRB 130603B independently and are 
reported in \citet{Sanchez-Ramrez2013}. 

\noindent
We obtained optical spectra with the GTC(+OSIRIS) starting at 23:58\,h. Observations
consisted of two 450\,s exposures, one with each of the R1000B and R500R grisms, using a
slit of width 1.2 arcsec. Data reduction was performed using standard
routines from the Image Reduction and Analysis Facility (IRAF). 
The afterglow spectrum shows Ca II in absorption, and we detect a significant
contribution from the underlying host galaxy (eg. [OII], [OIII], H-beta and
H-alpha emission lines about 1'' offset), together implying a redshift of z
= 0.356$\pm$0.002, consistent with the values provided by 
\citet{deUgartePostigo2013} and \citet{Foley2013}. The reduced spectrum obtained at the 
location of the afterglow along with the lines identified are shown in Fig. 2. 
Using our redshift value and the fluence published by
\citet{Golenetskii2013}, the isotropic-equivalent gamma-ray energy is 
E$_{\rm iso,\gamma} \sim 2.1\times10^{51}$ erg (20 to $10^4$ keV, rest-frame).

\begin{figure}
\centering
\includegraphics[height=7.5cm,width=5.0cm,angle=90]{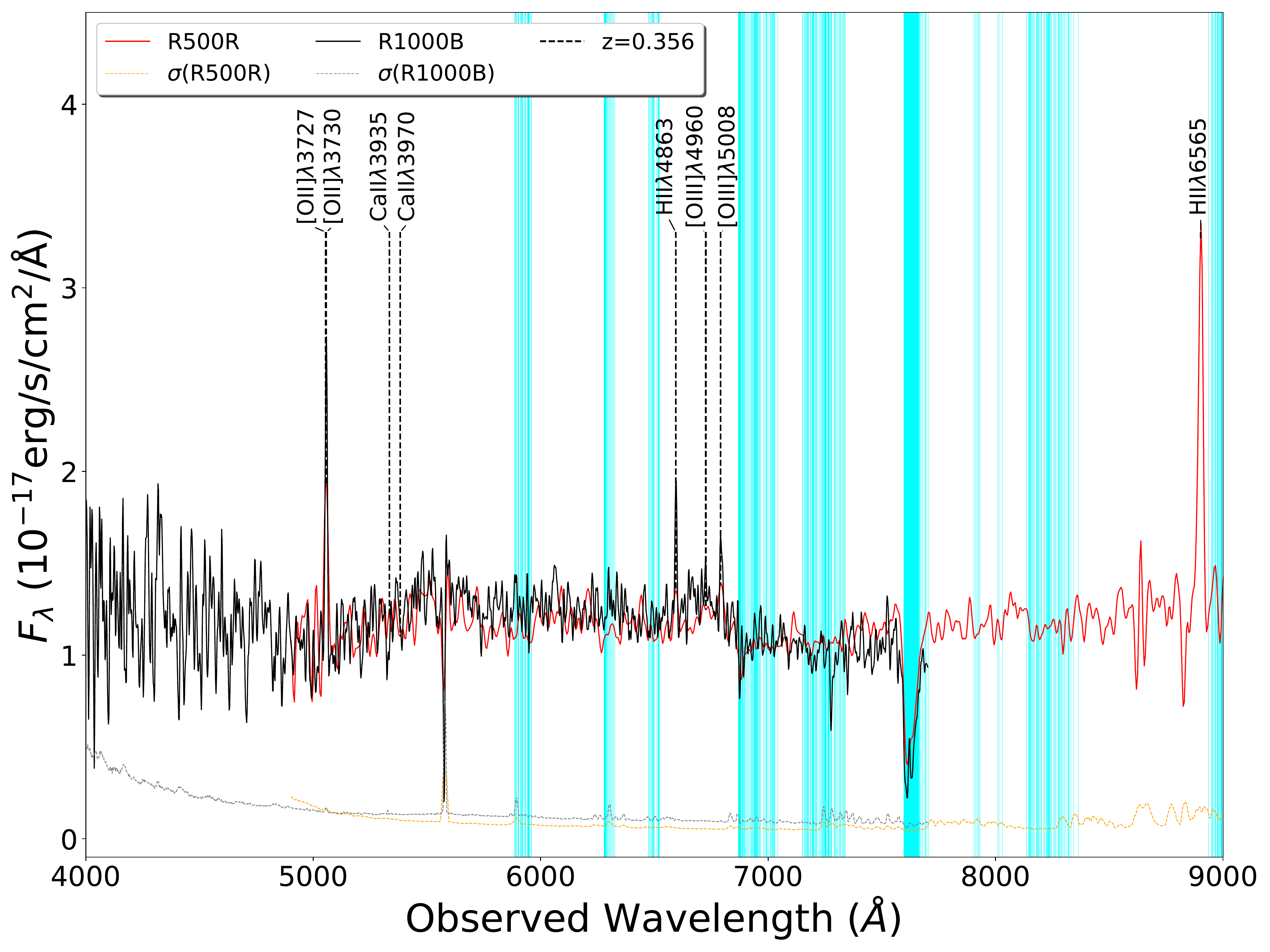}
\caption{\label{light} Spectroscopic observations of the sGRB 130603B at the location of the 
afterglow taken by the 10.4\,m GTC
(+OSIRIS) using grisms R1000B and R500R starting $\sim$ 8 hours after the burst 
\citep{Sanchez-Ramrez2013}. Telluric absorption bands are marked as cyan.}
\end{figure}

\begin{table}
\caption{Millimeter wave observations of the sGRB 130603B,  sGRB 140606A, sGRB 140622A and
sGRB 140903A (1-$\sigma$ upper limits) afterglows as 
observed by Plateau de Bure Interferometer (PdBI) and centimeter wave observations using 
RT-22 in Crimea. }

\begin{center}
\scriptsize
\begin{tabular}{ccccll} \hline \hline
Start & end & center & frequency & Flux & Telescopes \\
time&time&t-t0(d)&(GHz)&center (mJy)  & \\   \hline
\hline
&  &      {\bf sGRB 130603B}  &   &        \\
2013 June 03.844&03.926&03.901&86.743&+0.051$\pm$0.120 & PdBI \\
2013 June 04.826&03.908&04.867&86.743&-0.307$\pm$0.095  & PdBI\\
2013 June 12.721&12.828&12.775&86.743&-0.043$\pm$0.073  & PdBI\\
\hline
2013 June 04.730&04.801&04.765&36.0&1.6$\pm$0.9 & RT-22 \\
2013 June 05.703&05.732&05.717&36.0&1.9$\pm$1.2  & RT-22\\
2013 June 05.710&05.785&05.747&36.0&2.6$\pm$0.9  & RT-22\\
\hline
&  &      {\bf sGRB 140606A}  &   &        \\
2014 June 14.039&14.099&14.069&86.743&0.331$\pm$0.187  & PdBI\\
2014 June 15.039&15.099&15.069&86.743&-0.592$\pm$0.214  & PdBI\\
\hline
&  &      {\bf sGRB 140622A}  &   &        \\
2014 June 26.050&26.108&0.079&86.243&-0.376$\pm$0.123  & PdBI\\
\hline
&  &      {\bf sGRB 140903A}  &   &        \\
2014 Sep 05.617&05.705&02.661&86.743&0.120$\pm$0.130  & PdBI\\
\hline
\hline
\end{tabular}
\end{center}
\end{table}
\normalsize

\subsection{mm-wavelength Observations}

The afterglow of sGRB 130603B was observed with the Plateau
de Bure Interferometer \citep{Guilloteau1992}, one of the largest  
observatory in the Northern Hemisphere operating at 
millimetre wavelengths (1, 2 and 3 mm). Observations were performed
in a four-antenna extended configuration for the first epoch whereas a five-antenna 
configuration on the consecutive dates as listed in
Table 2. The data reduction was done with the standard CLIC and MAPPING 
software distributed by the Grenoble GILDAS group. Flux calibration 
includes a correction for atmospheric decorrelation which has been 
determined with a UV plane point source fit to the phase calibration quasar
1156+295. The carbon star MWC349 was used as the
primary flux calibrator due to its well-known millimeter
spectral properties (see e.g. \citealt{Schwarz1980}). The burst location was also followed-up 
using the RT-22 radio telescope of CrAO (Crimea) at 36 GHz and the data reduced using 
the standard software routines \citep{Villata2006} and used modulated radiometers in combination with the registration 
regime ``ON-ON'' for collecting data from the telescope \citep{Nesterov2000}. 
The upper limits based on these 
observations are also given in Table 2.
As a part of the present analysis, upper limits (1-$\sigma$) based on IRAM Plateau de Bure Interferometer 
observations of sGRB 140606A, sGRB 140622A and sGRB 140903A 
using the carbon star MWC349 as the primary flux calibrator are also tabulated in Table 2. \\

\noindent
Observations at mm-wavelengths are very important as they suffer negligible absorption or 
interstellar scintillation effects,  so sGRBs at high redshifts or highly-extinguished bursts 
could be observed. It is expected that emission at mm-wavelengths is normally above the 
self-absorption frequency and lies around peak of the GRB synchrotron spectrum, allowing to probe 
for possible reverse shock emission at early epochs and to constrain afterglow models 
observed recently in case of many lGRBs \citep{deUgarte2012, Perley2014}. 
  
\begin{figure}
\centering
\includegraphics[width=\columnwidth]{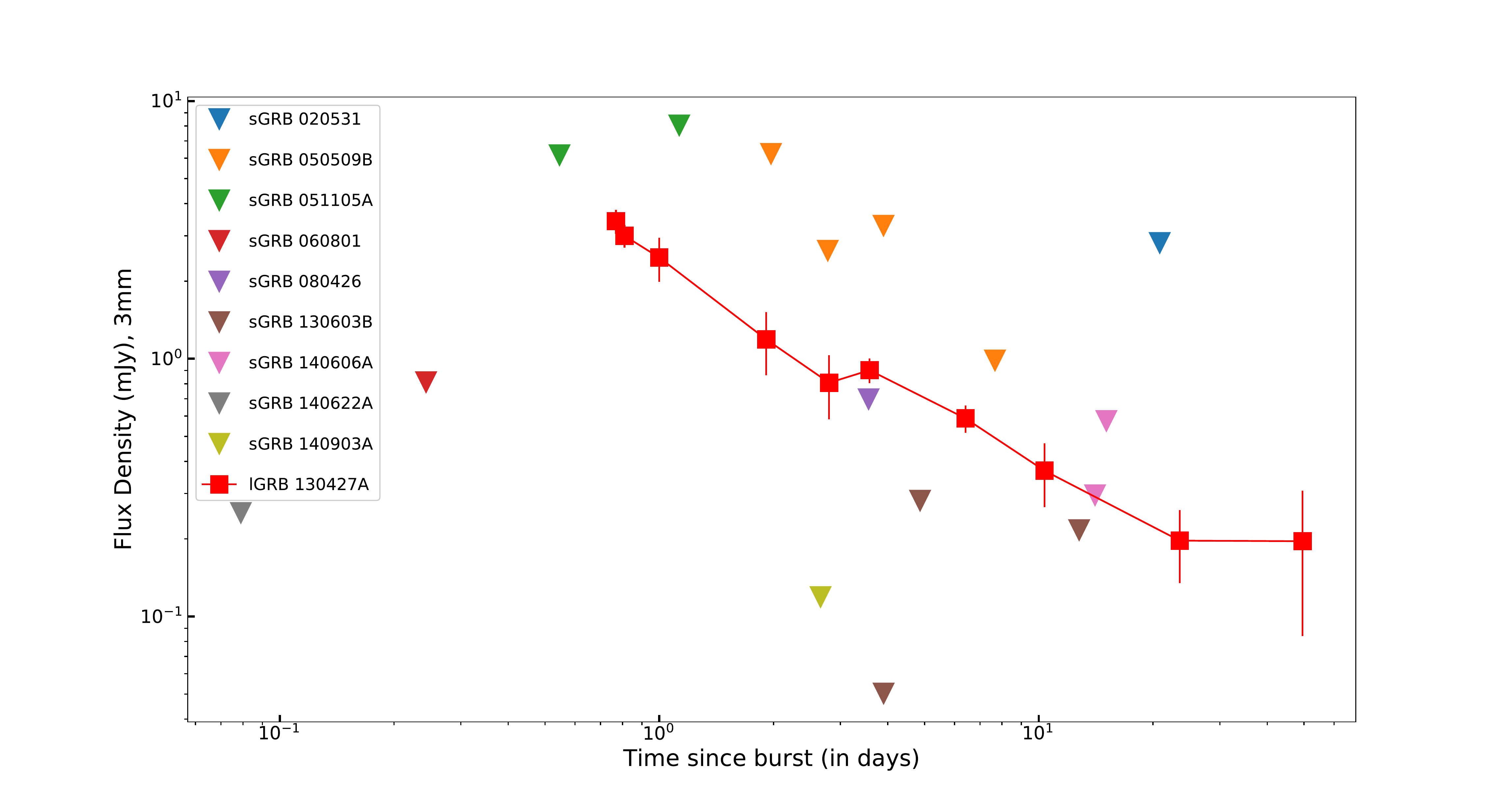}
\caption{\label{light} Comparison of the 3-mm afterglow light curve of nearby lGRB 130427A \citep{Perley2014} 
to the present set of mm-wavelength upper-limits (1-$\sigma$) of 4 sGRBs (from Table 2) along with 
another set of upper-limits of 5 sGRBs as discussed in \citet{Castro2019} placed at a common redhsift of z = 0.34.} 
\end{figure}

\noindent
In Fig. 3, observed mm-wavelength upper-limits of four sGRBs presented in Table 2 were
plotted along with previous observations of another 5 sGRBs (namely sGRB 020531, sGRB 050509B, sGRB 051105A, sGRB 060801 and 
sGRB 080426, data taken from \citet{Castro2019}) 
and were compared with the afterglow lightcurve of a well-known nearby and bright
lGRB 130427A observed at 3-mm \citep{Perley2014}. It is clear from Fig. 3
that using PdBI, we have been able to observe 9 sGRBs so far but none was detected at
mm-wavelengths in contrast with lGRBs which have been detected in many cases constraining various
physical models \citep{deUgarte2012, Perley2014}. Out of these 9 sGRBs, only sGRB 130603B \citep{Fong2014} 
and sGRB 140903A \citep{Troja2016} were detected at VLA radio frequencies so far. However, as discussed further 
in this work, the observed 3-mm PdBI 1-$\sigma$ upper-limits for these two bursts are consistent with 
those predicted by the forward shock afterglow models.  
The gamma-ray fluence and observed X-ray flux values for these 9 sGRBs are similar to those
observed in case of other sGRBs. Non-detections of these nine sGRBs at 3-mm in the last decade
using PdBI and other mm-wavelength facilities globally are helpful to constrain underlying
physics behind these energetic sources and demand for more sensitive and deeper follow-up observations. 
 
\section{Properties of sGRB 130603B}

\subsection{Afterglow light-curves and comparison to models}

Fig. 4 shows the $r$ and $i$ pass-band light curves of the sGRB 130603B 
afterglow including data from the present analysis and those published 
in the literature \citep{deUgartePostigo2014, Tanvir2013, Cucchiara2013b, 
Berger2013}. To plot the light-curves along with those published in the literature, the data were 
scaled to respective AB magnitudes in SDSS $r$ and $i$ bands (see Fig. 4).
The $R_c$ band data taken at $\sim$ 0.122d comprise of the earliest 
reported ground-based detection and the remaining data fill the temporal gap 
in the light curve for this interesting short-duration burst. 
From the present analysis, the number of new data points both in $r$ and $i$ bands are 
four each spread up to $\sim$ 2.3d post-burst. Careful image-subtraction and calibration of the afterglow data 
$<$ 0.23d post-burst indicates possible deviations from smooth power-law 
behavior during the first few hours. 

\noindent
To determine the temporal flux decay slopes and the break time, we fitted an 
empirical function representing a broken power-law, 
$F_\nu = A[(t/t_b)^{s\alpha_1} +(t/t_b)^{s\alpha_2}]^{-1/s}$  
\citep{Beuermann1999} to the $r$ band combined light curve. The quantities $\alpha_1$ and 
$\alpha_2$ are asymptotic power-law flux decay slopes at early and late 
times with $\alpha_1 < \alpha_2$. The parameter $s > 0$ controls the sharpness of the 
break and $t_b$ is the break time. The best fit of this broken power-law 
function to the $r$ band data including the very first data point taken at 
$\sim$ 0.122d gives : $\alpha_1 = 0.81\pm0.14$; 
$\alpha_2 = 2.75\pm0.28$ and $t_b = 0.41\pm0.04$ with $\tilde\chi^2$$/dof = 2.22$ 
for a value of the smoothing parameter $s=4$. The values of $t_b$ and $\alpha_2$
are similar to those derived by \citet{Fong2014}. 
Although the data from {\it Swift} XRT is consistent with a break occurring around 0.3 days, the later XMM-Newton 
observations suggest no turnover at X-ray frequencies and a continuing power law instead 
(this ``X-ray excess'' is also discussed by \citet{Fong2014}).
The present analysis also helped to constrain the value of $\alpha_1$ using a single band light curve and found to be
shallower in comparison to that derived by \citet{Fong2014}.
  
\noindent
The present data set has also been used to constrain the spectral energy 
distribution (SED) of the afterglow. The RATIR data taken simultaneously at $\sim$0.52d post-burst
(see Table 1), require an optical-NIR spectral index $\beta_{opt} \sim$ 0.7 once corrected 
for Galactic and considerable host extinction, 
similar to those measured by \citet{deUgartePostigo2014} at $\sim$ 0.35d and by 
\citet{Fong2014} at $\sim$ 0.6d post-burst. The optical-NIR spectral index, together with the 
published value of the XRT spectral index $\beta_X = 1.2\pm0.1$ are consistent with  
$\Delta \beta = \beta_X - \beta_{opt}$ = 0.5, as expected in the case of a slow-cooling synchrotron spectrum
\citep{Sari1998} where the optical and XRT 
frequencies lie in two different spectral regimes. 

\noindent
Additionally, the derived values of the temporal slope $\alpha_1$ and the spectral slope 
$\beta_{opt}$ 
above are consistent with the closure relation $\beta=3\alpha/2$ in the case of adiabatic 
deceleration in the interstellar medium {\em ISM} afterglow model for the spectral regime 
$\nu_m < \nu < \nu_c$, where $\nu_m$ is the break frequency corresponding to the 
minimum electron energy and $\nu_c$ is the cooling break frequency.
The temporal flux decay index $\alpha_2 = 2.75\pm0.28$, the break-time $t_b = 0.41\pm0.04$ 
and estimated slopes of the SEDs using the optical-NIR and XRT data are broadly consistent 
with the scenario described by \citet{Rhoads1999} where the edge of the relativistic outflow causes a 
steepening (jet-break) in the observed light curve by $t^{-p}$ \citep{Sari1999}, where $p$ is 
the electron energy index.
Also, for the observed XRT frequencies which lie above $\nu_c$, the temporal and spectral 
indices are consistent with the predictions made by the {\em ISM} model in case of the 
adiabatic deceleration for the data up to one day post-burst \citep{deUgartePostigo2014, Fong2014}. 

\noindent
Present afterglow data has made it possible to construct a 
single band afterglow light-curve and do the temporal fitting to derive parameters like 
temporal indices and jet-break time. The optical afterglow data in $r$ and $i$ bands from the present 
analysis has allowed to construct a better-sampled light-curve of the sGRB 130603B and to constrain 
the value of the pre jet-break temporal decay index $\alpha_1$ for the first time using data from 
a single band. This overall analysis supports the scenario that the
observed steepening in the optical light-curves is a jet-break as predicted 
theoretically by \citet{Sari1999} and \citet{Rhoads1999}. However, the 
observed X-ray excess emission \citep{Fong2014} for epochs $>$ 1d are not supported by the afterglow model.

\begin{figure}
\centering
\includegraphics[width=\columnwidth]{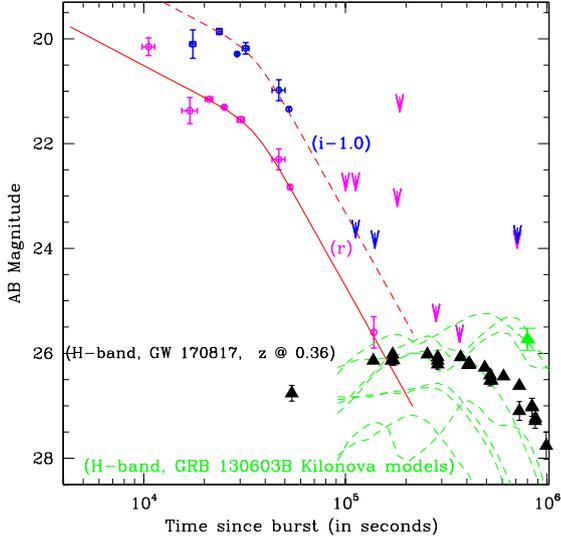}
\caption{\label{light} Afterglow optical $r$ (pink) and $i$ (blue) pass-band light-curves of the sGRB 130603B. The solid red curves are the best-fit broken power-law model to the $r$-band light curves as described above. The red dashed line is the model over-plotted on the $i$-band light curve to guide eyes. The green triangle in the right bottom corner is the single point detection of the underlying ``kilonova'' detection as described in \citet{Tanvir2013}. The green dashed lines are the H-band ``kilonova'' models at the redshift of $\sim$ 0.36 as taken from \citet{Tanaka2014}. The black triangles are the H-band light curve (at redshift z = 0.36) of the 
electromagnetic counterpart of the recently discovered GW170817 (sGRB 170817A/AT 2017gfo) for comparisons as compiled in \citet{Villar2017a}.}
\end{figure}

\subsection{Afterglow SED at the epoch of mm observations}

Based on the present analysis and using the afterglow data in X-ray, $r$, $i$ 
bands and the results published by \citet{deUgartePostigo2014} and 
\citet{Fong2014}, an afterglow SED was constructed for the epoch of 
our earliest millimeter observations i.e. 0.22d after the burst (see Fig. 5). 
We first built a time-sliced X-ray spectrum from the Leicester
XRT webpages \footnote{http://www.swift.ac.uk/xrt\_spectra/}, extracting data in the 
range 10ks - 18ks after the trigger. This tool provides the appropriate spectral and response
files that are compatible for use with the spectral fitting package
{\textsc XSPEC}. The source spectral file was normalized so that it has the same count rate as 
a single epoch spectrum measured at 0.22d (see \citet{Schady2010} for details). For the optical 
data, we created appropriate spectral and response files for each filter. The flux values at 0.22d for 
each spectral file were determined from an extrapolation/interpolation the data between 10ks and 30ks 
by fitting a powerlaw and fixing the slope as 0.81. This is the decay index found for the first
segment of the broken-powerlaw fit to the r-band data. The optical errors were estimated by taking 
the average error of the data between 10 and 30ks and adding a 5\% systematic error in quadrature.

\begin{figure}
\centering
\includegraphics[height=7.0cm,width=7.0cm]{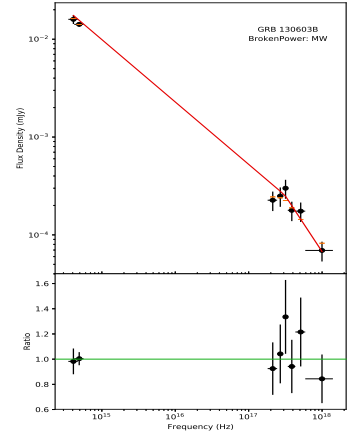}
\caption{\label{light} X-ray and optical SED of sGRB 130603B at the epoch of first millimeter
observations i.e. 0.22d after the burst. We plot the best-fitting absorption and extinction corrected 
spectral model (solid red lines, broken power-law model), as well as the host galaxy absorbed and 
extinguished spectral model (orange dash lines) and the data (black circles) using the method described 
in \citet{Schady2010}.}
\end{figure}

\noindent
The SEDs were fitted using {\textsc XSPEC}, following the procedure outlined in \citet{Schady2010, 
Schady2007}. We fit two different models, a power-law and broken power-law, which include Galactic 
and host galaxy absorption and extinction components (phabs, zphabs and zdust). The best-fit results 
obtained using the procedure mentioned above are plotted in Fig. 5 which supports broken power-law model
for Milky-way (MW) type of host extinction. Values of the best-fit borken power-law model and MW type
of host extinction are consistent with those derived by \citet{deUgartePostigo2014}. Assuming $\nu_m$
around mm-wavelengths, 86.7 GHz upper limits of the sGRB 130603B at 0.22d post-burst (see Table 2) 
are also consistent with the extrapolated modeled flux values.

\subsection{Broad-band modeling of sGRB 130603B afterglow}

The multi-band afterglow data of sGRB 130603B discussed above along with those published in \citet{Fong2014} were 
used to fit numerical-simulation-based model to constrain physical parameters of the jetted emission as described in 
\citet{Zhang2015}. The numerical modeling \citep{Zhang2015} calculates the flux density at any
frequency and observer time. The Monte Carlo method is used to determine the best parameter values (i.e.,
with the smallest $\chi^2$ value) utilizing the MultiNest algorithm from \citet{Feroz2009}. The optical-NIR data were 
corrected for the Galactic and host extinction values 
as constrained in \citet{Fong2014}. The XRT data was also corrected for absorption effects. Based on the literature,
it was decided to utilize the data 1000s after the burst for the modeling to avoid possible prompt emission effects
at early epochs as described in \citet{Zhang2015}.   

\begin{figure}
\centering
\includegraphics[width=\columnwidth]{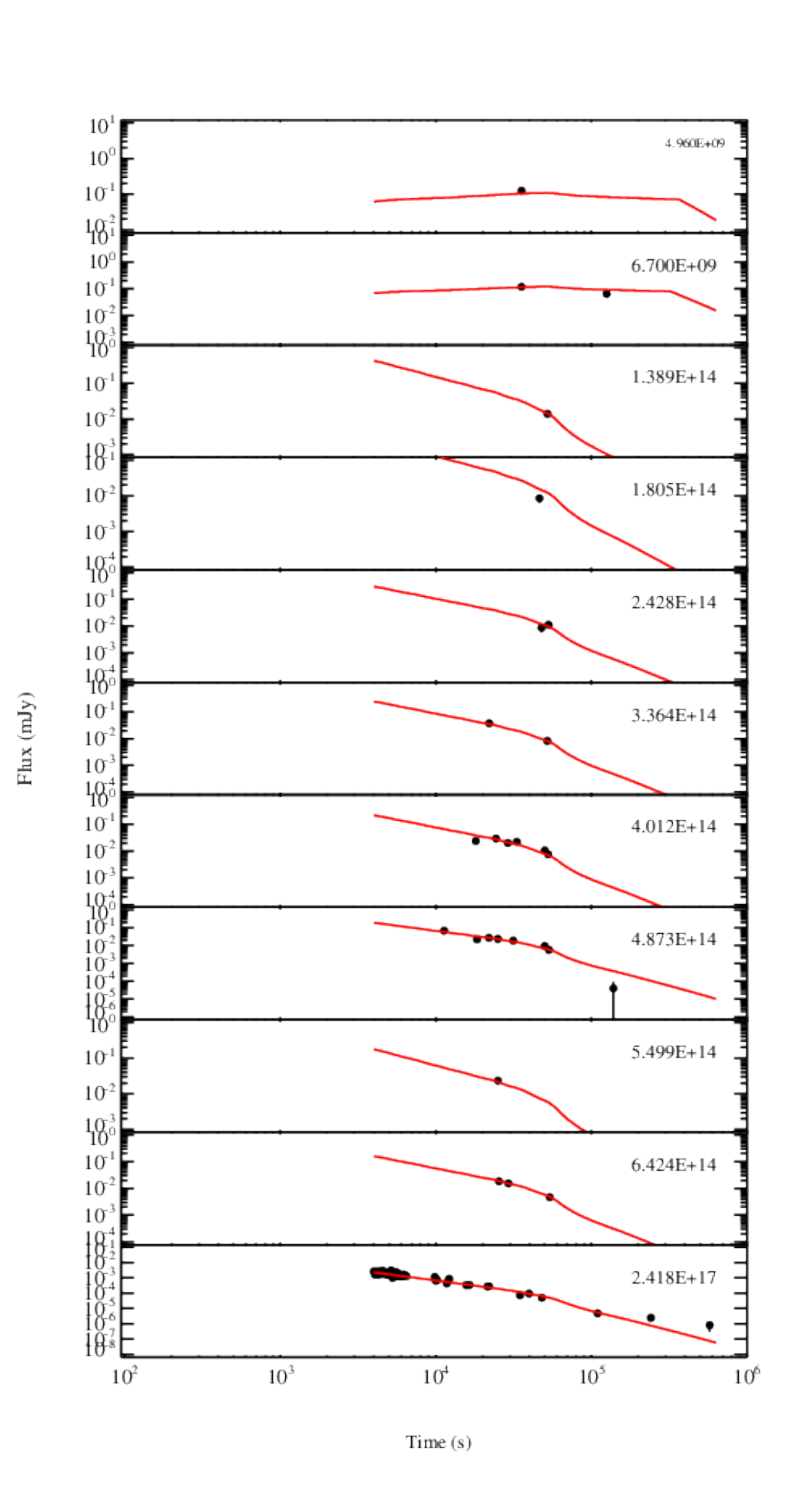}
\caption{\label{light} 
The best fit modeled multi-band light-curves determined from the numerical simulations as described 
above \citep{Zhang2015}. The corresponding
frequency is marked on the right corner in each panel in unit of Hz.
The x-axis is the time since trigger in units of seconds. The observed
flux density of each instrument is on the y-axis in units of mJy. All
data were corrected for MW and host galaxy absorption and extinction
effects before modeling. Red solid lines represent the modeled light
curves.} 
\end{figure}

\begin{figure}
\centering
\includegraphics[width=\columnwidth]{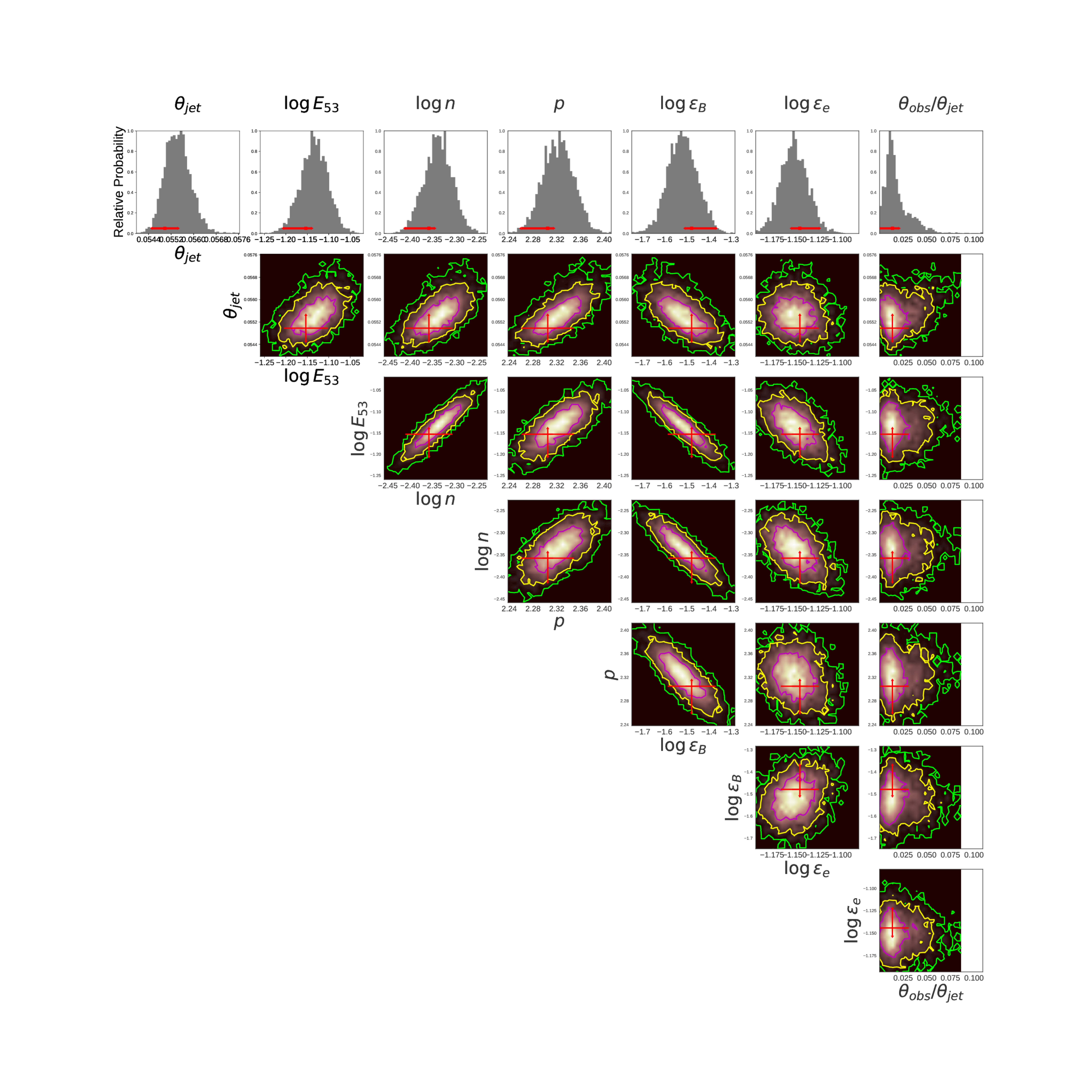}
\caption{\label{light} 
Triangle plot of the Monte Carlo fitting to our simulation-based model as described above \citep{Zhang2015}. It shows the posterior distribution and the correlation between the
parameters.} 
\end{figure}

\noindent
Using the model and initial guess values, following set of parameter values were determined: the blast wave total energy 
E$_{\rm iso,53}$ (in the units of $10^{53}$ ergs), the ambient number density $n$, the electron energy density fraction $\epsilon_e$ 
, the magnetic field energy density fraction $\epsilon_B$, the electron energy index $p$ and values of 
jet opening angle $\theta_{jet}$ and the observed angle $\theta_{obs}$. The best-fit light-curves obtained at 
different wavelengths are plotted in Fig. 6, the Monte Carlo parameter distributions are plotted in Fig. 7 and 
the resulting best-fit parameters and their uncertainties are listed 
in Table 3. A cross-check using an updated version of the {\it scalefit} package \citep{vanEerten2012,Ryan2015}, 
produces a similar jet opening angle and inferred energy.

\begin{table}
\caption{{\bf The best-fit parameters of the numerical simulation \citep{Zhang2015} to the
multi-wavelength afterglow data of sGRB 130603B.}}
\begin{center}
\scriptsize
\begin{tabular}{cccl} \hline 
Parameters & Value (-/+) \\ \hline
$p$   & $2.31^{-0.01}_{+0.04}$ \\ \\

log$n$& $-2.36^{-0.01}_{+0.05}$\\ \\

log$\epsilon_e$&$-1.14^{-0.02}_{+0.01}$ \\ \\

log$\epsilon_B$&$-1.47^{-0.11}_{+0.03}$ \\ \\

logE$_{\rm iso,53}$&$-1.15^{-0.01}_{+0.05}$ \\ \\

$\theta_{jet}$ &$0.055^{-0.001}_{+0.001}$  \\ \\

$\theta_{obs}/\theta_{jet}$ &$0.014^{-0.06}_{+0.017}$ \\
\hline
\end{tabular}
\end{center}
\end{table}
\normalsize

\noindent
Using the new data set discussed in this work, derived values of the physical parameters using present 
modeling method (Table 3)
are constrained better than those reported by \citet{Fong2014}. The derived value of observed jet opening angle, 
$\theta_{obs}$ is $\sim$ 3.2 degrees. This value of $\theta_{jet}$ gives rise to the beaming corrected 
E$_{\rm iso,53}$ is $\sim 1.4\times10^{49}$ erg. It is also clear from the present modeling that 
the best-fit model was unable to reproduce the very late time X-ray emission observed
in case of sGRB 130603B as noticed by using Chandra observations \citet{Fong2014}. 
It is also noted that 
values of the isotropic-equivalent gamma-ray energy is E$_{\rm iso}$ and the blast wave energy 
E$_{\rm iso,\gamma}$ are comparable, which in turn indicates the GRB radiative efficiency 
$\eta_{\gamma}$ to be $\sim$ 23\% (with an uncertainty of $\sim$ 4\%), one of the highest among the known sample of sGRBs \citep{Lloyd2004, Wang2015}. 

\subsection{sGRB 130603B and ``kilonovae'' connection}

The ``kilonova'' or ``macronova'' events are electromagnetic transients powered by the radioactive 
decay of r-process elements synthesized in dynamical ejecta, and in the accretion disk winds
during compact binary mergers where at least one component is a neutron star 
\citep{LiPaczynski1998, Kulkarni2005, Rosswog2005}. Compact binary mergers are also expected to be
sources of gravitational waves \citep{MetgzerBerger2012, TanakaHotokezaka2013, Nissanke2013, Siegel2016a, 
Abbott2017a, Abbott2017b}. 
For ``kilonovae'', ejection of radioactive material during the merging process of the compact 
binaries could lead to an excess emission at optical-infrared or ultra-voilet frequencies. The brightness, 
duration and spectrum of such emission is a function of the opacity, velocity, ejecta mass and viewing angle 
\citep{Metgzer2010, BarnesKasen2013, Piran2013, Grossman2014, Tanaka2014, Mooley2018, Radice2018}. 
In turn, the opacity depends crucially on the neutron richness of the ejecta, which determines 
how far any r-process nucleosynthesis proceeds. The high mass lanthanides, in particular, create 
heavy line-blanketing which is expected to largely block out light in the optical bands.
Recently, hydrodynamical modeling of such processes \citep{MetgzerFernandez2014, Kasen2015} has predicted a 
brief early blue emission component produced in the outer lanthanide-free ejecta and a rather longer infrared 
transient produced in the inner lanthanide-blanketed regions at later epochs \citep{Bulla2019}. Using their disk-wind model 
for a case with a non-spinning black-hole \citep{Kasen2015}, the optical bump observed in the case of sGRB 080503 
\citep{Perley2009} was interpreted in terms of an underlying ``kilonova'' 
emission for an assumed redshift of z=0.25. Their \citep{Kasen2015} models were, however, unable 
to explain the observed infrared excess in sGRB 130603B which required higher accretion disk mass and perhaps a 
rapidly spinning black hole \citep{Fan2013, Tanaka2014, Just2014}. In this section, we attempt to place some 
constraints on the possible blue-component of associated ``kilonova'' based on the observed prompt emission and
afterglow observations in bluer wavelengths for sGRB 130603B and their comparison with theoretical models. 

\noindent
It has been proposed by \citet{Barkov2011} that one should observe extended prompt emission
in the case of sGRBs initiating Blandford-Znajek (BZ) jets \citep{Blandford1977} due to
large accretion disk mass and high accretion rate. However, in the case of sGRB 130603B EE 
was not detected (see section 2 and Fig. 1). The absence of observable EE
may indicate either that the observer is located off-axis with respect to the narrow BZ-jet, or that 
the accretion disk mass is small. In general, accretion disk mass should correlate with the ejected 
mass and the presence of EE could be an indicator of the emerging ``kilonovae'' in sGRBs. 
Indeed, the plateau phase in X-ray emission observed in sGRB 130603B cannot be explained by a 
BZ-jet model \citep{Kisaka2015} if we assume a small accretion disk mass. The absence of the 
EE and the presence of a plateau phase could be explained by a low accretion rate which 
has still initiated BZ jet but with moderate bulk relativistic gamma-factor. Alternatively, the 
magnetar model could explain the plateau phase of sGRB 130603B and ``kilonovae'' features 
\citep{Fan2013, MetzgerPiro2014}.
Observing EE during the burst phase, along with the presence/absence of an early time X-ray plateau 
during afterglow phase for a larger sample of sGRBs, would allow discriminating among the possible 
progenitors as a sub-class of compact-binary mergers producing magnetars 
\citep{Zhang2011, Rowlinson2013, Siegel2016a, Siegel2016b} but would also allow predicting some 
of them as potential candidates like GW170817.

\noindent
In addition to the analysis described above, using published early time afterglow data of 
sGRB 130603B in {\it Swift-UVOT} $u$ and Gemini $g'$ bands around $\sim$ 1.5d post-burst 
\citep{deUgartePostigo2014}, we attempt to constrain the possible early time blue emission 
contributing to the underlying ``kilonova''. The observed limiting magnitude in 
$u >$ 22.3 mag and $g'$ > 25.7 mag place limits on the corresponding luminosities 
of $L_u < 3.5\times10^{27}$ erg/s/Hz and $L_{g'} < 0.3\times10^{27}$ erg/s/Hz respectively. 
Using the transformation equations (2) \& (3) given in \citet{Tanaka2016} (also see equations 
(7) \& (8) in \citealt{FernandezMetzger2016}), we tried to constrain the parameter called ejected mass 
$M_{ej}$. However, these limiting values of luminosities in the two bands are not sufficiently deep to constrain 
values of the ejected mass meaningfully ( $>$ 1.5 $M_{\odot}$) for the bluer component of ``kilonova'' at the given 
epoch for the assumed values of the standard parameters. Considering the $WIND$ models of ``kilonovae'' with rather
lower opacity and expansion velocities \citep{Tanaka2016, Kasen2015, MetgzerFernandez2014},
constraints for the ejected mass $M_{ej}$ are even weaker i.e. $M_{ej} >$ a few $M_{\odot}$ which is
un-physical. We caution that the placed limits on $M_{ej}$ could be shallower if there were 
some contribution from the afterglow at the epoch of observations, which is certainly plausible. 
It is also worth mentioning that the some of parameters in the ``kilonovae''
models like the range spin of the neutron star, f-parameter, neutron richness have not been well-constrained so far
\citep{Metgzer2010, Kasen2015}, causing large uncertainty when predicting the possible emission at UV, 
optical or IR frequencies. On the other hand, in case of recently observed under-luminous and nearby event sGRB 170817A/GW170817, 
lanthanide-poor observed blue-components were successfully modeled using a three-component ``kilonova'' 
model \citep{Villar2017a, Villar2017b} with more realistic value of $M_{ej} \sim$ 0.016 $M_{\odot}$. 
So, present constrain on $M_{ej}$ in case of sGRB 130603B indicate that either blue-component ``kilonova'' emission 
was absent/weaker in comparison to the observed blue-component in case of GW170817. These constraints further indicate that
it could be possible to get a range of blue-component of ``kilonovae'' emission due to possible effects caused by range 
of the dynamical ejecta, life-time and spin of the promptly formed magnetar/Black Hole, viewing angle effects etc. 
in case of some of the sGRBs.  
Early time deeper observations at bluer wavelengths for many such events at various distances are required to 
determine the range of properties like brightness, duration and possible diversity among these events. 

\subsection{Host galaxy SED modeling of sGRB 130603B}

Information about the host galaxy, such as the characteristic age
of the dominant stellar population and the average internal extinction,
were obtained by analyzing its broad-band SED (Table 4) using
stellar population synthesis models. 
The host galaxy of GRB~130603B is a perturbed spiral galaxy as seen in high-resolution HST image 
\citep{Tanvir2013} due to interaction with another galaxy. We combined our observational data in filters 
$B, g, r, R_C, i, z, J, H, K_s$ obtained with GTC, CAHA, and DOT telescopes (see Table 1) and combined 
them with ultra-violet data in $uvw2, uvm2, uvw1, U$ bands from \citet{deUgartePostigo2014} to construct the broad-band 
SED of the host galaxy. Taking into account a Galactic reddening along the line of sight of $E(B-V)=0.02$ mag, and 
fixing the redshift of $z=0.356$, we fitted the host SED using \textsc{Le~Phare} software 
package~\citep{lephar1,lephar2}. We used the \textsc{PEGASE2} population synthesis models 
library \citep{pegase} to obtain the best-fitted SED and the main physical parameters of the galaxy: 
type, age, mass, star-formation rate (SFR) etc. We tried different reddening laws: Milky Way \citep{seaton-mw}, 
LMC \citep{lmc}, SMC \citep{smc}, and the reddening law for starburst galaxies \citep{calzetti,calzetti_mod}.

\noindent
According to the best fit, the host is a type Sd galaxy with absolute magnitude in rest-frame $M_B = -20.9$, 
moderate bulk extinction of $E(B-V)=0.2$, and Milky Way dust extinction law. It is about $0.7$ Gyr years old, 
has a mass of $1.1\times10^{10}$M$_{\sun}$ and a low star-formation rate of SFR $\sim 6$M$_{\sun}/$yr. All the 
parameters are listed in the Table 4. The reduced $\chi^2$, galaxy morphological type, bulk extinction, absolute 
rest-frame $B$ magnitude, age, mass, star formation rate, and specific star formation rate (SSFR) per unit 
galaxy stellar mass are listed for all 4 tested extinction laws. Fig. 8 represents the best model 
corresponding to the Milky Way extinction law.

\noindent
These results confirm the previous host galaxy studies \citep{Cucchiara2013b,deUgartePostigo2014,chrimes2018} 
by independent observations and modeling, and adding new piece of information about the extinction law inside the host galaxy.
Our SED modeling results also constrain that SFR and mass of the host galaxy of sGRB 130603B are typical
to those observed in case of other short bursts as shown in Fig. 11. 
However, the resulting SFR is 5 times higher than that obtained by \citet{chrimes2018} using different population 
synthesis libraries.

\begin{table*}
\centering
  \caption{GRB~130603B host galaxy properties derived from the SED fitting using stellar population synthesis models.}
  \begin{tabular}{lllll}
  \hline
  \hline
Fitted                            & Starburst              & Milky Way              & LMC                    & SMC \\
parameters                        & model                  & model                  & model                  & model \\
  \hline
  \hline
$\chi^2$/DOF                      & 12.0/11                & 11.1/11                & 11.7/11                & 12.2/11 \\
Type                              & Sbc                    & Sd                     & Sd                     & Sc \\
$E(B-V)$, mag                     & 0.05                   & 0.20                   & 0.20                   & 0.00 \\
$M_B$, mag                        & $-20.05(\pm0.07)$      & $-20.86(\pm0.07)$      & $-20.06(\pm0.07)$      & $-20.83(\pm0.07)$ \\
Age, Gyr                          & $0.58^{+0.60}_{-0.42}$ & $0.72^{+0.84}_{-0.55}$ & $3.75^{+0.80}_{-2.25}$ & $7.50^{+0.44}_{-5.82}$ \\
Mass, $(\times10^{10})$M$_{\sun}$ & $1.4^{+0.4}_{-0.1}$    & $1.1^{+0.2}_{-0.7}$    & $0.2^{+1.1}_{-0.1}$    & $1.5^{+1.2}_{-0.9}$ \\
SFR, M$_{\sun}$/yr                & $8.3^{+16.8}_{-4.6}$   & $5.9^{+11.9}_{-1.8}$   & $7.6^{+16.4}_{-3.7}$   & $8.3^{+17.2}_{-4.3}$ \\
SSFR, $(\times10^{-10})$yr$^{-1}$ & $4.6^{+15.3}_{-2.1}$   & $5.3^{10.8}_{-1.0}$    & $5.3^{+19.5}_{-1.1}$   & $2.1^{+25.3}_{-3.7}$ \\
  \hline
  \hline
  \end{tabular}
  \label{130603bhost}
\end{table*}
%

\begin{figure}
\centering
\includegraphics[height=8.0cm,width=8.0cm,origin=c]{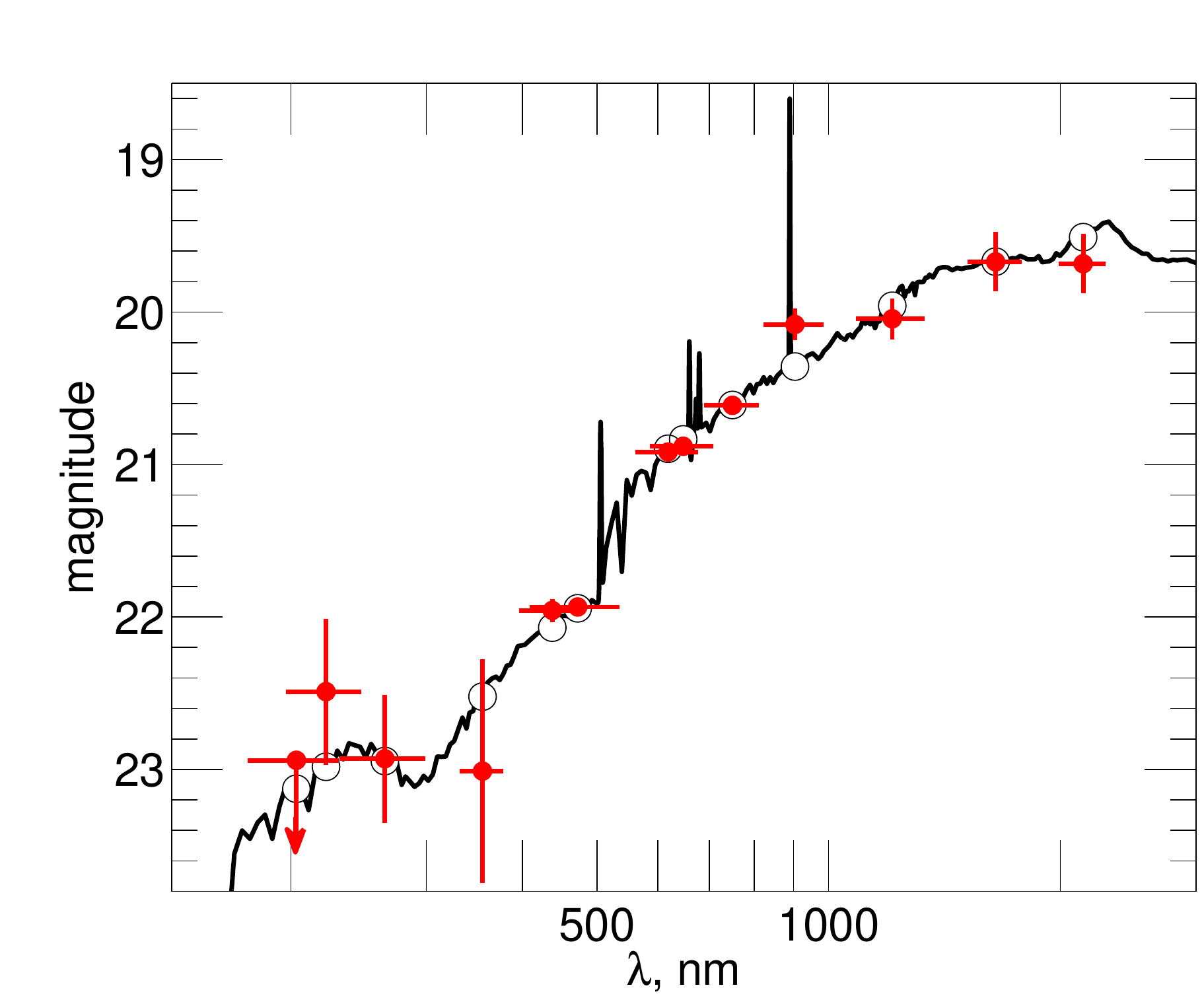}
\caption{\label{light} The SED of the host galaxy of sGRB~130603B fitted by the \textsc{Le~Phare} with fixed 
redshift $z=0.356$. Filled red circles depict respectively the data points in the filters $uvw2, uvm2, uvw1, U$, 
taken from \citet[][, Table 4]{deUgartePostigo2014}, and $B, g, r, R_C, i, z, J, H, K_s$ from original observations 
(see 2.2). Data points in $B$ and $R_C$ pass-bands were obtained using the 4K$\times$4K CCD Imager \citep{Pandey2018} mounted at
the axial port of the recently commissioned 3.6m DOT at Nainital India \citep{Kumar2018}. Open circles represent model magnitudes 
for each filter. All magnitudes are in AB system.}
\end{figure}

\section{Multi-wavelength observations of 8 sGRBs during 2012-2015}

During 2012-2015, a total of 45 sGRBs were localized by several space-missions.
Only 23/45 of these sGRBs were seen by
{\it Swift}-XRT. Out of those 23, only 9 were detected at optical bands, and, for 7 such events
redshifts were determined. In this section, details of the prompt emission and multi-band 
observations to detect optical afterglow and host-galaxy of eight events
(sGRB 121226A, sGRB 131224A, sGRB 140606A, sGRB 140622A, sGRB 140903A, sGRB 140930B,
sGRB 141212A and sGRB 151228A) besides sGRB 130603B are discussed. Out of these 8 sGRBs,
3 events namely sGRB 131224A, sGRB 140606A and sGRB 151228A were not detected by {\it Swift}-
XRT. However, sGRB 140606A and sGRB 151228A were seen by {\it Fermi}-Gamma-ray Burst Monitor (GBM) 
continuous Time-Tagged Event (TTE) data having detailed description in Appendix ``A''. Out of the 8 sGRBs from
the present sample during 2012-2015, late time follow-up observations using GTC 10.4\,m and Gemini-N 8.0\,m could be 
obtained for 4 {\it Swift}-XRT localized bursts i.e. for sGRB 121226A, sGRB 140622A, sGRB 140930B and sGRB 141212A,
useful to constrain late-time afterglow emission, placing limits on possible ``kilonovae'' emission and host galaxy
as described in respective sections of Appendix ``A''.  
 
\noindent
The {\em INTEGRAL} SPI-ACS having a stable background (see \citealt{Bisnovaty2011} and \citealt{Minaev2010} for details)
is particularly useful in the search for EE after the prompt emission
phase of sGRBs. As a part of the present analysis, prompt emission {\em INTEGRAL} SPI-ACS
observations of sGRB 121226A, sGRB 130603B, sGRB 140606A, sGRB 140930B, sGRB 141212A and sGRB 151228A were analyzed 
and compared with other contemporaneous observations with the {\it Swift}-BAT and {\it Fermi}-GBM, when available.
Details about the gamma-ray and X-ray data analysis are described in respective sub-sections of Appendix ``A''. 
The analysis of the sub-set of these events do not show any signature of extended emission except sGRB 121226A
and their spectral and temporal properties do not differ from those seen by {\it Swift}-BAT. 
Out of the eight sGRBs, for sGRB 140606A and sGRB 151228A, the characteristic photon peak energy 
E$_{\rm peak}$ could be determined using the prompt emission {\it Fermi}-GBM data. These two sGRBs along with
others discussed in this paper with presumed redshift values allowed us to construct the Amati diagram along with
published lGRBs (see Fig. 9). Based on this diagram, nature of these
four bursts (namely sGRB 140606A, sGRB 140622A, sGRB 140930B and sGRB 151228A) are clearly categorized as 
short bursts. 

\begin{figure}
\centering
\includegraphics[width=\columnwidth]{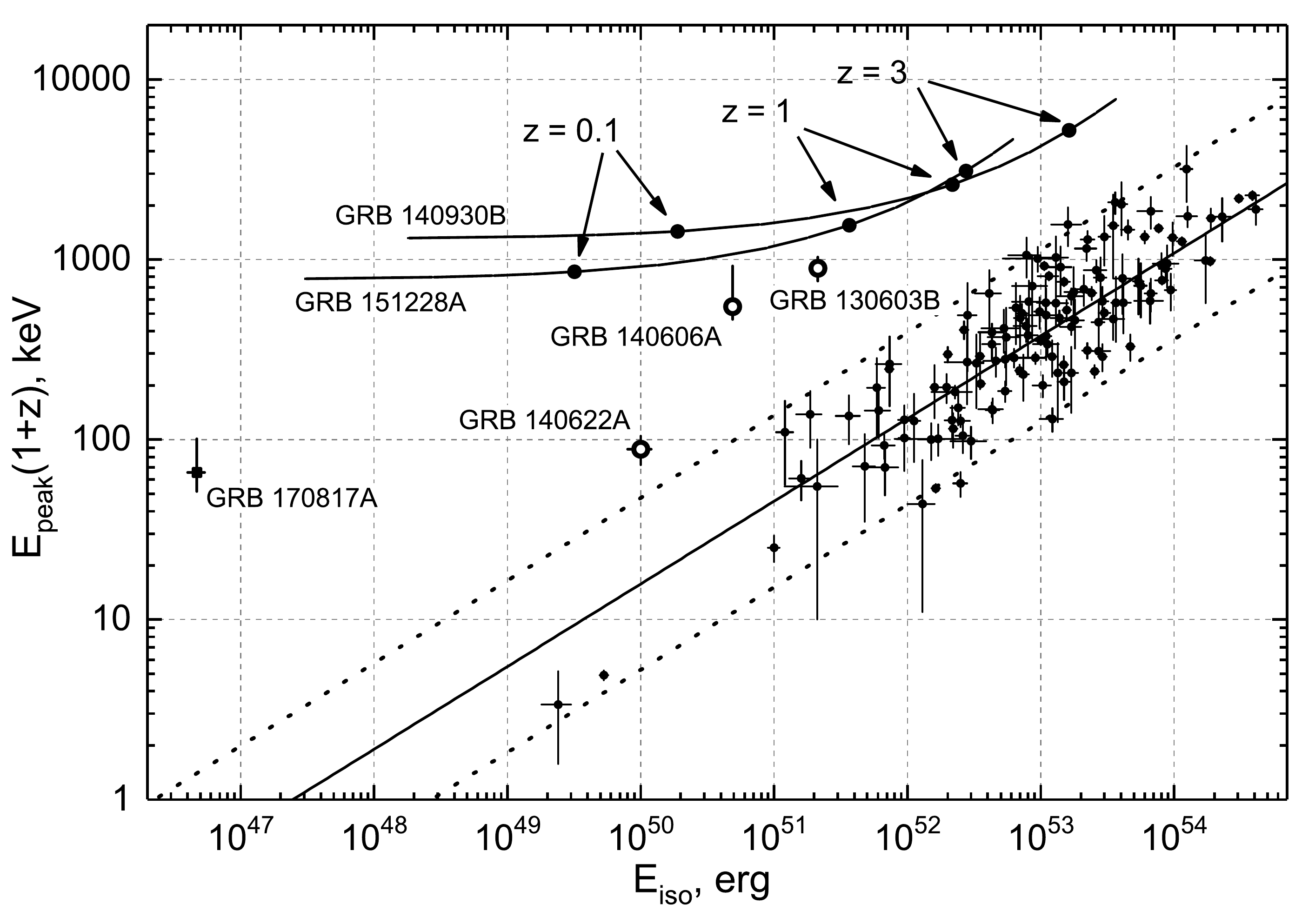}
\caption{\label{light} Amati diagram - a relation between equivalent isotropic energy 
emitted in the gamma-ray E$_{\rm iso}$ versus characteristic photon peak energy E$_{\rm peak}$(1+z) 
in the rest frame \citep{Amati2008}. The solid straight line indicates a power-law fit to 
the dependences for the long bursts; the dashed lines bound the 2$\sigma$ correlation region. 
The trajectories of sGRB 140930B and sGRB 151228A are plotted as a function of the 
presumed redshift z. Open circles indicate short bursts (sGRB 140606A, sGRB 140622A and sGRB 130603B) 
with measured values of E$_{\rm peak}$ and redshift. Parameters of sGRB 170817A/GW170817 are also over-plotted for comparisons.}
\end{figure}

\noindent
Follow-up observations of these eight sGRBs suggest that the afterglows of these events were
faint and were located either next to a bright star or embedded within the host galaxy, 
making the photometry complicated at the epoch of observations. Photometric results regarding 
the afterglow or host galaxies observed by the GTC 10.4\,m and other 
ground-based telescopes as a part of the present analysis are tabulated in Table 5.
Our optical-NIR observations indicate that for sGRB 141212A, the observed host galaxy
was relatively bright and had star formation activity.  
Deeper GTC 10.4\,m observations 
of the sGRB 140622A reveal that the burst could belong to a group of host-less bursts      
\citep{Tunnicliffe2014}. Follow-up optical observations of sGRB 140903A constrain
any underlying ``kilonovae'' emission down to a limiting magnitude of R $>$ 22 mag
at 10d after the burst. Our early to late time afterglow observations of sGRB 140930B using 
William Herschel Telescope (WHT) 4.2\,m and Gemini-N 8.0\,m observations along with those observed 
by {\it Swift}-XRT are able
to constrain the decay nature of the burst and late time 10.4\,m GTC observations places a deeper 
upper limit of $r \sim 24.8$ mag for any possible host galaxy.
Details about observations of the afterglows, host galaxies and their data analysis, 
calibrations etc. of each of the 8 individual bursts are described in the 
Appendix ``A'' below. A summary of the observed prompt emission and afterglow properties of all the
9 sGRBs are also listed in Table 6.


\section{GW170817 and the sample of sGRBs}

On 17 August 2017, 12:41:04.82 UT, the LIGO and Virgo interferometers detected a transient gravitational wave (GW)
signal from a source named GW170817 \citep{Abbott2017b}. The {\it Fermi}-GBM 
triggered and located a short burst named sGRB 170817A \citep{von2017} about 1.7 s after the 
GW signal spatially consistent with the GW event \citep{Blackburn2017}.
The error region was later followed-up extensively at lower frequencies to search for the underlying ``kilonova'' signature \citep{Coulter2017, Pian2017, Covino2017, Tanvir2017, Nora2017, Evans2017, 
Smartt2017, Cowperthwaite2017}. Discovery of this first GW event called GW170817/AT 2017gfo/SSS17a 
associated with the very nearby (host galaxy NGC 4993 at $\sim 40$ Mpc) sGRB 170817A and the 
underlying bright ``kilonova'' provides strong evidence favoring compact binary mergers as the 
progenitors for at least some of these events \citep[][and references therein]{Abbott2017a, Abbott2017b}. \\

\noindent
The T$_{90}$ duration of this GW170817  connected sGRB 170817A was 0.5$\pm$0.1 s (70-300 keV) having multiple 
emission episodes and had a relatively soft spectrum with E$_{\rm peak}$ = 65$^{+35}_{-14}$ keV \citep{Goldstein2017, 
Pozanenko2017}. The burst was also detected by SPI-ACS onboard {\it INTEGRAL} \citep{Savchenko2017} 
and morphology of the $\gamma$-ray light-curve is similar to that seen in the case of presently discussed 
sGRB 140930B i.e. having multiple episodes of emissions and belong to pattern-II class of bursts \citep{Lu2017}, 
suggesting a diverse set of progenitors and central engines \citep{Dichiara2013}. 
sGRB 170817A turned out to be the weakest detected sGRB having a soft spectrum with a thermal tail
and was under-luminous by a factor of $\sim$ 1000 in comparison to known sGRBs. So, observed properties like: 
harder pulse with multiple episodes of emissions and a softer tail emission in the spectra have attracted 
significant attention in an effort to understand the nature of the event in terms of various 
physical models \citep{Granot2017a, Granot2017b, Gottlieb2018, Pozanenko2017, Zhang2017}.
Except for resemblance with the duration T$_{90}$, all other 
observed prompt emission properties of the sGRB 170817A like the morphology of the $\gamma$-ray 
light-curve, E$_{\rm peak}$, E$_{\rm iso}$ etc. were outliers with 
the known set of sGRBs including those discussed in this paper as described in Fig. 9. \\

\noindent
sGRB 170817A counterparts at UV-optical-NIR frequencies are distinct to those expected for GRB 
afterglows \citep{Piran1999} and predominantly follow physical mechanisms suggested for
underlying ``kilonova'' emission \citep{Pian2017, Tanvir2017, Nora2017} 
consistent with a compact binary merger origin for this event. However, contrary to red ``Kilonova'' associated 
with the sGRB 130603B, sGRB 170817A UV-optical-NIR emission was explained well in terms of r-processed 
three-component sub-relativistic accretion disk powered ``kilonova'' model \citep{Villar2017a, Villar2017b}. 
In Fig. 4, the H-band light curve of the 
GW170817 counterpart (redshifted at z = 0.36) is compared along with ``kilonova''
detection and models for the sGRB 130603B \citep{Tanvir2013, Tanaka2014}. 
The H-band redshifted light curve of the GW170817 counterpart is fainter in comparison to the corresponding HST 
detection of the ``kilonova'' associated with the sGRB 130603B and exhibits distinct nature of the overall 
temporal decay. 

\noindent
Early time non-detection by the {\it Swift}-XRT  until 9d post-burst for sGRB 170817A 
compared to other known cases of X-ray detected sGRBs \citep{Fong2017}, places a constraint on the underlying 
emission mechanisms and supports a non-afterglow origin for the observed emission at lower frequencies. 
Recently, using deeper data-set of other bursts \citet{gompertz2017} have concluded that not all sGRBs 
are associated with ``kilonovae'' and share a diverse range of observed brightness. 
No detection of GW170817 like ``kilonova'' for a good number of well-studied sGRBs to a deeper 
limit also indicate a diverse set of progenitors for some of the bursts \citep{gompertz2017,Rossi2019}.\\

\noindent
As a part of the present study, sGRB 170817A/GW170817 was observed using GTC 10.4\,m in $i$-band starting around 05:47:40 UT 
on 19-01-2018 for a total exposure time of one hour (120sx30). The images were stacked and processed as per standard
techniques. A 3-$\sigma$ upper limit of the stacked image is $i \sim$ 25 mag whereas at the location of the optical
transient (see Fig. 10 and Table 5), rather shallow value of $i \sim$ 23.5 mag was estimated due to 
contamination of the host. 
As a part of the present analysis, second epoch of  GTC 10.4\,m observations of the host galaxy NGC 4993 were also on 
06-02-2019 around 5:10:00 UT in $i$-band (120sx30) and after image subtraction a deeper limit of $i \sim$ 24 mag was 
estimated at the location of the GW170817. This observed limiting magnitude ($\sim$ 154.7d post-burst) at the location of 
the optical transient is in agreement with the extrapolated at contemporaneous epochs by
\citet{margutti2018} and thus supports a non-thermal origin of the emission at the epoch of our observations. 
On the other hand, detections of the sGRB 170817A/GW170817/AT 2017gfo/SSS17a at X-ray \citep{Nora2017} and 
VLA radio frequencies \citep{alexander2017} $\sim$9d to 160d post-burst exhibit rising lightcurves both at X-ray 
and radio frequencies and are broadly consistent with non-thermal collimated emission viewed off-axis or structured 
outflow \citep{margutti2017, Fong2017, Hallinan2017, Haggard2017, Evans2017, Smartt2017, Lazzati2017, 
Nora2017, Granot2017a, Granot2002}. However, \citet{xie2018} and \citet{Lyman2018} have found that the late time 
multi-band data of the sGRB 170817A is well explained both by narrow and wide engine mild-relativistic models, 
though, early time non-detection at X-ray frequencies disfavors wide engine model. So, it is clear that none of 
the models have been able to re-produce the full set of multi-band data for this nearby event. 

\begin{figure}
\centering
\includegraphics[width=\columnwidth]{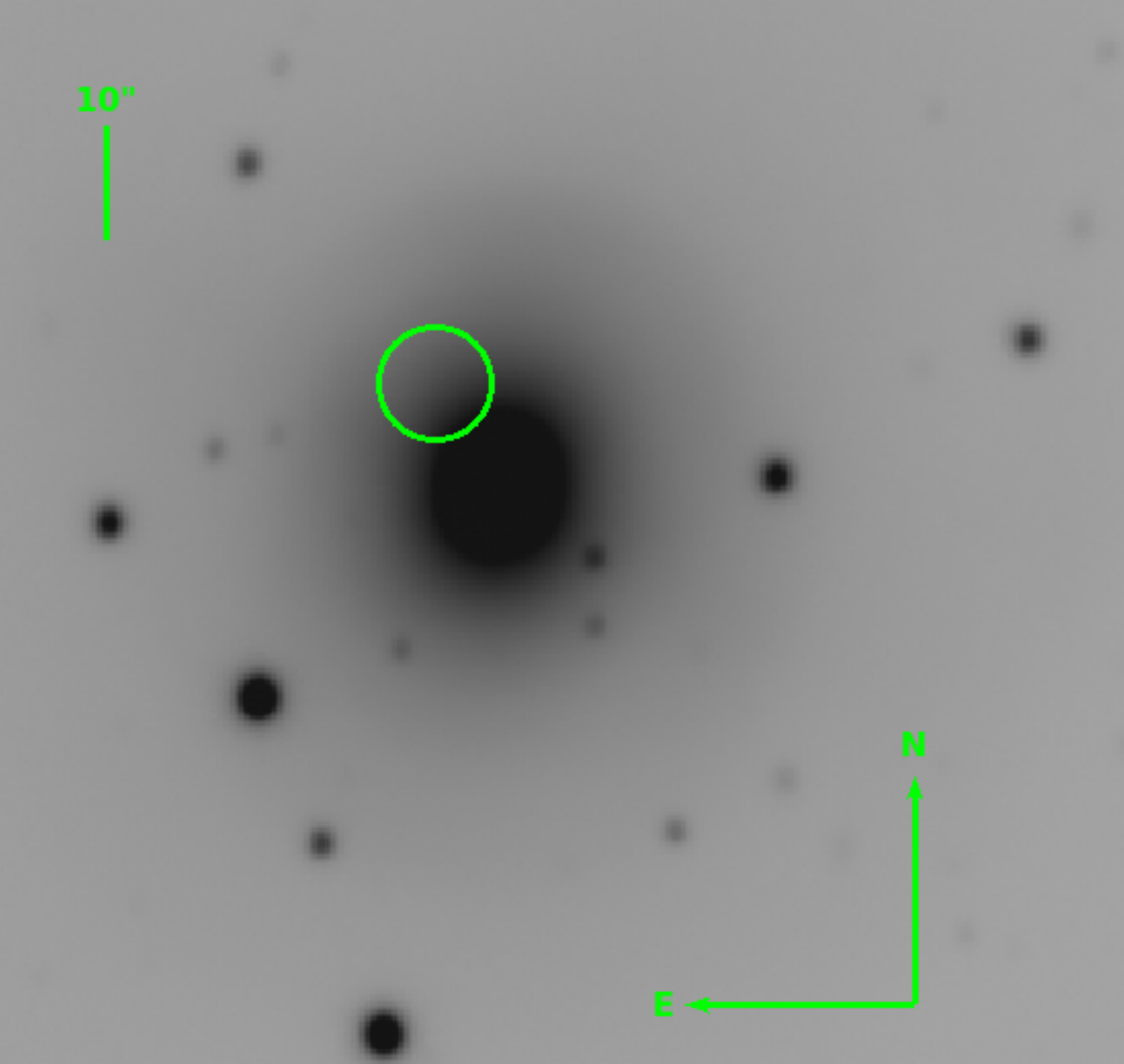}
\caption{\label{light} Finding chart of GW170817 (circle) in the stacked frame of $i$-band data obtained by the 
GTC 10.4\,m telescope obtained $\sim$ 154.7d post-burst as a part of the present analysis.} 
\end{figure}

\noindent
The host galaxy SED modeling of sGRB 130603B and sGRB 141212A from the present sample of bursts indicate that 
their respective hosts are young and bluer with moderate values of star formation activity. However, 
in case of sGRB 170817A, the host galaxy NGC 4993 is an old elliptical galaxy with
little star formation activity and the projected offset of the burst location is rather 
closer to what has been seen in case of other sGRBs \citep{Fong2017, Levan2017}. 
Fig. 11 shows the distribution of star formation rates versus stellar mass (top panel) and specific 
star formation rates versus stellar mass (bottom panel) for the known set of host galaxies of lGRBs and sGRBs
\citep{Savaglio2009} and GW170817 \citep{Blanchard2017}. In Fig. 11, corresponding values for the 
sGRB 170817A/GW170817 clearly indicate that the star formation rate for sGRB 170817A/GW170817 host 
galaxy is well below from those seen in case of normal population of GRBs.
Overall properties of the GRB 170817A/GW170817 and their comparison with other sGRBs indicate that 
we need multi-wavelength observations of a significantly larger number of nearby events to explore 
the full diversity of ``kilonovae'' and their association with sGRBs. \\

\begin{figure}
\centering
\includegraphics[width=\columnwidth]{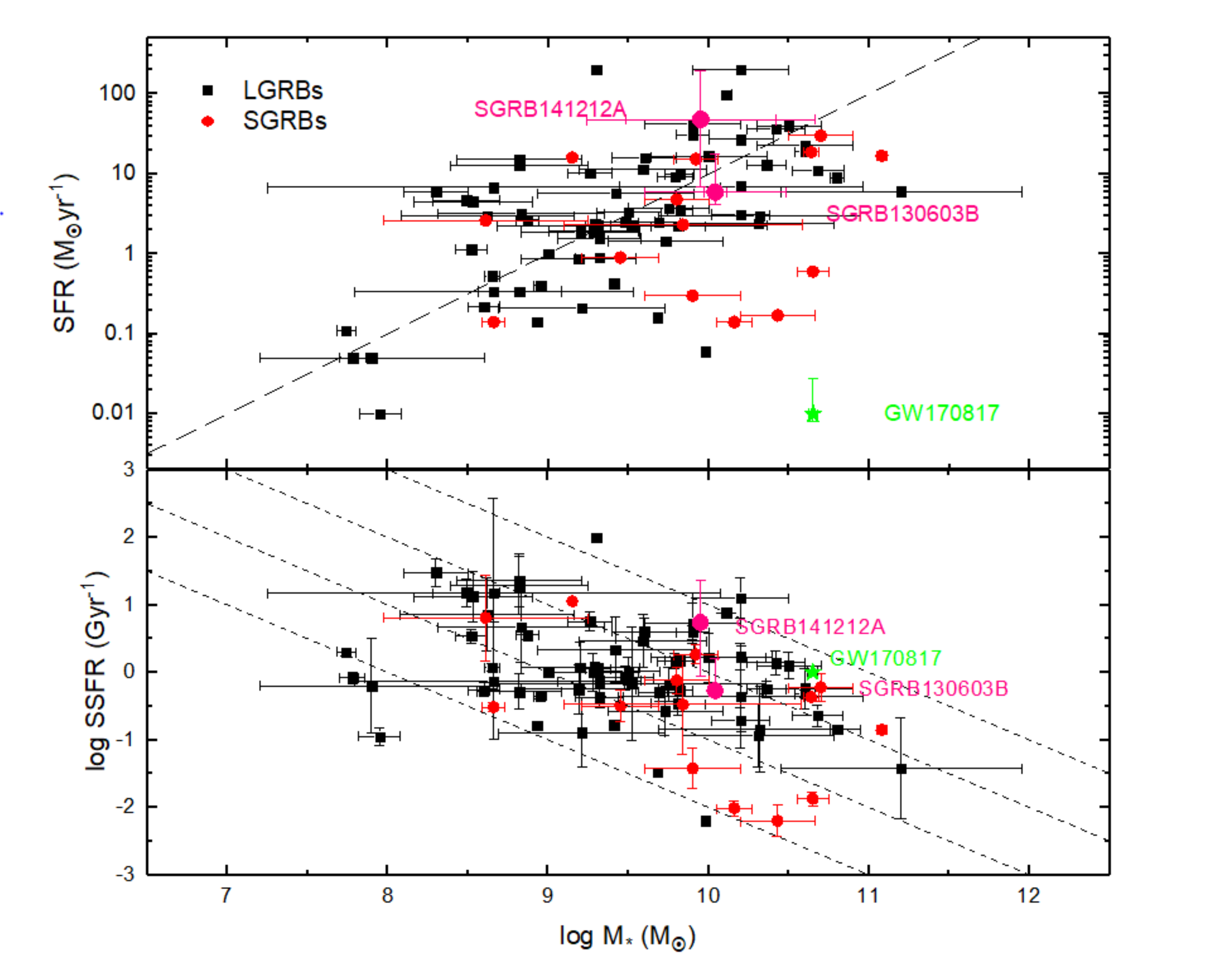}
\caption{\label{light} Plot of star formation rate versus stellar mass (top panel) and specific
star formation rate versus stellar mass (bottom panel) for the known set of host galaxies of lGRBs and sGRBs.
The dashed line marks a constant SFR of 1 Gyr$^{-1}$ (top panel). The dashed lines mark the constant
specific SFR of 0.1, 1, 10 and 100 Gyr$^{-1}$ from left to right (bottom panel).
The modeled values of star formation rates and mass of the hosts of sGRB 130603B and sGRB 141212A 
(date taken from the present analysis, Table 4 and Table A1) are plotted as pink circles. 
Corresponding values for the GW170817 are plotted as green star.}
\end{figure}

\normalsize
\begin{table}
\vspace{-5mm}
\caption {\scriptsize Summary of the optical photometric observations (AB-magnitudes) of the afterglows of the
eight sGRBs (2012-2015) and their host galaxies (h) using ground-based optical telescopes as a part of the present study. 
Recently observed GW170817 using GTC 10.4\,m are also appended to this table. The values of the magnitudes are in AB 
system (limiting magnitudes are 3$\sigma$) and no extinction corrections have been applied.} 
\begin{center}
\tiny
\begin{tabular}{cccll} \hline 
t-t0\,mid(d)&exp(s)&Afterglow/&pass-band&Telescopes\\
&&Host galaxy&& \\
\hline
&  &      {\bf sGRB 121226A}  &   &       \\ 
0.0042&5$\times$19&$>$19&clear&0.6\,m BOO-4 MET  \\
0.0833&300$\times$4&$>$18.8&I$_c$&1.04\,m ST ARIES \\
0.0833&300$\times$6&$>$19.5&R$_c$&1.04\,m ST ARIES \\
0.432&75$\times$2&23.65$\pm$0.37&$z$&GTC 10.4\,m    \\
0.441&85$\times$5&24.03$\pm$0.32&$i$&GTC 10.4\,m    \\
0.451&70$\times$8&24.30$\pm$0.30&$r$&GTC 10.4\,m    \\
53.25&50$\times$31&$>$23.79&$z$&GTC 10.4\,m    \\
53.27&70$\times$12&$>$24.47&$i$&GTC 10.4\,m   \\ 
\hline
&  &      {\bf sGRB 131224A}   &   &       \\ 
1.111&30$\times$1&$>$18.3&$r$&GTC 10.4\,m    \\
1.113&60$\times$3&$>$19.5&$i$&GTC 10.4\,m    \\
1.116&75$\times$3&$>$24.3&$z$&GTC 10.4\,m    \\
7.099&$5\times$4+10$\times$1&$>$23.6&$i$&GTC 10.4\,m    \\
7.105&$20\times$4+10$\times$1&$>$22.8&$z$&GTC 10.4\,m   \\ 
\hline
&  &      {\bf sGRB 140606A}  &   &       \\ 
0.3315&3600&$>$21.7&clear& Abastumani AS-32   \\
0.4292&4$\times$30+3$\times$120&$>$26.0&$R_c$&BTA6.0\,m    \\
0.3857&120+30&$>$24.2&$V$&BTA6.0\,m    \\
0.3893&120&$>$24.4&$B$&BTA6.0\,m    \\
271.642&60$\times$5&$>$25.36&$r$&GTC 10.4\,m   \\ 
\hline
&  &      {\bf sGRB 140622A}  &   &       \\ 
0.0687&4320&$>$23.64&r&RATIR 1.5\,m \\
0.0687&4320&$>$23.49&i&RATIR 1.5\,m \\
0.0687&1836&$>$19.41&Z&RATIR 1.5\,m \\
0.0687&1836&$>$18.73&Y&RATIR 1.5\,m \\
0.4752&4800&$>$22.5&R&TShAO Ziess-1000    \\
0.781&100$\times$6+5$\times$2&$>$25.8&$r$&GTC 10.4\,m   \\ 
\hline
&  &      {\bf sGRB 140903A}  &   &       \\ 
0.1406&100&$>$18.6&clear&ISON-Kislovodsk   \\
&  &       &   & SANTEL-400A       \\
1.0648&720&$>$22.0&R&Maidanak AZT-22  \\
3.0072&900&$>$22.0&R&Maidanak AZT-22 \\
4.0090&900&$>$22.0&R&Maidanak AZT-22  \\
10.0500&720&$>$22.0&R&Maidanak AZT-22 \\ 
\hline
&  &      {\bf sGRB 140930B}  &   &       \\ 
0.0291&3600&$>$20.4&clear&ISON-Kislovodsk,   \\
&  &        &   & SANTEL-400A       \\
0.0145&1200&$>$19.5&clear&ISON-Kislovodsk,     \\
&  &        &   & SANTEL-400A       \\
0.0309&2000&$>$19.6&clear&ISON-Krasnodar,     \\
&  &        &   & Astrosib       \\
0.0249&415&$>$16.1&clear&UAFO ORI-65 \\
0.133&300$\times$5&22.65$\pm$0.09&$r$&WHT 4.2\,m/ACAM    \\
0.153&300$\times$5&22.61$\pm$0.06&$i$&WHT 4.2\,m/ACAM    \\
0.172&400$\times$2&23.17$\pm$0.12&$g$&WHT 4.2\,m/ACAM    \\
0.677&150$\times$9&24.01$\pm$0.04&$r$&Gemini North/GMOS-N \\
1.656&150$\times$9&25.11$\pm$0.11&$r$&Gemini North/GMOS-N \\
3.141&60$\times$13&$>$24.5&$r$&GTC 10.4\,m   \\ 
1535.5&90$\times$34&$>$24.75&$r$&GTC 10.4\,m  \\  
\hline
&  &      {\bf sGRB 141212A}  &   &       \\ 
0.0189&60$\times$10&$>$22.2		&R&	Mondy AZT33-IK \\
0.0363&60$\times$60&22.73$\pm$0.26 (h)	&R&	Mondy AZT33-IK \\
0.0783&120$\times$30&22.75$\pm$0.28 (h)	&R&	Mondy AZT33-IK \\
0.0573&60$\times$60+120$\times$30&22.71$\pm$0.19 (h)	&R&	Mondy AZT33-IK \\
0.0242&60$\times$5&$>$18.5		&clear&	Khureltogot ORI-40  \\
0.0641&60$\times$74&$>$19.9		&clear& Khureltogot ORI-40  \\
0.6814&180$\times$5&22.13$\pm$0.04 (h)	&$i$&	Gemini North/GMOS-N \\
1.1563&300$\times$13&22.63$\pm$0.18 (h)	&R&	TShAO Ziess-1000 \\
1.7461&180$\times$5&22.23$\pm$0.04 (h)	&$i$&	Gemini North/GMOS-N \\
2.0544&120$\times$57&22.76$\pm$0.33 (h)	&R&	Mondy AZT33-IK \\
6.0676&120$\times$85&22.86$\pm$0.16 (h)	&R&	Mondy AZT33-IK \\
427.375&5$\times$3+120$\times$11&23.86$\pm$0.08 (h)&$g$&	GTC 10.4\,m \\
427.385&120$\times$7&22.80$\pm$0.06 (h)	&$r$&	GTC 10.4\,m \\
427.403&90$\times$6&22.32$\pm$ 0.05 (h)	&$i$& GTC 10.4\,m\\ 
\hline
&  &      {\bf sGRB 151228A}  &   &       \\ 
0.0011&60$\times$3+20$\times$2&$>$17.5&$R$& 0.60\,m T60 \\
1.1429&5$\times$60&$>$23.7&$i$&GTC 10.4\,m    \\
69.0036&7$\times$75&$>$24.8&$i$&GTC 10.4\,m   \\ 
\hline
&  &      {\bf sGRB 170817A/GW170817}  &   &       \\ 
154.7&120$\times$30&$>$24.0&$i$& GTC 10.4\,m \\ 
536.8&120$\times$10&$>$24.0&$i$& GTC 10.4\,m \\ 
\hline
\end{tabular}
\end{center}
\end{table}

\normalsize
\begin{table*} 
\centering
\caption{\scriptsize Summary of the prompt emission and afterglow properties of the 9 sGRBs discussed in this paper. The symbols used in the table have their usual meanings as discussed in the main text.} 
\tiny
\begin{tabular}{ccccccccccccc} \hline \hline
sGRB&Redshift&T$_{90}$&E$_{\rm iso}$$^{a}$&E$_{\rm peak}$&\multicolumn{4}{c}{Early first observation}&Afterglow detection &Host galaxy& Comments\\
\cline{6-9}
  & &sec&erg&keV&t-t0,mid(s)&Magnitude&Pass-band&Telescope& & & \\
\hline
121226A &       $-$     &       1.00$\pm$0.20$^{a}$     &       $-$     &       $-$     &       93      &       $>$17.5 &       R       &       Zadko$^{a}$     &       $+$     &       $>$24.6($r$)     &       EE      \\
130603B &       0.3564$\pm$0.0002$^{a}$   &       0.18$\pm$0.02$^{b}$     &       (2.1$\pm$0.2)$\times$10$^{51}$$^{b}$ &       660$\pm$100$^{a}$       &       137     &       $>$19.6 &       $white$ &       UVOT$^{b}$      &       $+$     &       22.13$\pm$0.05($r$)     &       KN      \\
131224A &       $-$     &       0.8$^{c}$       &       $-$     &       $-$     &       44      &       $>$15.5 &       $white$     &       MASTER II$^{c}$ &       $-$     &       $>$23.6($i$)    &       not a GRB \\
140606A &       1.96$\pm$0.1    &       0.34$\pm$0.09$^{d}$     &       4.9$\times$10$^{50}$   &       185.13$^{+126}_{-28}$$^{d}$     &       143     &       $>$21.1 &       $white$      &       UVOT$^{d}$      &       $-$     &       $>$25.3($r$)    &               \\
140622A &       0.959$^{b}$     &       0.13$\pm$0.04$^{e}$     &       (1.0$\pm$0.2)$\times$10$^{50}$  &       44$\pm$8$^{b}$  &       106     &       $>$17.5 &       R       &       TAROT$^{e}$     &       $-$     &       $>$25.8($r$)    &               \\
140903A &       0.351$^{c}$     &       0.30$\pm$0.03$^{f}$     &       (6.0$\pm$0.3)$\times$10$^{49}$$^{c}$   &       $-$     &       152     &       $>$20.0 &       $white$      &       UVOT$^{f}$      &       $+$     &       20.58$\pm$0.09($r'$)$^{a}$      &               \\
140930B &       $-$     &       0.84$\pm$0.12$^{g}$     &       $-$     &       1302$^{+2009}_{-495}$$^{c}$       &       44      &       $>$16.0 &       $white$     &       MASTER II$^{g}$ &       $+$     &      $>$ 25.1($r$)     &               \\
141212A &       0.596$^{d}$     &       0.30$\pm$0.08$^{h}$     &       (6.7$\pm$1.1)$\times$10$^{49}$   &       $-$     &       51      &       $>$16.8 &       $white$     &       MASTER II$^{h}$ &       $-$     &       22.80$\pm$0.06($r$)     &               \\
151228A &       $-$     &       0.27$\pm$0.01$^{i}$     &       $-$     &       261.18$^{+164.94}_{-58.28}$$^{d}$   &       95      &       $>$17.5 &       R       &       T60     &       $-$     &       $>$24.8($i$)    &               \\
\hline
\hline
\multicolumn{12}{l}{\it{References}}\\
\multicolumn{12}{l}{\it{Redshift: $^{a}$ Xu et al. (2012), $^{b}$ Hartoog et al. (2014), $^{c}$ Cucchiara et al. (2014), $^{d}$ Chornock et al. (2014)}}\\
\multicolumn{12}{l}{\it{T$_{90}$: $^{a}$ Baumgartner et al. (2012), $^{b}$ Barthelmy et al. (2013), $^{c}$ Mereghetti et al. (2013), $^{d}$ Cummings et al. (2014), $^{e}$ Sakamoto et al. (2014)}}\\
\multicolumn{12}{l}{\it{E$_{\rm iso}$: $^{a}$ for GRB 130603B and GRB 140622A in the range 1-10000 keV, for other GRBs in the range 15-150 keV, $^{b}$ Frederiks et al. (2013), $^{c}$ Troja et al. (2016)}}\\ 
\multicolumn{12}{l}{\it{$^{f}$ Palmer et al. (2014), $^{g}$ Baumgartner et al. (2014), $^{h}$ Palmer et al. (2014b), $^{i}$ Barthelmy et al. (2015)}}\\
\multicolumn{12}{l}{\it{E$_{\rm peak}$: $^{a}$ Golenetskii et al. (2013), $^{b}$ Sakamoto et al. (2014), $^{c}$ Golenetskii et al. (2014), $^{d}$ Present analysis}}\\
\multicolumn{12}{l}{\it{Early first observation: $^{a}$ Klotz et al. (2012), $^{b}$ Melandri et al. (2013), $^{c}$ Gorbovskoy et al. (2013), $^{d}$ Marshall and Stroh (2014), $^{e}$ Klotz et al. (2014)}}\\
\multicolumn{12}{l}{\it{$^{f}$ Breeveld and Cummings (2014), $^{g}$ Gorbovskoy et al. (2014), $^{h}$ Gres et al. (2014)}}\\
\multicolumn{12}{l}{\it{Host galaxy: $^{a}$ Troja et al. (2016)}}\\

\end{tabular}
\end{table*} 


\section{Conclusions}

\begin{enumerate}[I]

\item \renewcommand{\labelitemi}{$\bullet$}
In the present work, we have analyzed and reported prompt emission data of nine short bursts 
including sGRB 130603B as observed by {\em Swift}, {\em INTEGRAL} and {\it Fermi} observatories. 
The SPI-ACS {\em INTEGRAL} prompt emission observations of 
sGRB 130603B, sGRB 140930B, sGRB 141212A and sGRB 151228A in the energy range 0.1-10 MeV 
do not show any EE which is in agreement with those seen in the case of 
{\it Swift} observations. However, in case of sGRB 121226A, the EE was seen
as discussed in Appendix section A1. Using {\it Fermi}-GBM data, E$_{\rm peak}$ values were
determined for sGRB 140606A, sGRB 151228A and Amati diagram was constructed to establish the
nature of the five sGRBs from the present sample.  Also, analysis of the {\em INTEGRAL}/JEM-X 
observations indicates that sGRB 131224A may not be of a cosmological origin as discussed in the 
Appendix section A2. 

\item 
\renewcommand{\labelitemi}{$\bullet$}
Multi-wavelength afterglow observations for sGRB 130603B presented in this paper include 
the earliest ground-based optical detection and millimeter observations complementary to 
those published in the literature. Our $r$ and $i$-band data together with those previously 
published have helped to produce a well-sampled $r$ band light curve, made it possible to 
estimate the value of pre-jet break temporal index $\alpha_1 = 0.81\pm0.14$ robustly. 
The derived values of pre- and post-jet break temporal flux decay indices along with the 
X-ray and  optical-NIR spectral indices support the {\em ISM} afterglow model with cooling 
frequency $\nu_c$ between optical and X-ray frequencies. 

\item 
\renewcommand{\labelitemi}{$\bullet$}
Derived values of the jet-break time, electron energy index were used to model the afterglow 
data of sGRB 130603B using numerical simulation-based Monte Carlo model as described in \citet{Zhang2015}. 
Except at very early times ($<$ 1000s) and very late time ($>$ 100000s), largely the multi-band data of sGRB 130603B
are explained in terms of forward shock fireball model. The derived values of micro-physical parameters of 
the burst are better constrained than those reported in \citet{Fong2014}. The observed mm and cm-wavelength 
upper limits for sGRB 130603B are also consistent with forward-shock model predictions. 

\item 
\renewcommand{\labelitemi}{$\bullet$}
In this paper, using the reported values of photometric upper limits in bluer bands 
(i.e. {\it Swift}-UVOT $u$ and Gemini-N $g'$ bands at $\sim$ 1.5d after the burst), we attempted
to constrain the possible blue-component of ``kilonova'' emission in case of sGRB 130603B. 
Accordingly, the values of the ejected mass were calculated as proposed by \citet{Kasen2015} 
and \citet{Metgzer2010} for the possible blue emission. However, the shallower observed 
limits at early epochs in {\it Swift}-UVOT $u$ and Gemini-N $g'$ bands do not provide any 
meaningful constraint for the blue-component of ``kilonova'' emission for sGRB 130603B but indicate that
some of sGRBs may not have the predicted blue-component. 

\item 
\renewcommand{\labelitemi}{$\bullet$}
Deep afterglow observations of a further 8 sGRBs using GTC 10.4\,m and other telescopes reveal
the nature of the decay and the complex environments of some of sGRBs not well-studied so far. 
In case of sGRB 140930B, our early to late time afterglow observations using
4.2\,m WHT and 8.0\,m Gemini-N along with those observed by {\it Swift}-XRT are able
to constrain the decay nature of the burst and the late time 10.4\,m GTC observations 
places a deeper upper limit of $r \sim 24.8$mag for any possible host galaxy.
Whereas, in the case of sGRB 140622A, our optical observations using 10.4\,m GTC puts a deep limit of 
$\sim$25.6 mag for any afterglow or a host galaxy within XRT error-box. These deep 
observations by the GTC 10.4\,m also indicate that sGRB 140622A could belong to the 
category of known host-less bursts. 

\item 
\renewcommand{\labelitemi}{$\bullet$}
Observed limiting flux values at mm and cm-wavelengths for a set of 9 sGRBs using PdBI and their comparison with
published light-curve of lGRB 130427A at 3-mm place constraints on the possible underlying physical mechanisms and 
demand for much deeper observations at these wavelengths.

\item 
\renewcommand{\labelitemi}{$\bullet$}
Deeper optical-NIR follow-up observations of 4 {\it Swift}-XRT localized bursts sGRB 121226A, 
sGRB 140903A, sGRB 140930B and sGRB 141212A using GTC 10.4\,m, Gemini-N 8.0\,m and Maidanak AZT-22 1.5m
upto a few days post-burst constrain for any ``kilonova'' such as the one associated with the GW170817. 
Using prescription given in \citet{Rossi2019}, comparison of rest-frame luminosity of ``kilonova'' 
associated with GW170817 indicate that for sGRB 141212A, any such event would have been detected at
the epoch of our Gemini-N 8.0\,m observations. However, in cases of sGRB 121226A, sGRB 130603B, 
sGRB 140903A and sGRB 140930B the derived luminosity values were found to be dominated by 
afterglow i.e. brighter than the ``kilonova'' associated with the GW170817.  

\item 
\renewcommand{\labelitemi}{$\bullet$}
Upper limit derived using late time (154.7d post-burst) GTC 10.4\,m observations ($i \sim$ 23.5 mag) of the GW170817 
is in agreement with non-thermal origin of the emission as seen at other wavelengths. 
Comparison of prompt emission and properties of the host galaxy of the GW170817 discussed 
in the present work point towards diverse properties of associated ``kilonovae'' and in turn 
points towards possibly diverse classes of compact binary mergers producing normal sGRBs and those with 
associated ``kilonovae''. 

\item 
\renewcommand{\labelitemi}{$\bullet$}
Optical-NIR photometry of the host galaxy of sGRB 130603B was independently 
modeled using \textit{LePHARE} software. The modeling results support the Milky-way Galaxy 
model with a moderate value of the star formation activity in the host galaxy. We also conclude that
the SFR and mass of the host galaxy are typical of those seen in case of other GRBs. 
The host galaxy modeling of the sGRB 141212A indicates that the host is a 
MW type of Sc galaxy with a moderate value of star formation. 

\item 
\renewcommand{\labelitemi}{$\bullet$}
Our observations and analysis of the 8 sGRBs and sGRB 170817A/GW170817 (Table 5 and 6) demand for systematically deeper and more prompt 
multi-wavelength observations of many of these events to detect the afterglow or to constrain the
possible associated ``kilonovae'', host galaxies and their properties in more detail. 
In the future, $JWST$ and other upcoming ground-based optical-NIR facilities like TMT and E-ELT will facilitate the study of 
sGRBs and GW events with unprecedented sensitivity.      \\

\end{enumerate}

\section*{Acknowledgments}

{\it Swift} data/science center is thankfully acknowledged for the publically available data about GRBs. AJCT acknowledges support from the Junta de Andalucia (Project P07-TIC-03094)
and support from the Spanish Ministry Projects AYA2012-39727-C03-01 and 2015-
71718R. This work has been supported by the 
Spanish Science Ministry "Centro de Excelencia Severo Ochoa” Program under grant SEV-2017-0709.
The work is partly based on the observations made with the Gran Telescopio Canarias (GTC), installed in the
Spanish Observatorio del Roque de los Muchachos of the Instituto de Astrofisica de
Canarias, in the island of La Palma Based on observations collected at the Centro Astron\'omico Hispano Alem\'an (CAHA) at Calar Alto, operated jointly by the Max-Planck-Institut for Astronomie and the Instituto de Astrofisica de Andalucia (CSIC). This research was also partially based on observations carried out at the OSN operated by CSIC. FEDER funds are acknowledged. This work is partly based on observations carried out under project
numbers xa52, s14dd001, s14dd002 and s14dd006 with the IRAM NOEMA
Interferometer (http://www.iram-institute.org/EN/content-page-188-7-55-188-0-0.html). 
S.B.P. acknowledge BRICS grant number ``DST/IMRCD/BRICS/PilotCall1/ProFCheap/2017(G)''for the present work. IRAM is supported by INSU/CNRS (France), MPG (Germany) and IGN (Spain). SRO gratefully acknowledges the support of the Leverhulme Trust Early Career Fellowship.
SJ acknowledges support from Korean grants NRF-2014R1A6A3A03057484 and NRF-2015R1D1A4A01020961. 
E.S. acknowledges assistance from the Scientific and Technological Research Council of Turkey 
(TUBITAK) through project 112T224. We thank to TUBITAK for a partial support in using T100 telescope with project number 10CT100-95. S.B.P. acknowledge discussions with Dr. Masaomi Tanaka on kilonovae and related science. A.S.P acknowledges partial support grants RFBR 17-02-01388, 17-51-44018 and 17-52-80139.
E.D.M., A.A.V. and P.Yu.M. are grateful to RSCF grant 18-12-00522 for support.
B.-B.Z. acknowledges support from National Thousand Young
Talents program of China and National Key Research and
Development Program of China (2018YFA0404204).
R.Ya.I. is grateful for partial support by the grant RUSTAVELI/FR/379/6-300/14. 
We thank the RATIR project team and the staff of the Observatorio Astron\'omico 
Nacional on Sierra San Pedro M\'artir. RATIR is a collaboration between the University of California, the 
Universidad Nacional Auton\'oma de M\'exico, NASA Goddard Space Flight Center, and Arizona State University, 
benefiting from the loan of an H2RG detector and hardware and software support from Teledyne Scientific and 
Imaging. RATIR, the automation of the Harold L. Johnson Telescope of the Observatorio Astronomico Nacional on 
Sierra San Pedro Martir, and the operation of both are funded through NASA grants NNX09AH71G, NNX09AT02G, 
NNX10AI27G, and NNX12AE66G, CONACyT grants INFR-2009-01-122785 and CB-2008-101958 , UNAM PAPIIT grant 
IN113810, and UC MEXUS-CONACyT grant CN 09-283. R.S.R. acknowledges support from ASI (Italian Space Agency) 
through the Contract No. 2015-046-R.0 and from European Union Horizon 2020 Programme under the AHEAD project 
(grant agreement No. 654215). SJ acknowledges the support of the Korea Basic Science Research Program through NRF-2015R1D1A4A01020961. Mondy observations were performed with budgetary funding of Basic Research program II.16 and the data were obtained using the equipment of Center for Common Use "Angara" (http://ckp-rf.ru/ckp/3056/). We are also thankful to I. M\'arquez for useful discussions.

\appendix

\section{Multi-wavelength observations of sGRBs in 2012-2015}
\subsection{sGRB 121226A}

{\it Swift} discovered sGRB 121226A (trigger=544027) on 2012 December 26 at 19:09:43 UT 
\citep{Krimm2012} which had a duration of $T_{90}$ = 1.00$\pm$0.20s and 
a hard spectrum, i.e. energy fluence ratio 50-100 keV/25-50 keV = 1.4, 
classified as a short-hard burst \citep{Baumgartner2012}. The light curve of the burst in 
{\it Swift}-BAT data has a complex structure with negligible spectral lag, which is also in good 
agreement with the phenomenology of short-hard bursts. 
The light curve of the burst in the energy range of 100-350 keV has a feature 
of $\sim$2s duration at approximately 25s after the trigger with a statistical 
significance of 3$\sigma$. This feature was also found in the light 
curve obtained by SPI-ACS {\em INTEGRAL} ($>$ 100 keV) at a significance of 
2.5 sigma. The off-axis angle of the SPI-ACS detector is 58 degrees and the detector has 
no in-flight IBAS trigger at the time of sGRB 121226A. Taking into account simultaneous detection of 
the {\it Swift}-BAT and {\em INTEGRAL} SPI-ACS of the feature 25s after the burst onset, we 
can classify it as EE. The corresponding fluence of EE component in SPI-ACS is S$_{EE} \sim$ 
$2.4\times10^{-7}$ ~erg ~cm$^{-2}$  in the (75, 1000) keV range.

\noindent
Starting at $\sim$ 36s, 62.8s and 104s after the burst, respectively, the 
0.6\,m BOOTES-4/MET robotic telescope at the Lijiang Astronomical Observatory 
(China), 1.0m Zadko robotic telescope located at the observatory at Gingin, Australia and {\it Swift}-UVOT 
responded automatically to the trigger and did not find any optical afterglow down to  
a limiting magnitude of 19-20 mag \citep{Guziy2012, Klotz2012, BreeveldKrimm2012}. 
Ground-based optical follow-up observations taken with 1.04m ST at ARIES Nainital $\sim$ 2 hours \citep{Bhatt2012} 
to 11.5 hours \citep{Xu2012} after the burst did not detect any optical source at the XRT location \citep{Littlejohns2012}. 
However, GTC 10.4\,m multi-band observations taken 10.2--10.8 hour 
post-burst \citep{Castro-Tirado2012} show a faint optical source consistent with the XRT position. 
The finding chart locating the XRT error circle is shown in Fig. A1 based on the data taken by the 
GTC 10.4\,m as a part of the present analysis. Magnitudes of the optical source detected by the GTC 10.4\,m in the $r,i,z$ 
bands are reported in Table 5. Observations at the same location using the 3.6\,m TNG $\sim$ 15.4d 
after the burst also detect an object \citep{Malesani2012} which did not appear to have faded  
in comparison to the detection in the $r$ band taken much earlier by the 10.4\,m GTC. However, the
$(r - i)$ and $(z - r)$ colors of the 10.4\,m GTC data is similar to those of other optical 
afterglows, though with large photometric errors. Our follow-up observations by the 10.4\,m GTC taken
around 53d post-burst in $i$ ($>$ 24.5 mag) and $z$ ($>$ 23.8 mag) pass-bands place deep 
limits for any possible host galaxy or possible underlying ``kilonova'' emission in the 
observed pass-bands. However, the 10.4\,m GTC multi-band data from the present analysis 
together with those observed by \citet{Malesani2012} do not firmly establish afterglow decay nature of the 
optical source coincident with the {\it Swift}-XRT \citep{Littlejohns2012} and VLA \citep{Fong2014n} 
detections. Considering that the optical source is not the host galaxy, flatter behaviour of the source between 0.5d to 
15.4d post-burst has a luminosity of $L_r < 1.2\times10^{27}$ erg/s/Hz for an assumed redshift z $\sim$ 0.5. This 
luminosity corresponds to 5 times brighter than the rest-frame luminosity of any possible GW170817 like ``kilonova'' 
at similar epochs and indicates, infered value of luminosity to be afterglow dominated as seen
in case of some of the sGRBs \citep{Rossi2019}.
It is also notable that the {\it Swift}-XRT spectral analysis favors a higher 
Galactic absorption column density towards the burst direction \citep{Littlejohns2012} having a 
steeper photon index. Further deeper observations would be required to look for any possible blue 
dwarf galaxy within the XRT error circle.  

\begin{figure}
\centering
\includegraphics[width=\columnwidth]{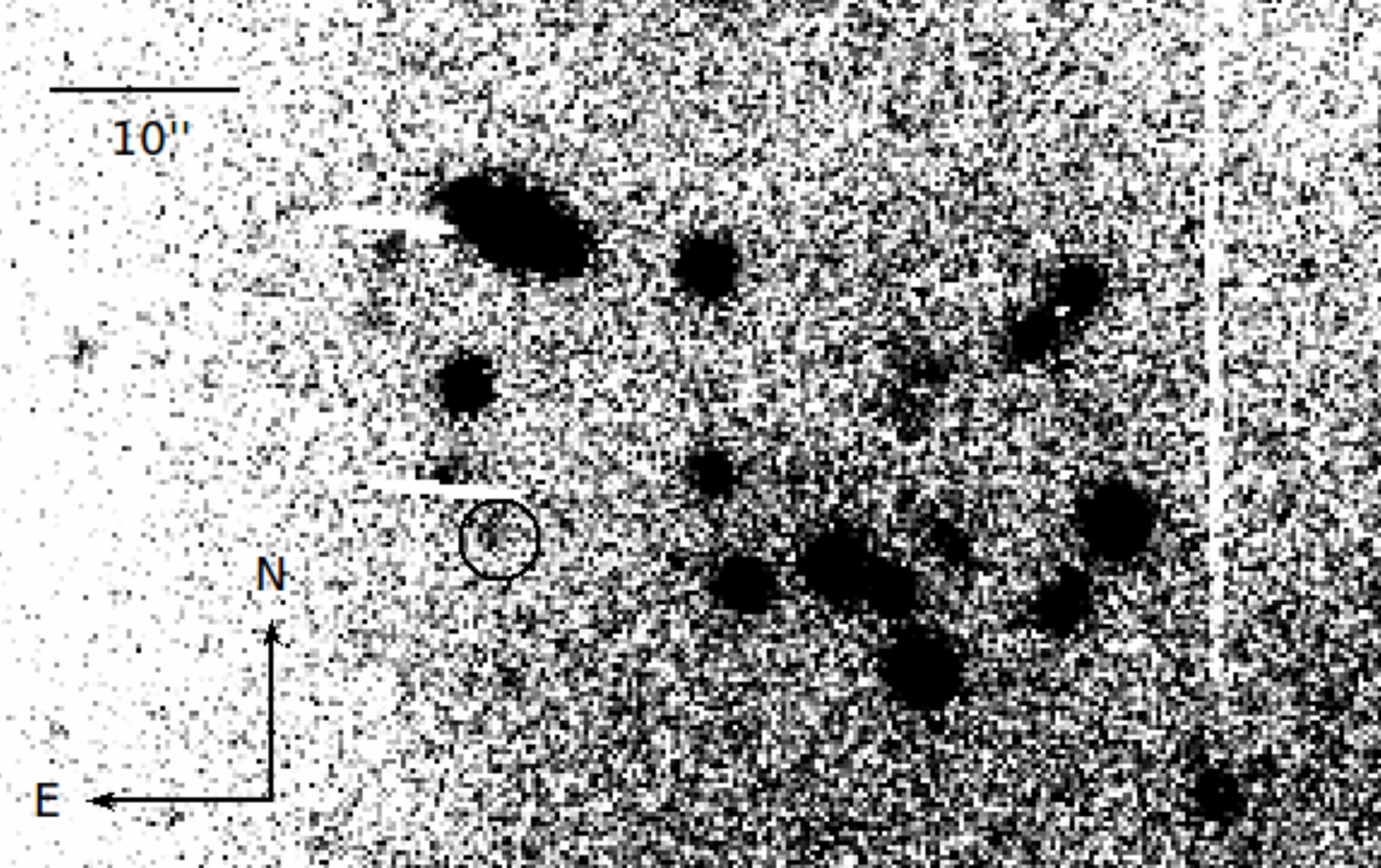}
\caption{\label{light} Finding chart of sGRB 121226A in the stacked frame of $r$ band data obtained by the GTC 10.4\,m telescope. The optical afterglow candidate within the XRT error box 
reported in \citet{Castro-Tirado2012} is circled in the chart.}
\end{figure}

\subsection{sGRB 131224A} 

sGRB 131224A was discovered on 2013 December 24 at 16:54:37 UT by the Imager on Board of the 
{\em INTEGRAL} Satellite (IBIS/ISGRI) with a fluence in the energy range 20- 200 keV of about 
$\sim 3\times10^{-8}\rm  ~ erg ~ cm^{-2} s^{-1}$  and duration of $T_{90}$ $\sim$ 0.8s \citep{Mereghetti2013}. The burst position is 2.7 degrees off axis and was also found by the Joint European X-Ray Monitor  (JEM-X),
X-ray telescope on-board {\em INTEGRAL}. The refined coordinates (J2000) are: R.A.= 296.821 deg, DEC.= +31.663 deg with an uncertainty of 1 arcmin (90\% c.l.). The burst is located (in projection) in the Galactic plane. Spectral lag between the light curves in energy ranges 3-35 keV and 20-200 keV is negligible. The burst consists of a single FRED pulse in the 3-35 keV energy range, emission is visible up to 4s after the trigger and nearly symmetric in the hard IBIS/ISGRI channels as derived in the analysis presented in this paper (see Fig. A2). 
Further, we analyzed {\it Fermi}/GBM data and found that sGRB 131224A was within the field of view but didn't 
trigger {\it Fermi}/GBM. In the temporal analysis, we found nothing significant in the {\it Fermi} daily Time-Tagged Event 
(TTE) data.
\noindent
Optical observations of the {\em INTEGRAL} error-box by the MASTER-II  
robotic telescope starting $\sim$ 39s after the burst trigger do not reveal any counterpart down
to a limiting magnitude of $\sim$15.5 mag \citep{Gorbovskoy2013}. {\it Swift-} XRT and UVOT observations
starting around 2.9h after the burst do not reveal any X-ray counterpart 
down to a limiting flux of $\sim 1.4\times10^{-13}\rm  erg ~ cm^{-2} s^{-1}$ 
\citep{Gompertz2013} or to a 
limiting magnitude of $\sim$ 21.1 mag in the UVOT $u$-band \citep{BreeveldDePasquale2013}, consistent with those
seen in the case of other sGRBs.

\noindent
It could also be discussed whether the event 131224A genuinely is a GRB event. The burst energy and morphology is very similar to 
type-I X-ray bursts which are thermonuclear flashes on the surfaces of weakly magnetic accreting neutron stars in low-mass X-ray 
binaries (LMXBs, for reviews see e.g. \citep{Lewin1995, Bildsten2000}). The burst is unusually soft for a short GRB and 
is not detected above 70 keV. The duration of the event in the soft (3-20 keV) energy band is 10s longer than in 
the hard (20-70 keV) energy band. The burst came from the direction of the Galactic plane, where the greatest number of known LMXBs are located. If the event is a type-I X-ray burst and taking into account no detection of any persistent X-ray emission 
in the follow-up XRT observation then this source is a new member of the rare class of X-ray bursters 
with very low ($<10^{35}$ erg/s) luminosity, the so-called ``burst-only'' sources (see e.g. \citet{Cornelisse2004} and references therein). 

\noindent
Deeper observations of this burst were performed under our program using the 10.4\,m GTC starting 1.11d and around 7d after the burst in $i$ and $z$ filters. Within the JEM-X {\em INTEGRAL} error-box no new fading 
source was revealed down to a limiting magnitude of $\sim$ 23.6 mag in $i$ band. 
The photometric results based on our analysis of the GTC data are tabulated in Table 5.   

\begin{figure}
\centering
\includegraphics[width=\columnwidth]{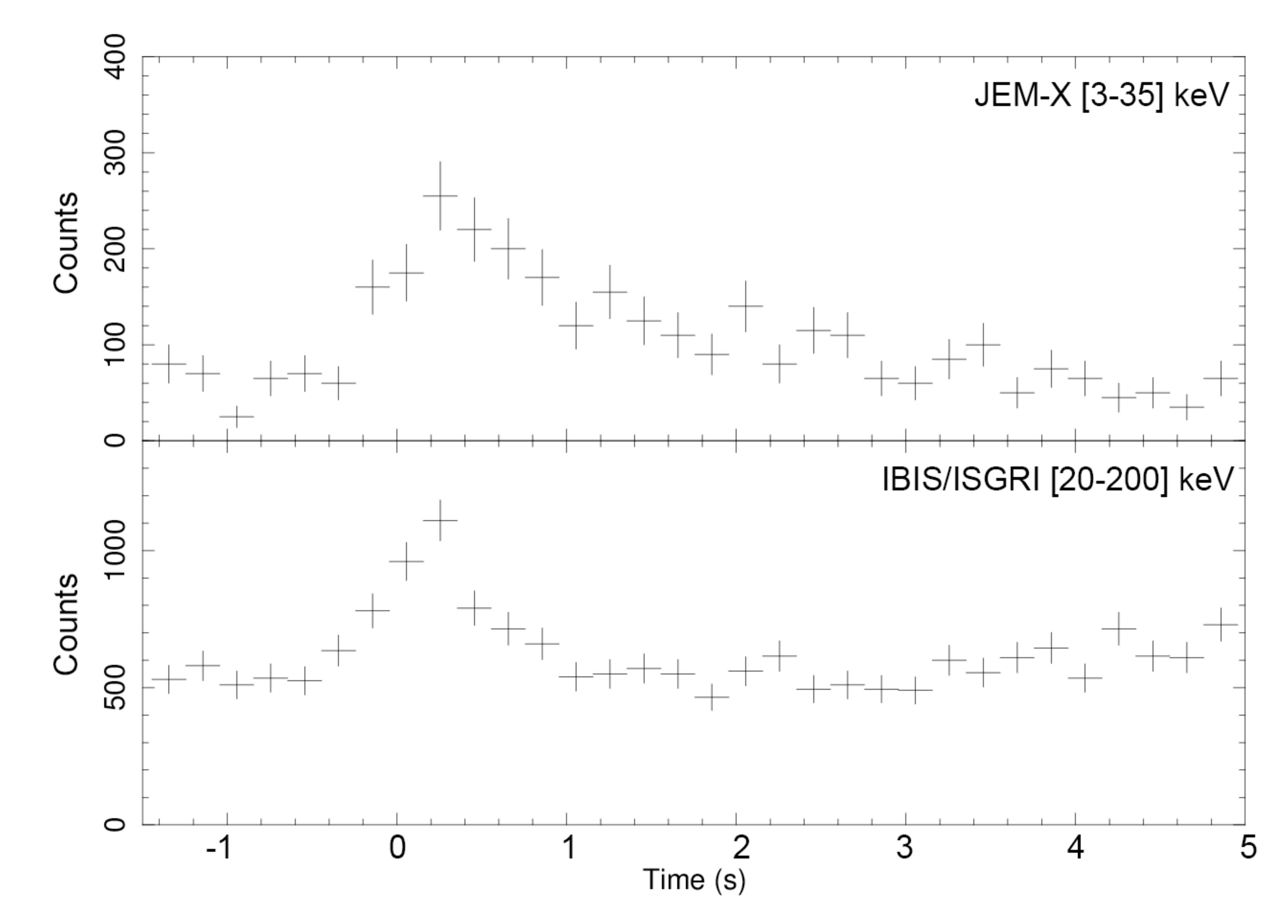}
\caption{\label{light} Light curve of sGRB 131224A obtained by JEM-X (top) and IBIS/ISGRI (bottom) on-board the {\em INTEGRAL} observatory with time resolution of 0.2 sec.}
\end{figure}
  
\subsection{sGRB 140606A}

{\em Swift} discovered sGRB 140606A (trigger=600951) on 2014 June 06 at 10:58:13 UT
which had a duration of $T_{90}$ = 0.34$\pm$0.09 s \citep{Stroh2014, Cummings2014}. The time-averaged spectrum from T-0.04 to T+0.35 is best fit by a simple power-law model. The burst is not visible in the soft energy channel (15-25 keV) and has negligible spectral lag. This confirms the short nature of the burst. 
{\it Fermi}/GBM data of the sGRB 140606A show that the burst was seen within the field of view but didn't trigger 
{\it Fermi}/GBM. However, significant gamma-ray emission in the {\it Fermi} Daily continuous Time-Tagged Event (TTE) data archive. 
We fit the spectrum of NaI n4 between T0-0.04 and T0+0.8s and found that cutoff-PL model is the best fit to the data. 
The low-energy photon index = 0.82$^{+1.34}_{-0.97}$ and E$_{\rm peak}$ = 185.13$^{+126}_{-28}$ keV. The corresponding GBM flux 
is $\sim$ 6.0 $\times10^{-7}\rm ~erg ~ cm^{-2}~ s^{-1}$  in 1-10$^{4}$ keV. The spectral fitting plot with cutoff-PL model
is shown in Fig. A3 (top panel).  
The burst was detected by IBAS in SPI-ACS {\em INTEGRAL} (off-axis angle is 40 deg) as a 0.25s single pulse and we do not detect EE
 (for details of SPI-ACS data analysis see, \citet{Minaev2010}). 
At a time scale of 50s, the upper limit on EE activity in SPI-ACS for sGRB 140606A is $\sim$ 7000 counts i.e.
fluence S$_{EE} \sim$ $(7.0\times10^{-7}$ ~erg ~cm$^{-2}$) at the 3$\sigma$ significance level 
in the (75, 1000) keV range.

No XRT counterpart of this burst could be observed due to an observing anomaly \citep{BurrowsKennea2014}. 
{\it Swift} UVOT observations, starting $\sim$ 68 sec after the BAT trigger, do not detect any
new optical source within the error circle \citep{MarshallStroh2014} down to a limiting magnitude of 
$\sim$ 20 mag. Further optical observations by \citet{Xu2014},
also do not find any new optical source within the BAT error circle. Optical observations using the Abastumani AS-32 telescope starting 0.332d after the burst do not find any optical afterglow down to a limiting magnitude of $\sim$ 21.7 in a clear filter as reported by \citet{Volnova2014}. 

\noindent
The field of sGRB 140606A was observed in B, V and Rc bands with the 6\,m BTA/Scorpio-I (SAO RAS, Russia) on the night of June, 7 2014. The observations started 10 hours after the trigger \citep{Moskvitin2014}. The first BTA image covers 100\% of the BAT refined error circle. In the stacked R-band image we detected a few hundred objects down to the limiting magnitude $R \sim 24.1$ mag (total exposure of 150 seconds). 
The stacked image combined from all obtained frames (total exposure of 480 
seconds) covers 14.7 square minutes, 82\% of the BAT circle. The limiting magnitude of this image is $R \sim 26$ mag. The field was also observed with the 10.4\,m GTC/Osiris (ORM, Spain) on February 26 2015, almost 9 months after the burst. 
The stacked image combined from $5 \times 60 + 10$ seconds frames in $r^{\prime}$ band covers 13.2 square minutes, 73\% of the BAT circle. We detected a few hundred faint objects down to the same limiting magnitude $R \sim 26$ mag. The brightest galaxies in the BAT circle are USNO 1275-0258796 and 1275-0258743 with magnitudes of about $R \approx 18$. Due to the large number of objects in the BAT circle we can not suggest a single candidate for the host galaxy or any possible flaring activity by an active galaxy in the observed 
error circle.  
As a part of the present analysis, mm-wavelength observations using the IRAM Plateau de Bure Interferometer for the full BAT error circle do not result any detection down to 
a limiting flux of 0.33$\pm$0.19 mJy around 4-15d post burst. The details of the mm observations of the sGRB 140606A taken at 86.74 GHz are tabulated in Table 2. 

\noindent
A blue object within the sGRB 140606A BAT error box at coordinates RA=13 27 07.9, Dec=+37 37 10.8 (1 arcmin error) with magnitude R =20.60$\pm$0.04 was found to be a quasar at z = 1.96 (see Fig. A3, bottom panel). The expected chance of finding a quasar within the BTA field of view is $\sim$ 0.08 (following the QSO surface number from \citep{KooKron1982}) but the lack of variability between the initial BTA frame and the late-time GTC image does not support a relationship. As mentioned above, due to lack of full coverage of the BAT error circle, the chance coincidence of the QSO gamma-ray flaring with the observed sGRB 140606A can not be established.  

\begin{figure}
\centering
\includegraphics[height=7.5cm,width=7.5cm]{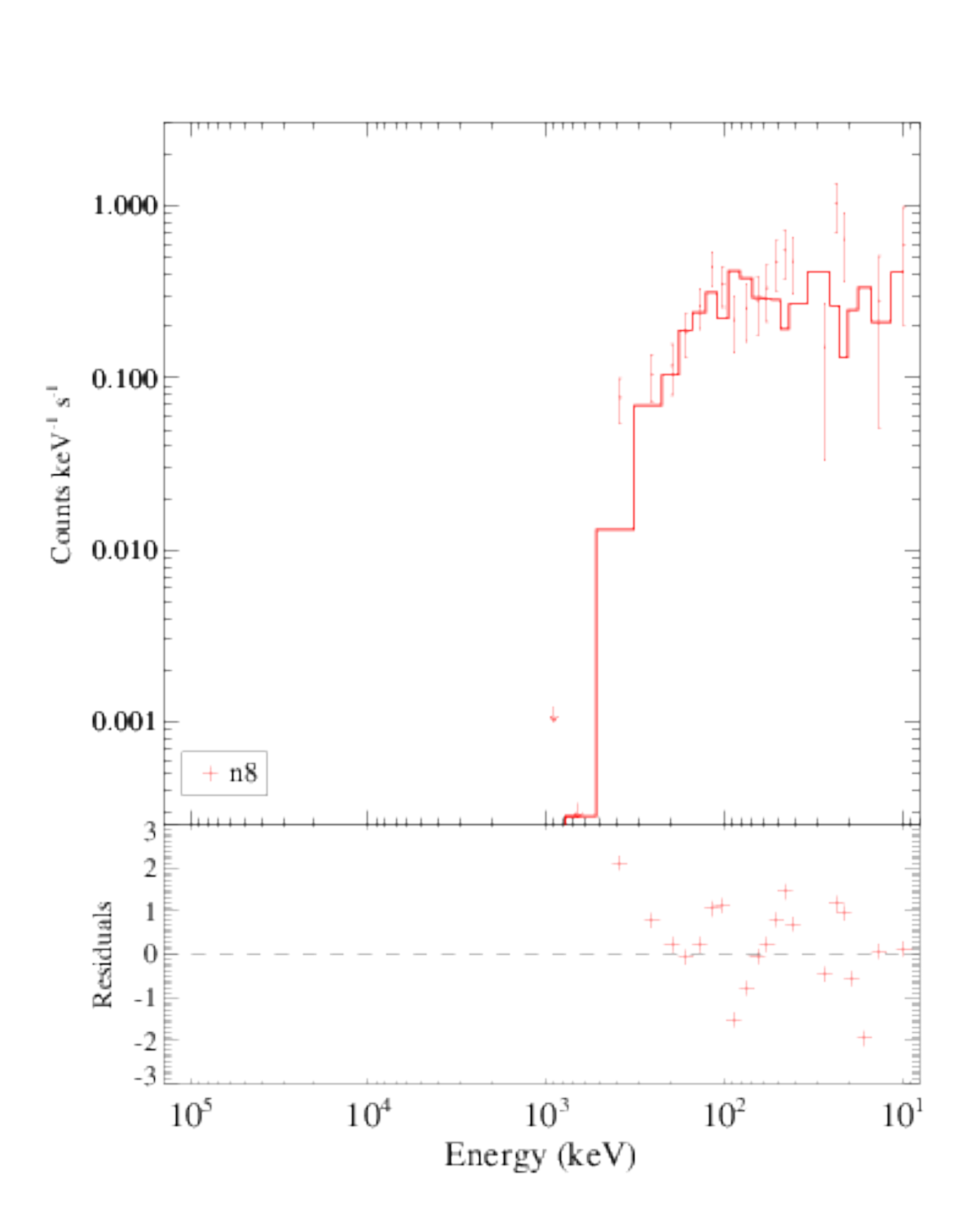}
\includegraphics[height=6.0cm,width=7.5cm]{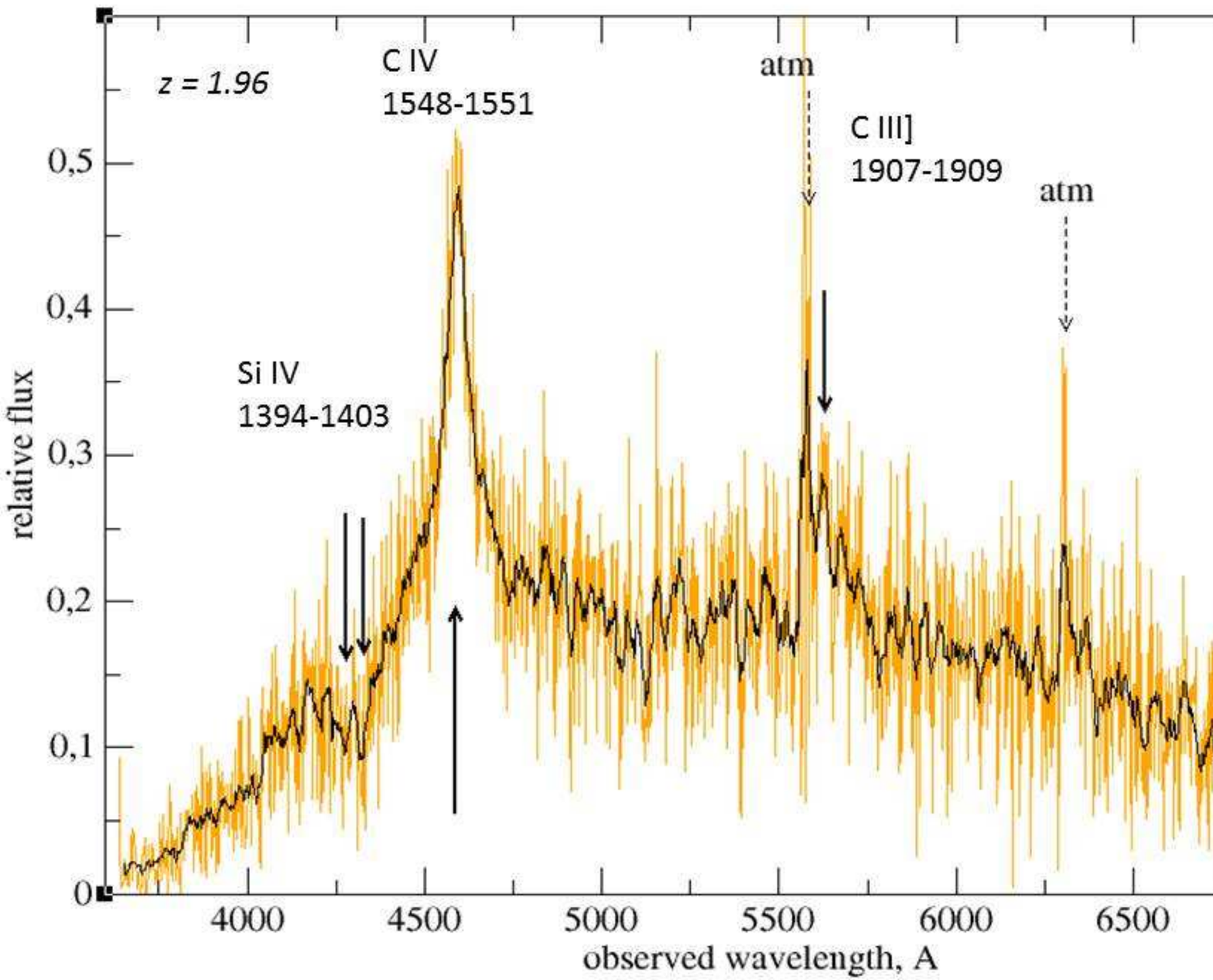}
\caption{\label{light}The best fit model of the prompt emission spectra of the {\it Fermi}/GBM (top panel) data of sGRB 140606A. The 6.0\,m BTA (+SCORPIO) spectrum (4$\times$900s) taken on 07 June 2014 of the new QSO (RA=13 27 07.9, Dec=+37 37 10.8 discovered within the sGRB 140606A BAT error box showing 
the typical QSO emission lines at a redshift z = 1.96 $\pm$ 0.1 (bottom panel).}
\end{figure}

\subsection{sGRB 140622A}

{\it Swift} discovered sGRB 140622A (trigger=602278) on 2014 June 22 at 09:36:04 UT with a duration of $T_{90}$ = 0.13$\pm$0.04 s \citep{D'Elia2014, Sakamoto2014}. The mask-weighted light curve shows a weak single FRED peak with a soft 
spectrum, which is best fit by a black-body  with kT = 11.6$\pm$1.8 
keV which is not typical for the class of short bursts \citep{Sakamoto2014}. The 
quickly fading X-ray light curve (temporal decay index, 7.1$\pm$0.9 and 
mostly taken in photon counting mode) does, 
however, appear consistent with a short burst model, and does not appear 
to be similar to the light curves of SGRs or other Galactic sources 
\citep{Burrows2014}. The burst was not detected by SPI-ACS {\em INTEGRAL} most 
probably due to the soft spectrum. The SPI-ACS {\em INTEGRAL} off--axis is 125 degrees. The early optical 
observations by 0.25m TAROT \citep{Klotz2014} $\sim$ 23.2 s post-burst,
by {\it Swift} UVOT $\sim$ 97 s post-burst \citep{MarshallDElia2014} and 
by 0.76\,m KAIT $\sim$ 198 s post-burst \citep{Zheng2014} do not reveal any optical source down to a 
limiting magnitude of $\sim$ 18, 21 and 19 mag respectively. 
However, optical observations taken by the TSHAO Zeiss-1000 (East) 
telescope starting 0.475d after the burst in $R_c$ filters with an exposure 
time of 60$\times$60s+5$\times$240s marginally detect a source at 
RA=21 08 41.69 Dec= -14 25 08.7 ($\pm$ 0.22'') at a magnitude of 22.5$\pm$0.3 mag. 
In the light of other non-detection to deeper limits from the data taken before and after 
the epoch of observations by TSHAO Zeiss-1000 (East), it seems that this 
marginal detection could be false one. So, an upper limit of $\sim$ 22.5 mag is reported in Table 5.
The 2.2m GROND observations taken $\sim$252 s after the burst do not 
reveal any optical counterpart within
the XRT error-box down to a limiting magnitude of $\sim$ 24.3 mag, however 
they do detect an optical source just outside the XRT error circle 
\citep{Tanga2014} at a measured redshift of z $\sim$ 0.959 using VLT observations 
\citep{Hartoog2014}. At this redshift, the host distance from the XRT 
error circle would be around 21 kpc which could easily rule out the suspected 
galaxy as a potential host for sGRB 140622A. The XRT error-box was also observed by the RATIR camera at the 
1.5m telescope starting $\sim$ 1.2 min after the burst in several filters and no counterpart could be 
detected to deeper limits \citep{Butler2014}. 
As a part of the present analysis, mm-wavelength observations using the IRAM Plateau de Bure Interferometer for the full BAT error circle do not result any detection down to 
a limiting flux of -0.37$\pm$0.12 mJy within a few hours post burst. The details of mm observations of the sGRB 140622A taken at 86.74 GHz are tabulated in Table 2. 

\noindent
So, to search for the potential host galaxy/counterpart, we triggered our
proposal on the 10.4\,m GTC. The analysis of the GTC $r$-band data 
(6$\times$100+5$\times$2 s) reveal that there is no optical counterpart down to 
a limiting magnitude of $\sim$ 25.8 mag at around 0.78d post-burst. So, it is clear from the above
observations that the host galaxy of this burst is fainter than $\sim$ 25.8 mag.
It is worth mentioning that no detection of any host galaxy down to 
a deep limit of $r \sim$ 25.8 mag indicates sGRB 140622A to be a candidate belonging to the 
sub-set of other host-less events \citep{Berger2010, Tunnicliffe2014}. 
The {\it Swift}-BAT fluence in the 15--150 keV band is 2.7$\pm$0.5$\times10^{-08}\rm ~ erg ~ cm^{-2}$
along with a $<$ 0.3 micro-Jansky limit at optical frequencies place a very crude limit for this burst
as a possible high redshift one \citep{Berger2010}. Early epoch deeper observational 
limits at optical wavelengths and along with unusual {\it Swift}-BAT and XRT spectra \citep{Sakamoto2014, 
Burrows2014} also indicate the peculiar nature of this burst. The finding chart locating the XRT 
error-circle is shown in Fig. A4 based on the data taken by the GTC 10.4\,m.
  
\begin{figure}
\centering
\includegraphics[width=\columnwidth]{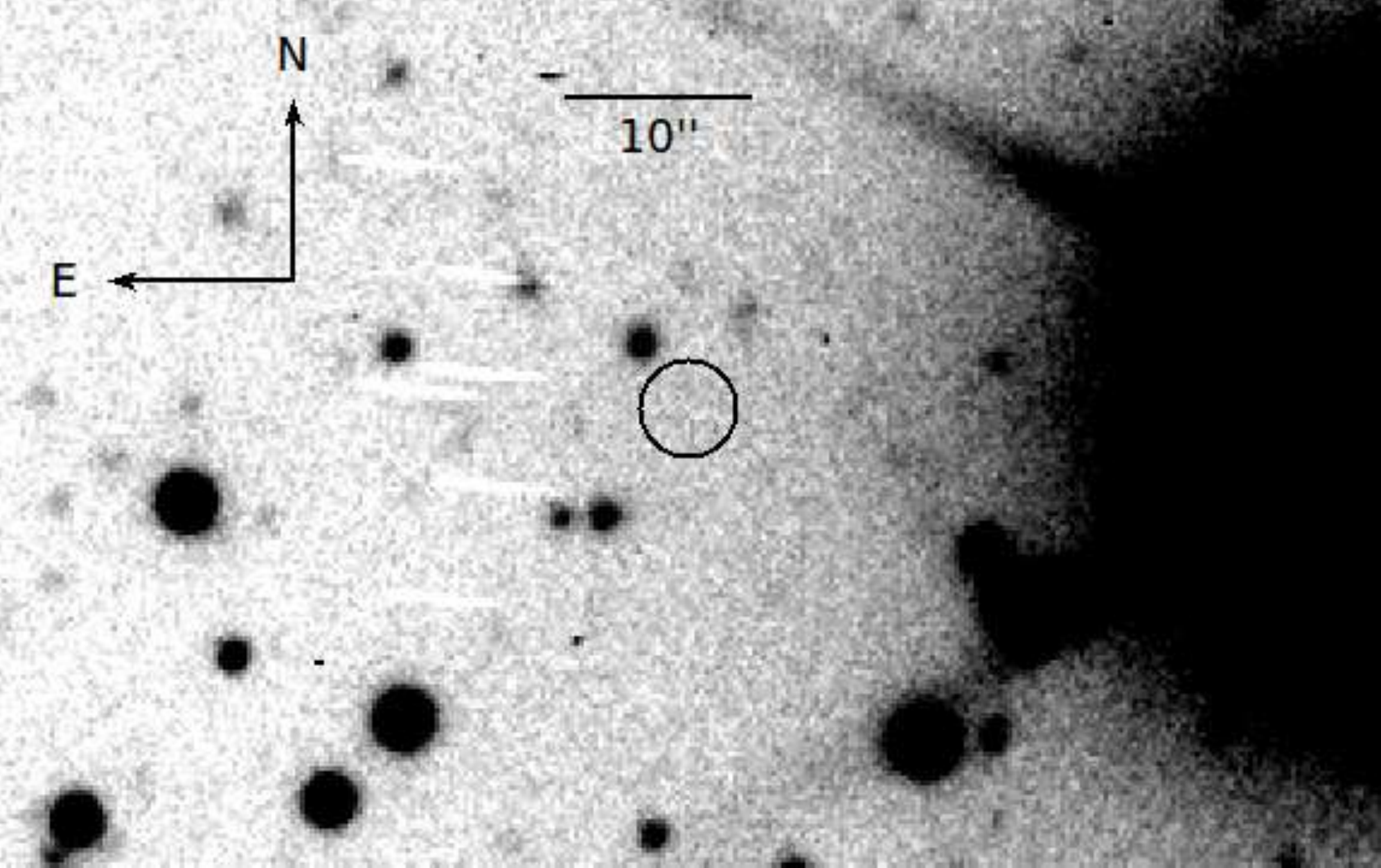}
\caption{\label{light} Finding chart of sGRB 140622A in the stacked frame of $r$ band observed by the GTC 10.4\,m 
telescope. The black circle is the XRT error box, having no signature of the optical afterglow down to a 
limiting magnitude of $\sim$ 25.8 mag $\sim$ 0.78d after the burst.}
\end{figure}

\subsection{sGRB 140903A}

{\it Swift-} BAT triggered on a possible GRB on 2014 September 03 at 15:00:30 UT. Due to a TDRSS telemetry gap, 
the XRT localization was performed $\sim$ 2.5 hours post-burst and ultimately the burst was 
found to be a duration of $T_{90}$ = 0.30$\pm$0.03s \citep{Cummings2014, 
Palmer2014}. The BAT and XRT data indicated a soft burst spectrum and 
an excess column density was observed \citep{DePasquale2014}, not very common in the case of sGRBs. The time-averaged spectrum from T-0.01 to T+0.35s was best fitted by a simple power-law model. The power law index of the time-averaged spectrum is 1.99$\pm$0.12. Extended emission was not found 
\citep{Sakamoto2014a, Serino2014} in the prompt emission light curve of this burst and the mask-weighted light curve shows a single FRED peak. 
The SPI-ACS {\em INTEGRAL} detector was switched off at the time of the burst. The spectral-lag analysis was 
performed by \citet{Sakamoto2014a} found that: the 
spectral lag for the 50-100 keV to 100-350 keV bands is 16$\pm$7ms, and 
21$\pm$ 7 ms for the 15-25 keV to 50-100 keV bands. According to \citet{Sakamoto2014a} these lag values indicate that GRB 140903A belongs to the long GRB population. This interpretation contradicts results obtained for individual pulses of BATSE bursts 
by \citet{Hakkila2011}. According to \citet{Hakkila2011}, 
short and long bursts show the same spectral evolution behavior if spectral 
lag analysis is performed for individual pulses of bursts instead of analyzing the 
whole burst structure. Similar results were also noted by \citet{Minaev2014} in their analysis of several other {\em INTEGRAL} bursts. 
sGRB 140903A is single-pulsed and belongs to the bottom-left region of the lag duration correlation constructed for individual pulses of 
BATSE bursts (Figure 3 in \citet{Hakkila2011}), which means that this burst belongs to the short GRB population. A low E$_{\rm iso}$ value 
(0.04$\times$10$^{51}$ erg, see below) is also more common for short bursts than for long ones. Recently, \citet{Troja2016} have shown that the burst has 
negligible lag and other prompt emission properties are very typical of those in case of other sGRBs. It was also noticed that this burst is located within 2.5 arc-min of the center of the galaxy cluster NSC J155202+273349 at a photometric redshift of $\sim$ 0.295 \citep{FoxCummings2014, Gal2003}. However, 
\citet{Troja2016} have established that the burst was not associated with the 
galaxy cluster. 

\noindent
The optical afterglow of this sGRB was discovered by the 4.3m Discovery 
Channel Telescope (DCT) within the XRT error circle around 12 hours after the 
burst \citep{Capone2014, Troja2016}. The optical afterglow candidate was also 
seen in further follow-up observations \citep{CenkoPerley2014, Dichiara2014, 
Xu2014a}. \citet{Furchter2014} noticed that the candidate optical afterglow 
was present in archival images of the Pan STARRS survey and was later 
suspected to be the host galaxy candidate. \citet{Troja2016} measured the 
redshift of the afterglow as $\sim$ 0.351 using the Gemini-N 8.0\,m
telescope equipped with 
the Gemini Multi-Object Spectrographs (GMOS) camera. The fading behavior of the optical afterglow candidate was 
established in further observations by \citet{Levan2014} and \citet{Cenko2014}. 
The radio afterglow of the burst were also observed by JVLA at 6 GHz 
\citep{Fong2014a, Troja2016} and by GMRT at 1390 MHz \citep{NayanaChandra2014}.
However, mm-wavelength observations using the IRAM Plateau de Bure Interferometer at the XRT location do not 
result any detection down to a limiting flux of 0.12$\pm$0.13 mJy within a few days post burst. The afterglow
modeling of the multi-band data by \citet{Troja2016} indicates that our mm-wavelength IRAM observations were
shallower in comparison to detected signals at the level of a few micro Jy at JVLA and GMRT frequencies.
The details of our mm observations of the sGRB 140903A taken at 86.74 GHz are tabulated in Table 2. 
Spectroscopy of the afterglow was also performed using the 10.4\,m GTC and the 
redshift value determined was $\sim$ 0.351 \citep{Troja2016} consistent with 
that reported by 
\citet{Cucchiara2014}. Using the measured redshift of this burst
\citep{Troja2016} and the $\gamma$-ray fluence by \citet{Palmer2014},  
the isotropic-equivalent gamma-ray energy is
E$_{\rm iso} \sim 0.04\times10^{51}$ erg (20 to $10^4$ keV, rest-frame).

\noindent
As a part of the present work, ISON-Kislovodsk SANTEL-400A optical telescope 
started observations $\sim$ 0.141d after the burst and did not see any 
afterglow down to a limiting magnitude of $\sim$ 18.6 mag \citep{Pozanenko2014}. 
To search further for the optical afterglow or for any
possible ``kilonova'' emission for this nearby sGRB, we observed the field of GRB 140903A with 
the 1.5\,m AZT-22 telescope of Maidanak astronomical observatory on 2014 September 4, 6, 7, and 13, 
taking 12-15 images of 60 s exposure in the $R$-filter.  
All images were processed using NOAO's IRAF software package. The position of 
the optical source is in the wing of a bright star SDSS J155202.58+273611.7 
(R = 12.9 mag). The limiting magnitude for every epoch far away from the bright 
star were obtained using nearby SDSS stars. To find a possible afterglow we 
subtracted the combined image obtained on September 13, 2014 from that of 
September 4, 2014. At the position of the afterglow in the residual image we do 
not find any source implying an equivalent upper limit variability of the source 
less than 0.5 magnitudes ($3\sigma$) between the two epochs.  This is in 
agreement with observations by \citet{Xu2014a} and confirms the absence 
of an afterglow signature 30 hours after the burst trigger.  Based on our 
present observations we can also exclude the possibility of an underlying ``kilonova''  brighter than R $\sim$ 
22.0 ($3\sigma$) at 10d associated with sGRB 140903A. 
The corresponding limiting value of the luminosity for the given redshift $L_R < 6.5\times10^{27}$ erg/s/Hz
seems afterglow dominated and brighter by a factor of 6 than any GW170817 like associated ``kilonova'' at 
similar epochs \citep{Rossi2019}.  

\begin{figure}
\centering
\includegraphics[width=\columnwidth]{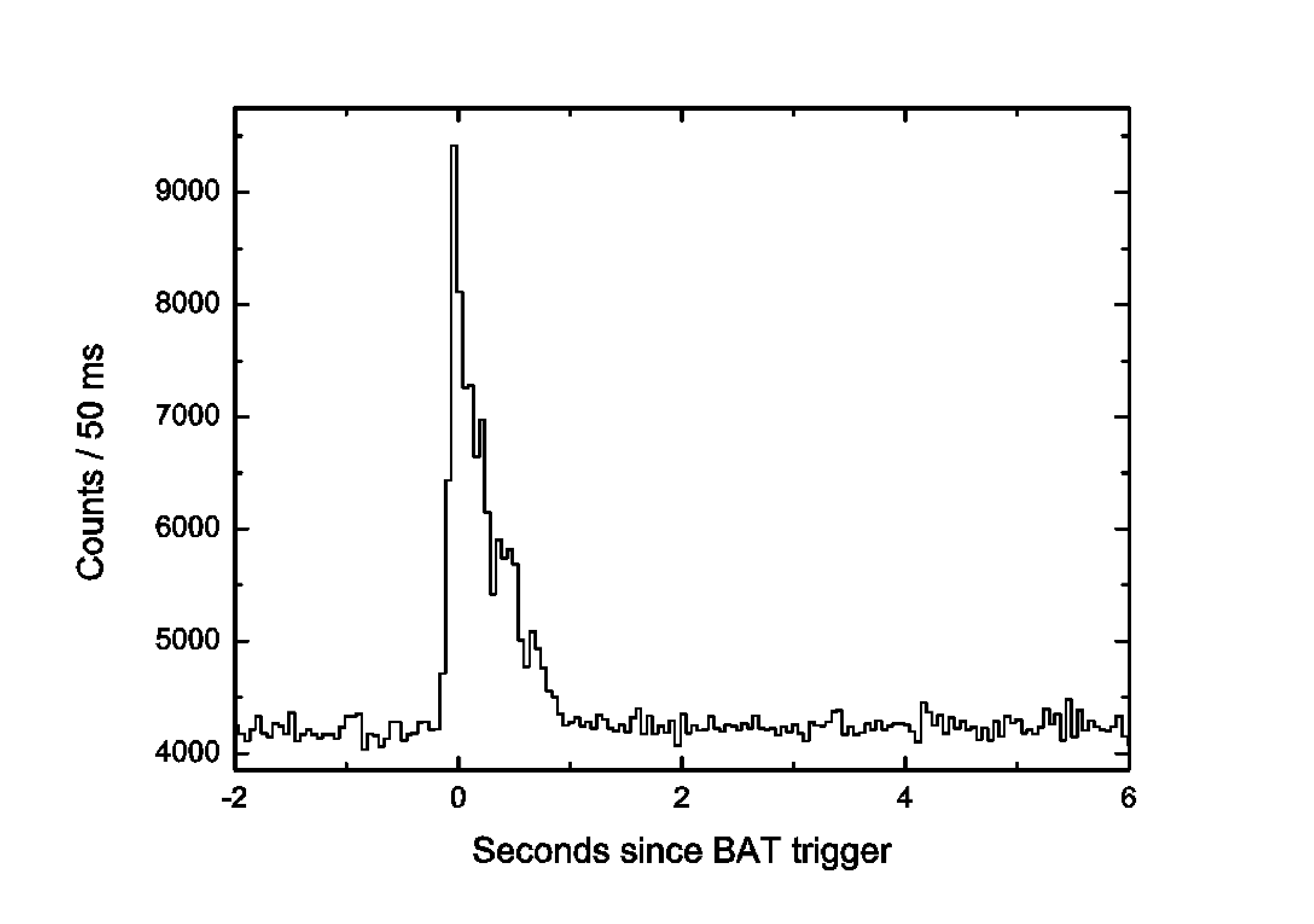}
\caption{\label{light} Light curve of sGRB 140930B obtained by SPI-ACS {\em INTEGRAL} in energy range 0.1-10 MeV with 50 ms time bins as a part of the present study. On x-axis time since BAT trigger is shown, on y-axis counts per 50 ms are presented.}
\end{figure}

\begin{figure}
\centering
\includegraphics[width=\columnwidth]{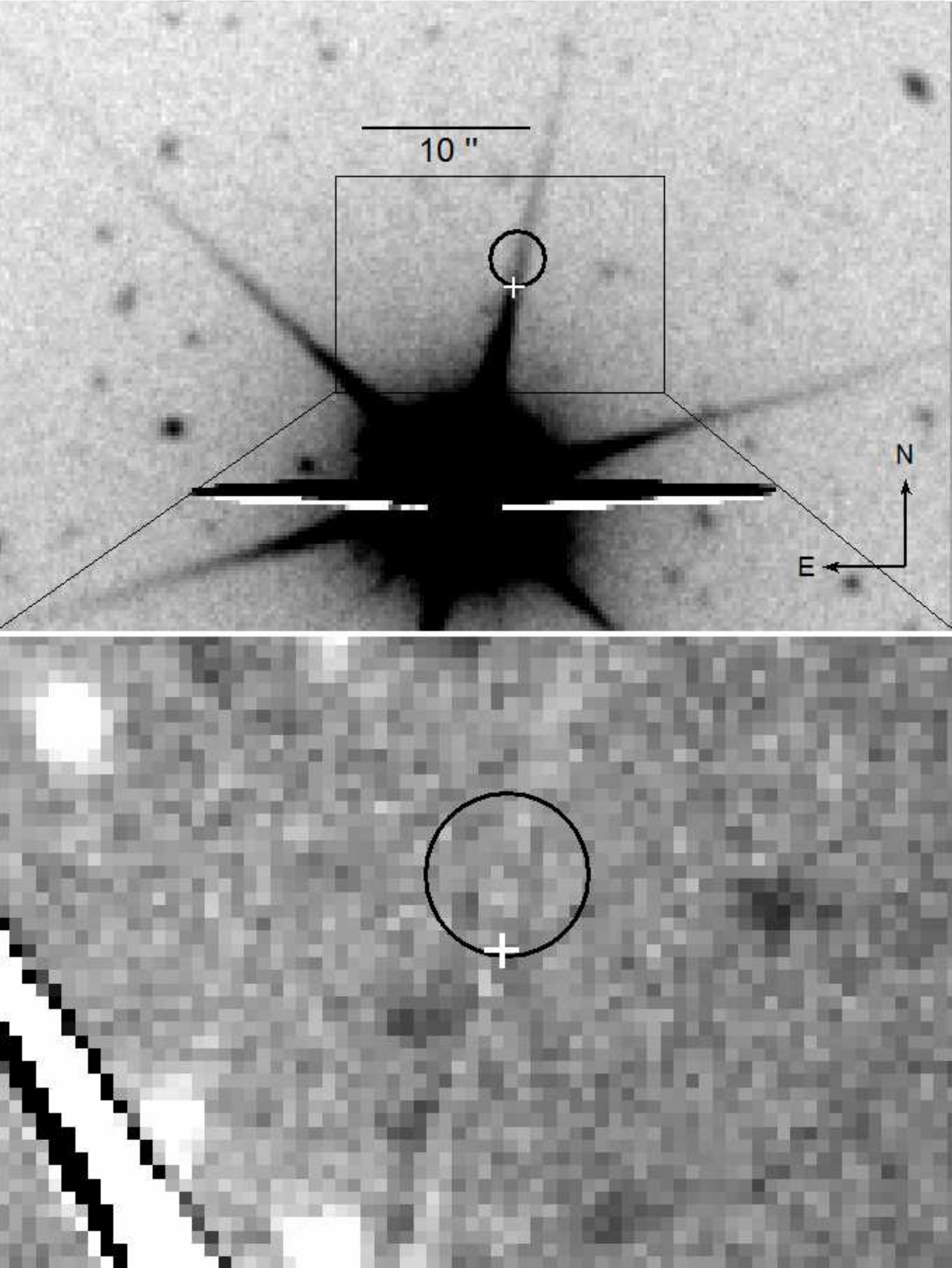}
\caption{\label{light} Finding chart of sGRB 140930B in the stacked frame of $r$ band observed by GTC 10.4\,m telescope. The XRT error box 
shown in black circle is overlapped with one of the spikes of the nearby bright star. In the bottom panel, the zoomed portion  
(inset) is shown after applying image subtraction and the ``+'' sign marks the position of the afterglow reported by \citet{Tanvir2014}} 
\end{figure}

\begin{figure}
\centering
\includegraphics[width=\columnwidth]{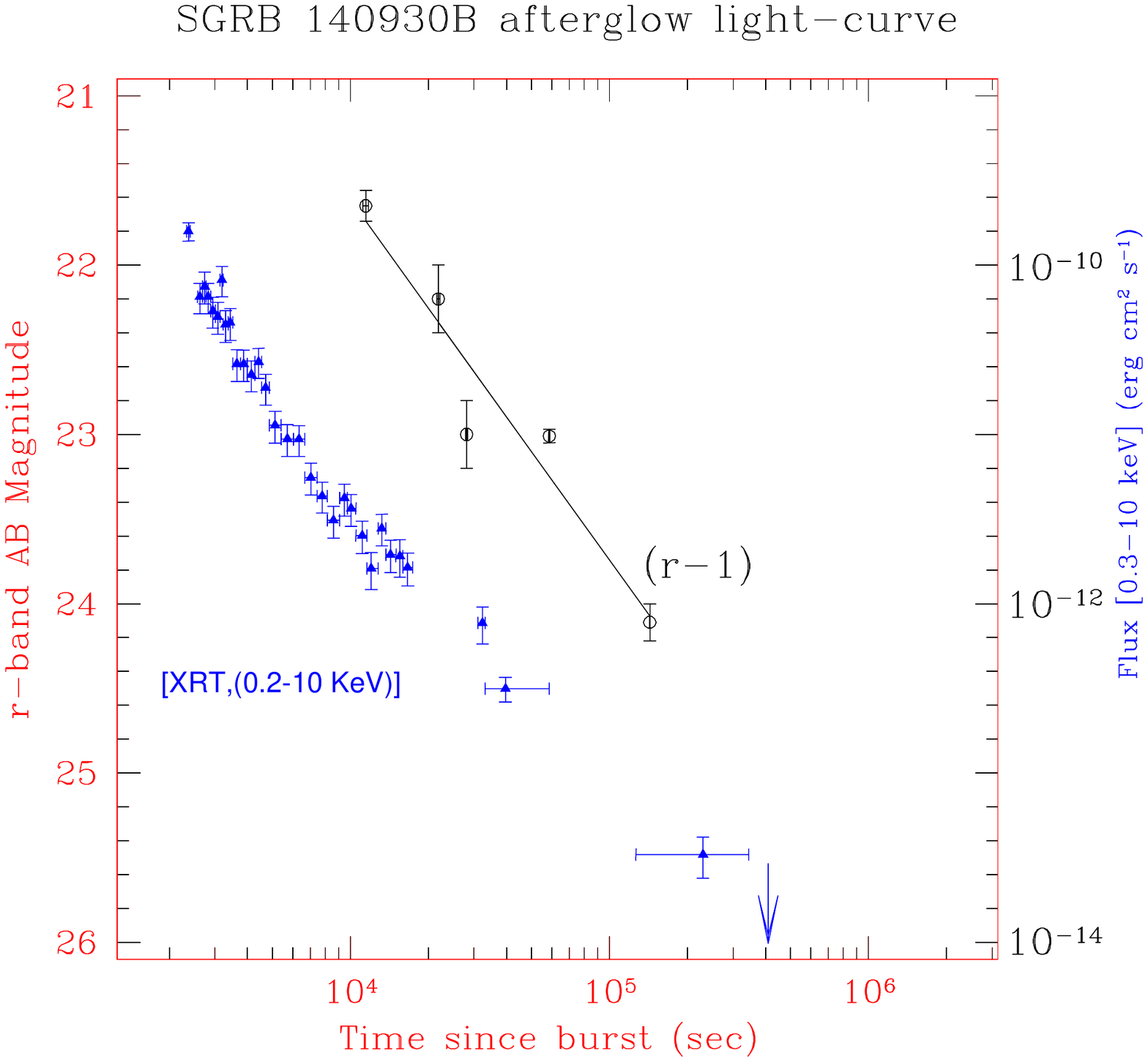}
\caption{\label{light} sGRB 140930B afterglow optical $r$ band afterglow light-curve. 
The solid black curve is the best-fit power-law model to the $r$-band light curve. 
The two $r$ band data points around 2$\times$10$^{4}$ s \citep{Fong2014b} and 
3$\times$10$^{4}$ s \citep{Graham2014} post-burst are from GCN circular archive, considered 
while fitting the power-law to derive the temporal decay index. For comparison, {\it Swift} XRT
light curves are also plotted in blue color.}
\end{figure}

\subsection{sGRB 140930B}

{\it Swift} detected sGRB 140930B (trigger=614094) on 2014 September 30 at 
19:41:42 UT with a duration of $T_{90}$ = 0.84$\pm$0.12s 
\citep{DePasquale2014, Baumgartner2014}. The burst was also observed by 
$Konus-Wind$ with the light curve having a complex multi-pulsed structure with a duration of 
$\sim$ 1.0s and the emission was seen up to $\sim$ 10 MeV \citep{Golenetskii2014}. 
The time-averaged spectrum of the burst (measured by $Konus-Wind$ from T0 to T0+8.448s) 
had a best fit in the 20 keV - 15 MeV range by a power-law with exponential cutoff model 
with E$_{\rm peak}$ = 1302$^{+2009}_{-459}$ keV and total fluence of
8.1$^{+5.1}_{-2.5}$ $\times10^{-6}\rm ~erg ~ cm^{-2}~$ \citep{Golenetskii2014}.
Since the redshift z of the sGRB 140930B is unknown, the trajectory of sGRB 140930B on the 
Amati diagram as a function of z (Fig. 9, see also \citet{Minaev2012}) can be constructed 
using the fluence and E$_{\rm peak}$(1+z) estimates. It follows from Fig. 9 that the trajectory 
does not cross the correlation region and lies above those drawn for lGRBs, which may suggest 
that sGRB 140930B belongs to the class of  short bursts. 
The higher E$_{\rm peak}$ value confirms that the burst is spectrally hard.
Overall a FRED light curve with three pulses after the
main peak is visible in SPI-ACS {\em INTEGRAL} (Fig. A5). The SPI-ACS {\em INTEGRAL} 
off-set is 67 degrees. There is no EE in either BAT 
\citep{Baumgartner2014} or SPI-ACS {\em INTEGRAL} light curves. 
At a time scale of 50 s, the upper limit on EE activity in SPI-ACS for sGRB 140930B is $\sim$ 7300 counts i.e.
S$_{EE} \sim$ $(7.3\times10^{-7}$ ~erg ~cm$^{-2}$) at the 3$\sigma$ significance level 
in the (75, 1000) keV range.
{\it Fermi-}/GBM could not observe the burst as the satellite was passing in its South Atlantic Anomaly. 

\noindent
Early time optical observations using {\it Swift} UVOT \citep{BreeveldDePasquale2014}, 
MASTER-II \citep{Gorbovskoy2014} and 1.23m CAHA \citep{Gorosabel2014} do not 
reveal any optical afterglow down to a limiting magnitude of R $\sim$ 21.1 mag. 
UAFO ORI-65 and ISON-Kislovodsk  SANTEL-400A telescopes started observations 
around 0.025 and 0.029d after the burst and did not see any afterglow down 
to a limiting magnitude of $\sim$ 16.1 mag and 20.4 mag respectively 
\citep{Polyakov2014}.  However, starting $\sim$ 3 hours after the burst 4.2\,m 
WHT found an optical source \citep{Tanvir2014} that decayed in later 
images obtained by the 6.5m MMT \citep{Fong2014b} and the 2.2m GROND \citep{Graham2014} 
telescopes. The spectroscopic observations using Gemini-south were reported by 
\citet{Cenko2014} and the afterglow candidate was also observed in $J$ and 
$K_s$ bands using Keck-MOSFIRE \citep{PerleyJencson2014}. 

\noindent
We started to observe the field of GRB 140930B on October 3, 2014 at 
22:58:33 UT, i.e. $\sim$ 3.1d after the trigger taking 13 frames with 
an exposure of 60 seconds in the $r$ filter under mean FWHM of 0.8 arcsec using GTC 10.4\,m. 
The refined position of the optical and infrared afterglow is strongly affected by a spike 
from nearby bright star S1 (J002523.61+241727.0, $r \sim 13.1$ mag). 
All bright stars in the frames from GTC have six symmetrical spikes from a 
secondary mirror mount. We found the central position of the S1 star and 
then we rotated the combined image around this position 60 degrees clockwise, 
to use a rotated image as a template for subtraction of the spike contaminating 
the position of the afterglow. In the resulting image we do not find any 
source at the position of the optical afterglow down to limiting magnitude of 
$r \sim 24.5$ mag. The finding chart locating the XRT error circle is shown in Fig. A6 
based on the late time data taken by the GTC 10.4\,m. At the epoch of our GTC observations,
limiting value of afterglow luminosity would be $L_r < 1.3\times10^{27}$ erg/s/Hz for an assumed 
redshift of z $\sim$ 0.5. This value is nearly similar to the expected luminosity of GW170817 like 
``kilonova'' at similar epochs \citep{Rossi2019}. 

\noindent
Also, as a part of the present study, the 4.2m WHT/ACAM and Gemini North/GMOS-N 
photometric data of the optical afterglow \citep{Tanvir2014} were analysed.
The results based on our multi-band photometry using the 4.2\,m WHT, the Gemini North and GTC 10.4\,m 
are reported in Table 5. The photometry was performed using NOAO's IRAF software package and calibrated 
using nearby SDSS stars. 
The $r$ band photometry from the present study along with those given in the 
GCN \citep{Fong2014b, Graham2014} were used to produce the afterglow light as shown
in Fig. A7. The temporal flux decay index using the $r$ band lightcurve was derived 
as $\alpha_o = 0.85\pm0.26$ during 0.13 to 1.65d after the burst. The contemporaneous 
{\it Swift} XRT light curve decay index is $\alpha_X = 1.6\pm0.1$ where as X-ray spectral index
$\beta_X = -0.71\pm0.15$. Assuming the cooling break frequency $\nu_c$ lying between
the two observed bands, the closure relations in case of the {\em ISM}
afterglow model \citep{Sari1998} are broadly consistent with the observed values of 
temporal decay at optical bands whereas the temporal decay index at X-rays are steeper than
the expected model predictions. 
GTC 10.4\,m was further triggered to search for any possible host galaxy on 10th Dec 2018 and
a total of 30 images of 120s each were acquired (see Table 5) in $r$-band. In the stacked image, we 
do not see any object down to a limiting magnitude of $\sim$ 24.8 mag at the location of the afterglow
after accounting far the possible effects of the nearby bright star.
So, it is concluded that the host galaxy of the sGRB 140930B would be fainter than $r \sim$ 24.8 mag. 
 
\subsection{sGRB 141212A}

sGRB 141212A was discovered on 2014 December 12 at 12:14:01 UT by the {\it Swift}-BAT \citep{Ukwatta2014}. The BAT light curve 
shows a single spike with a duration of about 0.1 sec in the energy range (25--350) keV. In the soft energy channel 15--25 keV a second 
pulse is clearly visible with a duration of 0.1s at 0.3s after the trigger. The duration parameter $T_{90}$ in the 15--350 keV energy 
range is 0.30$\pm$0.08s \citep{Palmer2014a}. The time-averaged spectrum from T+0.00 to T+0.34s is best fit by a simple 
power-law model with power-law index of 1.61$\pm$0.23. The fluence in the 15--150 keV band is 7.2$\pm$1.2$\times10^{-08}\rm ~ erg ~ cm^{-2}$
\citep{Palmer2014a}.
GRB 141212A was also found in {\em INTEGRAL} SPI-ACS data (there was no IBAS trigger) as a single pulse with duration of  0.15 sec and 
statistical significance of 7.3 sigma (Fig. A8). The second soft pulse is not visible in SPI-ACS  which is sensitive above $\sim$ 80 keV.
At a time scale of 50s, the upper limit on EE activity in SPI-ACS for sGRB 141212A is $\sim$ 7300 counts i.e.
S$_{EE} \sim$ $(7.3\times10^{-7}$ ~erg ~cm$^{-2}$) at the 3$\sigma$ significance level 
in the (75, 1000) keV range.
Ground based MITSuME \citep{Fujiwara2014}, MASTER network of telescopes \citep{Gres2014} and UVOT on-board {\it Swift} did not find
any new optical source within the XRT error-box in the images taken around 31s, 46s and 72s after the BAT trigger respectively
down to a limiting magnitude of V $\sim$ 19 mag.    

\noindent
We started observation of the sGRB~141212A with the 1.5\,m AZT-33-IK telescope at Mondy observatory on 12 
December 2014 at 12:36:10.7650 UT i.e. 
22 minutes after the trigger. We also observed it later with the same telescope on December 14 and December 18. We also observed the 
field with the 0.4m telescope at Khureltogot observatory and 1.0m telescope at Tien Shan observatory (see Table 5 for the complete log of observations).  The host galaxy suggested 
by \citet{Malesani2014} was also detected from our observations using 1.0 -- 1.5m telescopes. We did not find any evidence for the optical afterglow signature in our observations taken in $R$ filter. 
As a part of the present analysis, a deeper photometric data using Gemini-N/GMOS-N 8.0\,m (Gemini program ID = GN-2014B-Q-10) 
data in $i$ band was analyzed and the bright host galaxy candidate was clearly detected in the data taken at the two epochs as 
listed in Table 5. Using the Gemini-N/GMOS-N $i$ band data, the possibility of any point source in the vicinity of the host 
galaxy candidate \citep{Malesani2014} is ruled-out up to limiting magnitude of $i \sim 26$ mag (3-sigma) at 0.68d post-burst.
This deep limiting value translates to a luminosity of $L_i < 5\times10^{26}$ erg/s/Hz (a factor of 3 deeper than rest-frame luminosity of GW170817 like ``kilonova'' at contemporaneous epochs), further implies that at the epoch of our 
observations in $i$ band, any associated GW170817 like ``kilonova'' with the burst would have been detected as seen in a few 
cases of sGRBs in \citet{Rossi2019}. 

\noindent
As a part of the present study, multi-band photometry with the 10.4\,m GTC in $gri$-filters was performed 
at late epochs i.e. around 427.3d post burst to investigate properties of the host 
galaxy (see Table 5). The finding chart with the XRT error circle superimposed on the 
data taken by the GTC 10.4\,m is shown in Fig. A9. 
The observed flux of the host galaxy of sGRB~141212A obtained by 10.4\,m GTC in different filters (see Table 5) 
and the suggested redshift of the burst $z=0.596$ \citep{Chornok2014} allowed us to model the SED of the host galaxy. 
We also added upper limits in filters $u$ and $b$ from {\it Swift}-UVOT data \citep{Oates2014}. To build the SED of 
the host galaxy of sGRB 141212A and to estimate parameters we used the \textsc{Le~Phare} software package 
\citep{lephar1,lephar2} with fixed redshift. We used the PEGASE2 population synthesis models library 
to obtain the best-fit SED, the mass and the age of the galaxy, and star formation rate. We tested four different 
reddening laws: the Milky Way extinction law by \citet{seaton-mw}, LMC \citep{lmc}, SMC \citep{smc}, and the reddening 
law for starburst galaxies \citep{calzetti,calzetti_mod}. The reduced $\chi^2$, galaxy morphological type, bulk 
extinction, absolute rest-frame $B$ magnitude, age, mass, star formation rate, and specific star formation rate 
(SSFR) per unit galaxy stellar mass are listed in the Table~ A.1 for all 4 tested extinction laws. 
Fig. A10 represents the best model corresponding to the Milky Way extinction law.

\noindent
The best fit shows that the host is a galaxy of elliptic type with $M_B = -19.9$ mag and a moderate linear size 
along the major axis about 13 kpc. The major axis is oriented 45 degrees North-West. Age of the host galaxy is 
$\sim 2$ Gyr, and the average internal extinction in the galaxy is rather high, $E(B-V) = 0.50$ mag. The host galaxy has 
a mass of $\sim 9\times10^{9}$M$_{\sun}$, and a high star formation rate of SFR $\sim 50$M$_{\sun}$/yr. All 
obtained parameters are in a good agreement with previous studies by \citet{chrimes2018} except for SFR which 
is two orders higher in our results.

%
\begin{table*}
  \caption{sGRB~141212A host galaxy properties derived from the SED fitting.}
  \begin{tabular}{lllll}
  \hline
  \hline
Fitted                            & Starburst              & Milky Way              & LMC                    & SMC \\
parameters                        & model                  & model                  & model                  & model \\
  \hline
  \hline
$\chi^2$/DOF                      & 2.8/3                  & 2.7/3                  & 2.8/3                  & 5.9/3 \\
Type                              & E                      & E                      & E                      & S0 \\
$E(B-V)$, mag                     & 0.50                   & 0.50                   & 0.50                   & 0.00 \\
$M_B$, mag                        & $-19.9$                & $-19.9$                & $-19.9$                & $-19.7$ \\
Age, Gyr                          & $2.65^{+2.50}_{-0.11}$ & $2.23^{+1.70}_{-0.09}$ & $2.05^{+3.44}_{-0.13}$ & $3.10^{+0.09}_{-0.25}$ \\
Mass, $(\times10^{10})$M$_{\sun}$ & $1.0^{+5.4}_{-0.7}$    & $0.9^{+3.7}_{-0.6}$    & $0.9^{+8.5}_{-0.7}$    & $1.4^{+12.6}_{-0.8}$ \\
SFR, M$_{\sun}$/yr                & $87^{+343}_{-70}$      & $48^{+147}_{-41}$      & $49^{+155}_{-36}$      & $4.2^{+85.5}_{-1.5}$ \\
SSFR, $(\times10^{-10})$yr$^{-1}$ & $88^{+270}_{-65}$      & $55^{+173}_{-46}$      & $56^{+215}_{-49}$      & $2.9^{+73.6}_{-0.4}$ \\
  \hline
  \hline
  \end{tabular}
  \label{141212ahost}
\end{table*}
%

\begin{figure}
\centering
\includegraphics[width=\columnwidth]{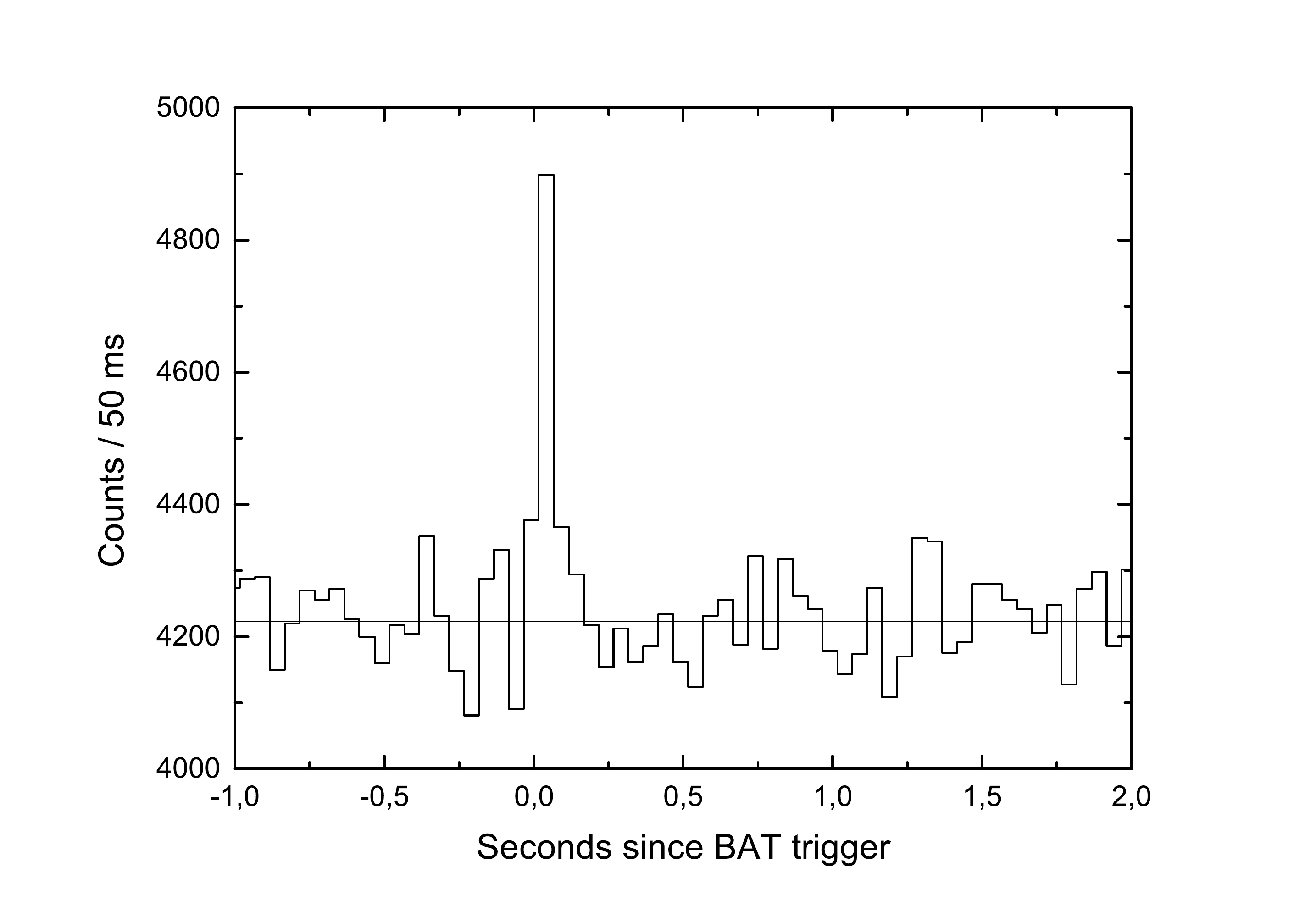}
\caption{\label{light} Light curve of sGRB 141212A from {\em INTEGRAL} SPI-ACS data in the energy range 0.1-10 MeV with 50 ms time resolution. The X-axis is time since BAT trigger, and the Y-axis is counts in 50 ms time bins. The thin horizontal line represents the background level.}
\end{figure}

\begin{figure}
\centering
\includegraphics[width=\columnwidth]{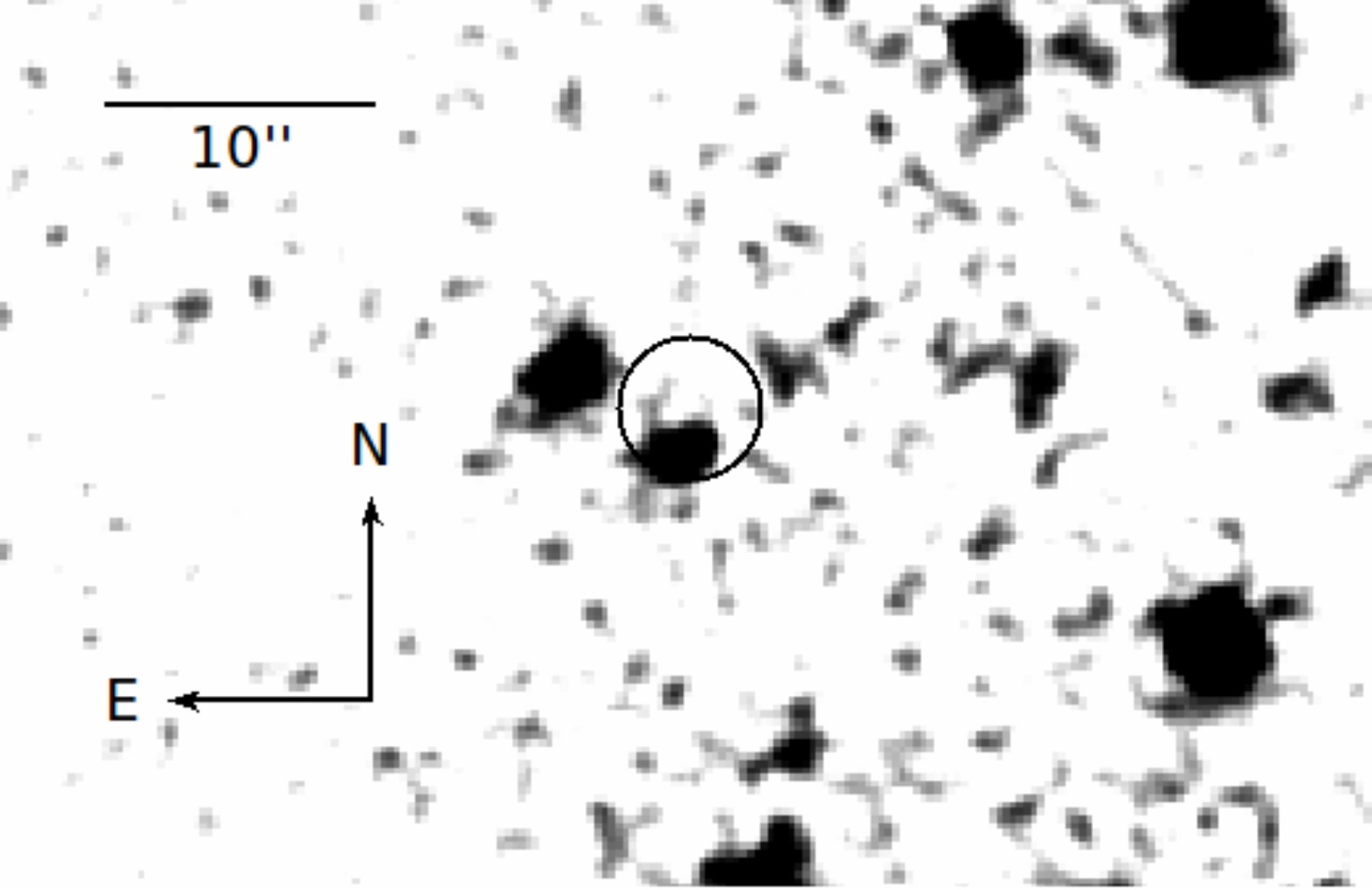}
\caption{\label{light} Finding chart of sGRB 141212A in the stacked frame of $r$ band data obtained with the GTC 10.4\,m telescope. The XRT error box is
shown as a black circle. The bright host galaxy is also visible within the XRT error circle.}   
\end{figure}

\begin{figure}
\centering
\includegraphics[height=8.0cm,width=8.0cm,origin=c]{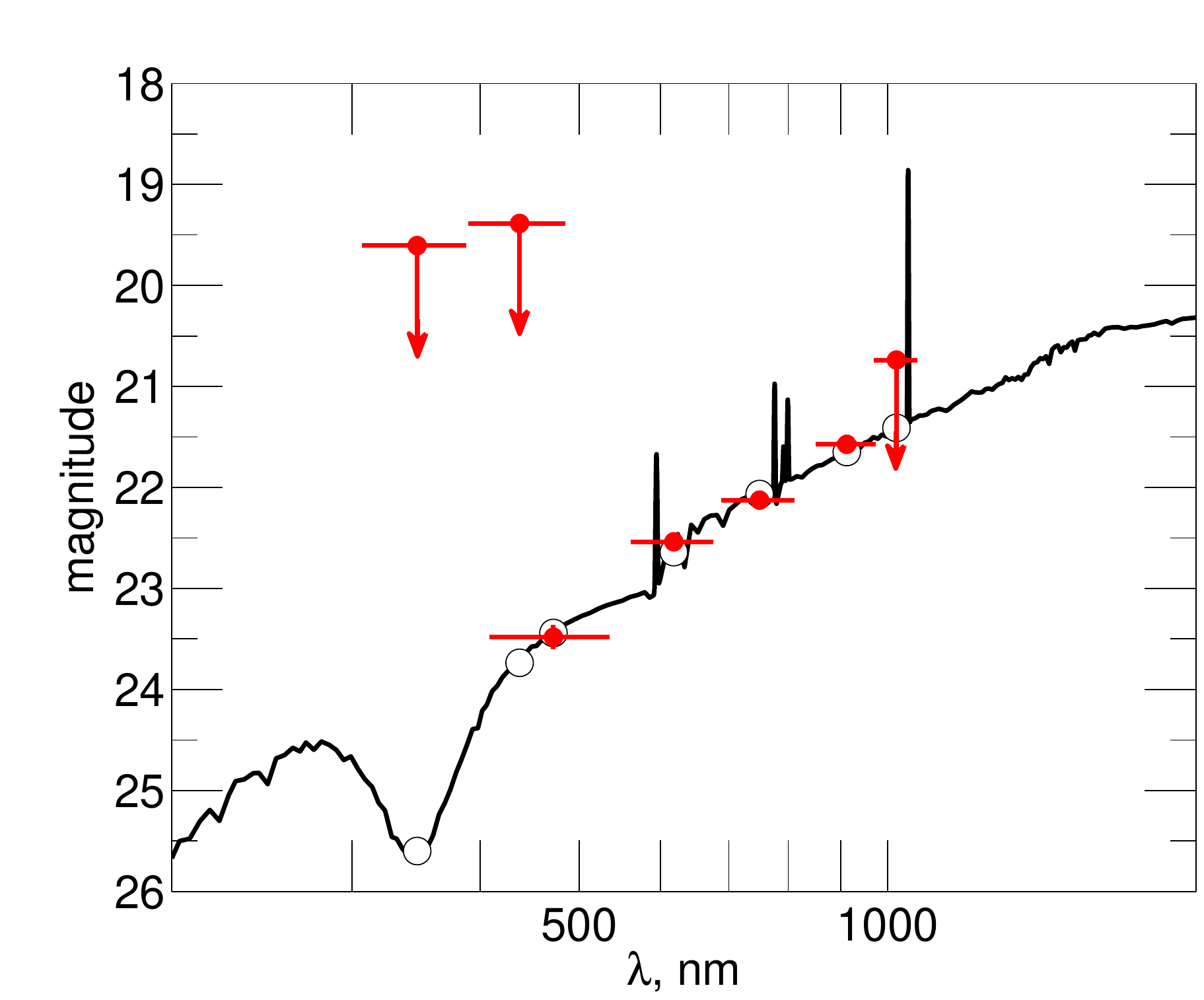}
\caption{\label{light} The SED (line) of the host galaxy of sGRB~141212A fitted by the \textsc{Le~Phare} with fixed redshift $z=0.596$. Filled red circles depict respectively the data points in the \textit{Swift}/UVOT filters $u, b$, taken from \citet{Oates2014}, $g, r, i$ from original observations (see Table 5), and $z, y$ from \citet[][, Table A1]{chrimes2018}. Open circles represent model magnitudes for each filter. All magnitudes are in AB system.
}
\end{figure}

\subsection{sGRB 151228A}

sGRB 151228A (trigger=668543) was discovered by {\it Swift} on 2015 December 28 at 
03:05:12 UT with a duration of $T_{90}$ = 0.27$\pm$0.01s \citep{Ukwatta2015, Barthelmy2015}. 
The burst was also detected by {\it Fermi}-GBM \citep{Bissaldi2015} but there was no {\it Swift}-XRT 
localization \citep{Page2015} due to an observing constraint. 
The burst was also detected by {\em INTEGRAL} SPI-ACS and triggered its IBAS system. The SPI-ACS 
light curve of sGRB 151228A is presented in Fig. A11 (top panel) and shows two overlapping pulses with 
a total duration of about $\sim$ 0.3 sec.
At a time scale of 50s, the upper limit on EE activity in SPI-ACS for sGRB 151228A is $\sim$ 7700 counts i.e.
S$_{EE} \sim$ $(7.7\times10^{-7}$ ~erg ~cm$^{-2}$) at the 3$\sigma$ significance level 
in the (75, 1000) keV range. As a part of the present analysis, {\it Fermi}-GBM data was fitted for the 
time-averaged spectrum of the NaI n4 data and was found that cutoff-PL model as the best fit. 
The low-energy photon index = 0.72$\pm$0.84 and E$_{\rm peak}$ = 261.18$^{+164.94}_{-58.28}$ keV, much lower than reported in \citet{Bissaldi2015}. The corresponding GBM flux
is (1.4$^{+1.39}_{-0.61}$) $\times10^{-6}\rm ~erg ~ cm^{-2}~ s^{-1}$  in 1-10$^{4}$ keV. 
The lightcurve of {\it Fermi}-GBM also have two overlapping pulses with a total duration of about $\sim$ 0.4 sec.
The spectral fitting plot with cutoff-PL model is shown in Fig. A11 
(bottom panel) \footnote{https: //fermi.gsfc.nasa.gov/ssc/data/analysis/rmfit/}.
As estimated in case of sGRB 140930B, we constructed the trajectory for sGRB 151228A on the 
Amati diagram (see Fig. 9), because the redshift z for sGRB 151228A was unknown. The trajectory 
lies above the main correlation at any z, which may suggest that sGRB 151228A belongs to the 
class of the short bursts. Since the burst does not fall into the E$_{\rm peak}$(1+z)/E$_{\rm iso}$ 
correlation region at any z, the redshift and E$_{\rm iso}$ of this burst cannot be estimated.

\noindent
Early optical searches within the BAT error circle 
do not find any new optical source down to a limiting magnitude of $\sim$17 mag using the 0.60m T60 telescope (TUBITAK National 
Observatory, Antalya - Turkey) starting 90 sec after the burst \citep{Sonbas2015}.  The GTC 10.4\,m was triggered
around $\sim$ 1.143d after the burst and covered the full error box in $i$ filter with a total exposure 
time of 5$\times$60 sec. The GTC observations cover the full BAT error circle, except for a gap 
between chips of a CCD camera (the gap covers $\sim$ 7.4\% of the total error box). 
The BAT error-circle was again observed by the GTC 10.4m in $i$ filter around 69d after the burst with 
a total exposure of 7$\times$75 sec. Due to different limiting magnitude, FWHM and inadequate flat-fielding 
for the whole FOV of the CCD camera we could not use image subtraction method to search for the source at 
the first epoch. Instead, we performed a catalog extraction at S/N = 3 for each epoch. We did not find any 
new object at the first epoch down to a limiting magnitude of $>$23.7 mag comparing with the second epoch 
(limiting magnitude for the second epoch was 24.8 mag). The results of our photometry and values of the limiting magnitude
for sGRB 151228A are reported in Table 5.

\begin{figure}
\centering
\includegraphics[width=\columnwidth]{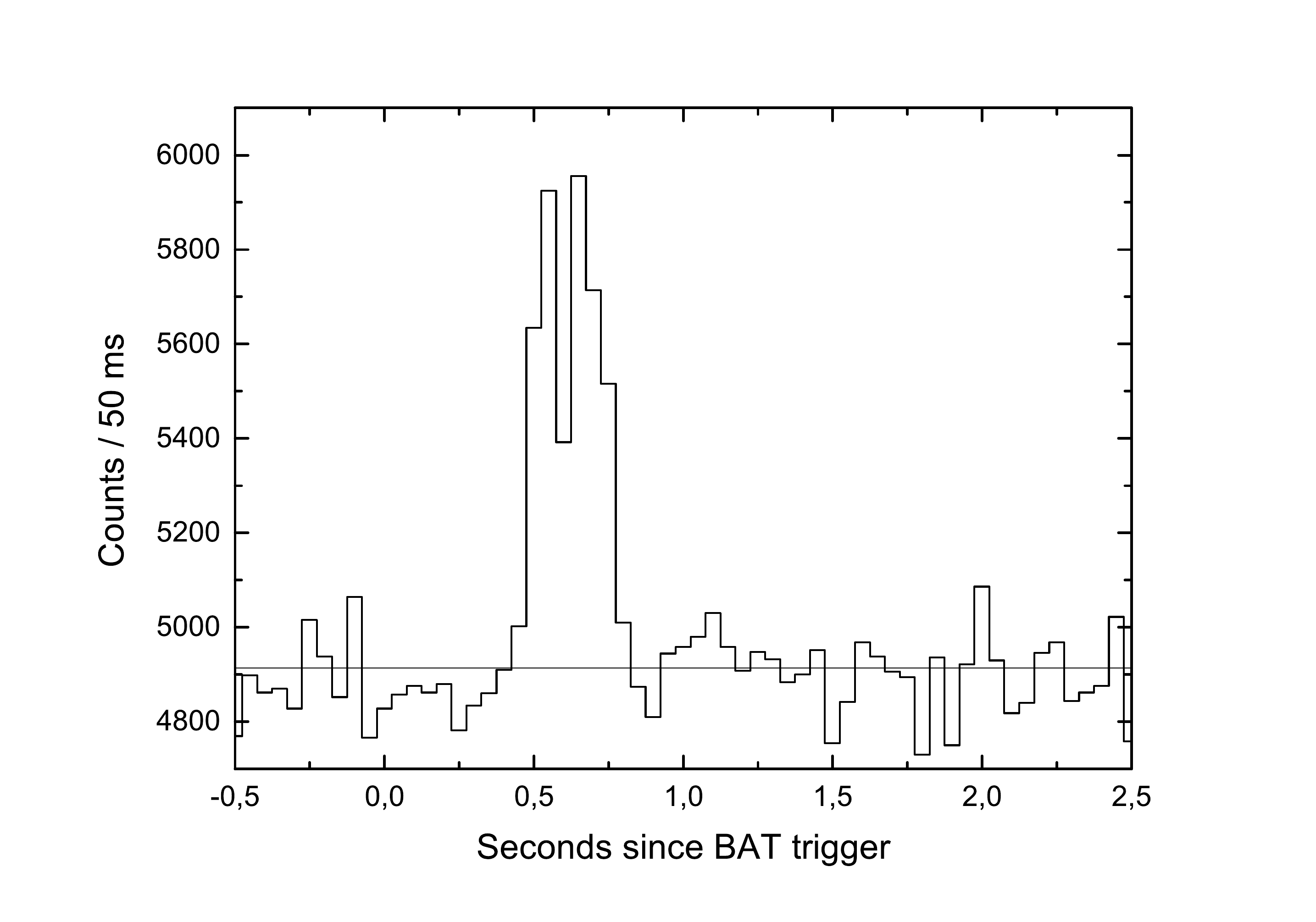}
\includegraphics[height=7.0cm,width=7.0cm]{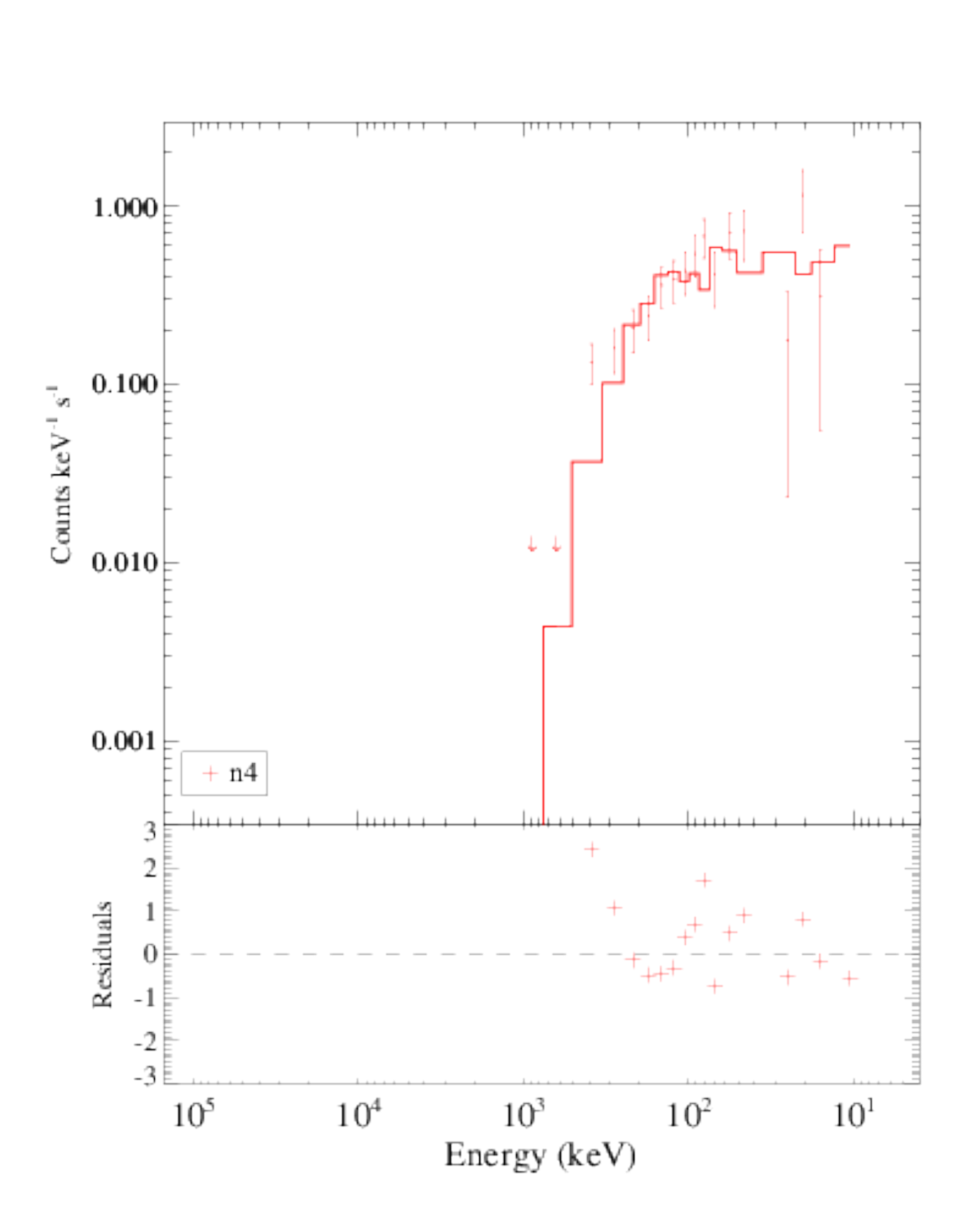}
\caption{\label{light} Light curve of sGRB 151228A from {\em INTEGRAL} SPI-ACS in the energy range 0.1-10 MeV
with 50 ms time resolution. The X-axis is time since BAT trigger, and the Y-axis is counts in 50 ms time bins
(top panel). The thin horizontal line represents the background level. The best fit model of the prompt
emission spectra of the {\it Fermi}-GBM (bottom panel) data of sGRB 151228A in counts.}
\end{figure}


\section*{Affiliations}
\small{
$^{1}$Aryabhatta Research Institute of Observational Sciences, Manora Peak, Nainital - 263 002, India.\\
$^{2}$Instituto de Astrofisica de Andalucia (IAA-CSIC), Glorieta de la Astronomia s/n, E-18008, Granada, Spain.\\
$^{3}$Unidad Asociada Departamento de Ingenier\'ia de Sistemas y Autom\'atica, E.T.S.  de Ingenieros Industriales, Universidad de Málaga, Spain.\\
$^{4}$Unidad Asociada Grupo Ciencias Planetarias UPV/EHU-IAA/CSIC, Departamento de Fisica Aplicada I, E.T.S., Universidad del Pais Vasco UPV/EHU, Bilbao, Spain.\\
$^{5}$Ikerbasque, Basque Foundation for Science, Bilbao, Spain.\\
$^{6}$Nikolaev National University, Nikolska 24, Nikolaev 54030, Ukraine.\\
$^{7}$Nikolaev Astronomical Observatory, Nikolaev, Ukraine.\\
$^{8}$Institute for Science and Technology in Space, SungKyunKwan University, Suwon 16419, Korea.\\
$^{9}$Astronomical Institute of the Academy of Sciences, Bo\v{c}n\'{\i} II 1401, CZ-14100 Praha 4, Czech Republic\\
$^{10}$Universidad de Ja\'en, Campus Las Lagunillas, s/n, Ja\'en, Spain.\\
$^{11}$Institute de Radioastronomie Millim\'etrique (IRAM), 300 rue de la Piscine, 38406 Saint Martin d'H\`eres, France.\\
$^{12}$Space Research institute, Moscow, Russia.\\
$^{13}$National Research Nuclear University MEPhI (Moscow Engineering Physics Institute), Moscow, Russia.\\
$^{14}$Thuringer Landessternwarte, Tautenburg, Germany.\\
$^{15}$Special Astrophysical Observatory, Nizhniy Arkhyz, Zelenchukskiy region, Karachai-Cherkessian Respublic, 369167, Russia.\\
$^{16}$Department of Physics, University of Adiyaman, 02040 Adiyaman, Turkey.\\
$^{17}$Instituto de Astrof\'sica de Canarias (IAC), V\'ia L'actea s/n, 38205 Santa Cruz de La Laguna, Tenerife, Spain.\\
$^{18}$School of Earth and Space Exploration, Arizona State University, Tempe, AZ 85287, USA.\\
$^{19}$NASA, Goddard Space Flight Center, Greenbelt, MD 20771, USA.\\
$^{20}$Radio Astronomy Laboratory of the Crimean Astrophysical Observatory, Katsiveli, Crimea. \\
$^{21}$Kharadze Abastumani Astrophysical Observatory of Ilya State University,
K. Cholokashvili Ave. 3/5, Tbilisi 0162, Georgia.\\
$^{22}$Ulugh Beg Astronomical Institute of the Uzbek Academy of Sciences, Tashkent, Usbekistan.\\
$^{23}$Fesenkov Astrophysical Institute, Almaty, Kazakhstan.\\
$^{24}$Astronomical Scientific Center,  Moscow, Russia.\\
$^{25}$Ussuriysk Astrophysical observatory of Far-East Department of Russian Academy of Sciences, Ussuriysk, Russia.\\
$^{26}$Kuban State University, Krasnodar, Russia.\\
$^{27}$Istanbul University Science Faculty, Department of Astronomy and Space Sciences, 34119, University-Istanbul, Turkey.\\
$^{28}$Instituto de Astronomia, UNAM, Unidad Academica en Ensenada. Ensenada 22860 Mexico.\\
$^{29}$University of Leicester, Department of Physics \& Astronomy and Leicester Institute of Space \& Earth Observation, University Road, Leicester, LE1 7RH, UK.\\
$^{30}$Astronomy Department, University of California, Berkeley, CA 94720-7450, USA.\\
$^{31}$Department of Astronomy and Astrophysics, University of California, 1156 High Street, Santa Cruz, CA 95064, USA.\\
$^{32}$Institute of Solar-Terrestrial Physics, Irkutsk, Russia.\\
$^{33}$Yunnan National Astronomical Observatory, Chinese Academy of Sciences, Phoenix Hill, 650011 Kunming, Yunnan, China. \\
$^{34}$Beijing National Astronomical Observatory, Chinese Academy of Sciences, 20A Datun Road, Chaoyang District Beijing 100012, China.\\
$^{35}$Center of Astronomy and Geophysics Mongolian Academy of Sciences, Ulaanbaatar, Mongolia.\\
$^{36}$National Research University Higher School of Economics, Moscow, 101000, Russia.\\
$^{37}$Keldysh Institute of Applied Mathematics, Moscow, Russia.\\
$^{38}$INAF - Istituto di Astrofisica e Planetologia Spaziali, Via Fosso del Cavaliere 100, 00133 Roma, Italy.\\
$^{39}$ Key Laboratory of Modern Astronomy and Astrophysics, Nanjing University, Ministry of Education, Nanjing, 210093, China.\\
$^{40}$School of Astronomy and Space Science, Nanjing University, 210093, Nanjing, China.\\
$^{41}$Department of Physics, University of Warwick, Coventry, CV4 7AL, UK.\\
$^{42}$Department of Physics, University of Bath, Claverton Down, Bath BA2 7AY, UK.\\
$^{43}$Samtskhe Javakheti State university, Rustaveli st.113, Akhaltsikhe, 0080, Georgia.\\
$^{44}$Instituto de Astronom\'ia, Universidad Nacional Aut\'onoma de M\`exico, Apartado Postal 70-264, 04510 M\'exico, CDMX, Mexico.\\
}

\end{document}